# The State of AI Ethics Report

# Volume 4

MAI3I

April 2021

This report was prepared by the **Montreal AI Ethics Institute (MAIEI)** — an international non-profit organization democratizing AI ethics literacy.  **Learn more on our [website](#) or subscribe to our weekly newsletter [The AI Ethics Brief](#).**

This work is licensed open-access under a **[Creative Commons Attribution 4.0 International License](#)**.

Primary contact for the report: **Abhishek Gupta ([abhishek@montrealethics.ai](#))**

**Full team behind the report:**

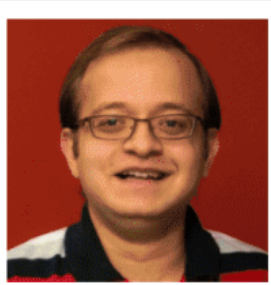
Abhishek Gupta
FOUNDER, PRINCIPAL RESEARCHER, AND DIRECTOR

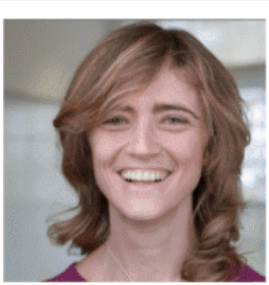
Marianna Ganapini, PhD
FACULTY DIRECTOR

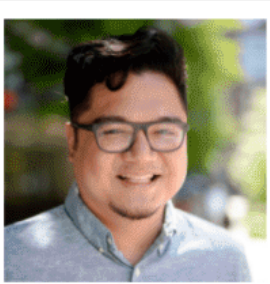
Renjie Butalid
CO-FOUNDER AND DIRECTOR

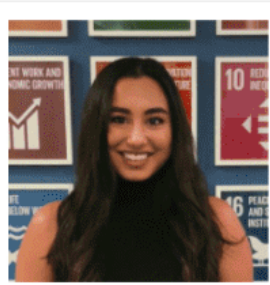
Muriam Fancy
NETWORK ENGAGEMENT MANAGER

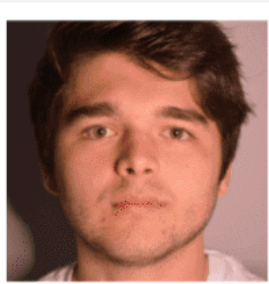
Connor Wright
PARTNERSHIPS MANAGER

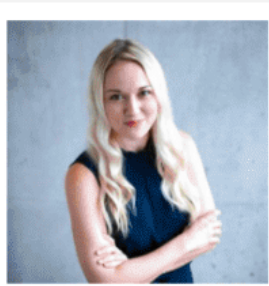
Victoria Heath
ASSOCIATE DIRECTOR, GOVERNANCE & STRATEGY

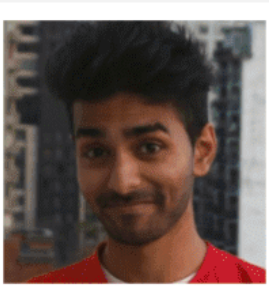
Mo Akif
DIRECTOR OF COMMUNICATIONS

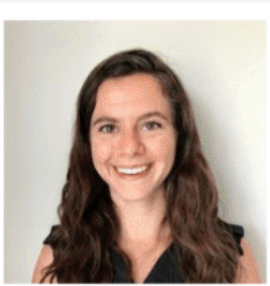
Alexandrine Royer
EDUCATIONAL PROGRAM MANAGER

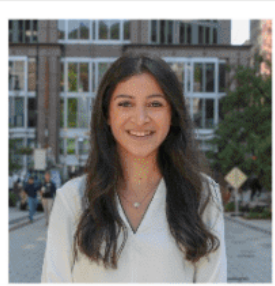
Masa Sweidan
BUSINESS DEVELOPMENT MANAGER

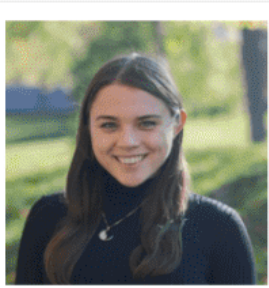
Shannon Egan
RESEARCH INTERN (QM)

**Special thanks to external research summary contributors:**
Nga Than, Iga Kozlowska, Sarah Grant, Charlotte Stix, Matthijs Maas, Falaah Arif Khan, Erick Galinkin, Andrea Pedeferri

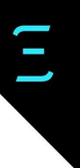

# Table of Contents

---

**\*Note:** The original sources are linked under the title of each piece. The work in the following pages combines summaries of the original material supplemented with insights from MAIEI research staff, unless otherwise indicated.

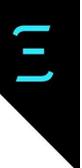

# Founder's Note

Welcome back to another edition of the State of AI Ethics Report from the Montreal AI Ethics Institute! It seems like yesterday that we were conversing through these pages on the most impactful reporting and research in the field of AI ethics. Yet, a quarter has already whizzed past us. This last quarter was filled with many exciting developments, some of which reflect the call to move from principles to practice. In fact, we're in the midst of a monumental moment: not only is there growing convergence on the universal principles for AI ethics but there is also tremendous progress in operationalizing those principles.

In this edition, we open with a broad view of "Ethical AI", addressing topics like "ethics owners" as a way of thinking about organizational responsibility, technical and political challenges in algorithmic content moderation (certainly an important notion given all that has transpired online this past quarter), and tackling questions like whether algorithmic audits can help eliminate bias. We dive deeper into algorithmic bias with the chapter on "Fairness and Justice", exploring areas like how AI systems might be designed for the Global South (akin to an article I co-authored with Victoria Heath, titled [AI ethics groups are repeating one of society's classic mistakes](#)), a framework for undoing technology's gender troubles, and a comprehensive overview on "bad" algorithmic behaviour from The Markup.

Let us pause for a moment and consider what the purpose of technology is. And why we tend to forget the people for whom we are building technology? In my recent TEDx talk titled [Civic Competence Against the Invisible Hand of AI](#), I call on industry and non-industry members alike to actively question the technology that surrounds us by examining its societal impacts. In the "Humans and Tech" chapter of this report, we do just that. It looks at risk-taking behaviours and human-robot interactions, as well as ideas like what Buddhism can do for AI ethics, with all contributions shedding light on the intersection of humans and technology.

There hasn't yet been an edition of the report where privacy hasn't been an important issue. The chapter on "Privacy" encompasses a wide-range of privacy-related issues, including the frothy Clubhouse app, the use of facial recognition technology in the workplace, and the complex global technology flows that empower authoritarian regimes in monitoring their populace.

The harms posed by the use of AI manifest in different ways. Our final chapter titled "Outside the Boxes" explores a myriad of tech-related issues like the environmental impact of algorithms,

AI safety, security, and stability among the great powers, algorithmic management among gig workers, and the notion of *unadversarial* examples as a way to strengthen machine vision.

You, members of our community , are addressing these challenges in insightful ways, and we wanted to shine a spotlight on some of your work in this report. Our featured community members, drawing on their own work, touch on the complexities of aligning AI to human values, how to bring in more people from marginalized groups into the AI workforce, and finally what it means to build AI for the right purpose.

We continue to work in the interest of our global community and invite you to comb through these pages. The chapters reveal encouraging developments in addressing ethical challenges in designing, developing, and deploying AI systems, despite the new problems rearing their ugly heads. We hope you can carry the ideas and messages from these pages to your colleagues, spark discussions with the community around you, and ultimately take action, no matter where you work, to better the State of AI Ethics! *Carpe Diem!*

---

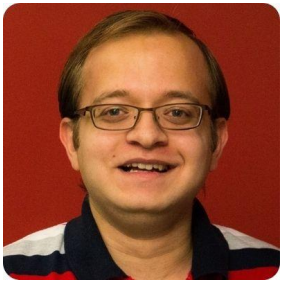

**Abhishek Gupta (@atg_abhishek)**
Founder, Director, & Principal Researcher
Montreal AI Ethics Institute

Abhishek Gupta is the founder, director, and principal researcher at the Montreal AI Ethics Institute. He is also a machine learning engineer at Microsoft, where he serves on the CSE Responsible AI Board.

# 1. AI and the Face: A Historian's View by Edward Higgs (Professor of History, University of Essex)

**Introduction**

Viewing the contributions to the briefings of the Montreal AI Ethics Institute, it is interesting to see the different disciplines that are involved in the debates regarding the pros and cons of artificial intelligence applications – anthropologists, sociologists, ethicists, as well as computer scientists. Contributions from historians are mostly conspicuous by their absence.

At first sight, the obvious reason for this is that historians have little to contribute. After all, aren't AIs part of the digital Brave New World sweeping away former ways of doing things, and the necessity to refer to a dusty, redundant past? What interests me as a historian, however, is the way that some AIs making decisions about people and the psychometric research on which they are based, repeat the biases and failures of past human systems.

As a historian of the face, and how it has been used in former centuries to identify people and to ascribe characteristics and emotions to them, recent problems identified with facial recognition systems, algorithms to detect criminal physiognomies, affect technologies, and so on, all sound depressingly familiar.

What I have to say on these subjects mostly reflects the history and ideas of the West. Although this is a limitation, it can be justified because the roots of the design of many AI systems lie in Western science and technology, as does much of the discussion around the ethics of artificial intelligence. Hopefully, people from other societies can use my comments as a starting point for their own cross-cultural comparisons.

**Data issues in the contemporary field of AI and in the past**

Some of the problems with AI systems can be seen, at least from one perspective, due to the deficiencies in data inputs. For example, Joy Buolamwini and Timnit Gebru, amongst others, have pointed to the problems that commercial facial recognition systems trained on white faces have in handling data from the faces of people with darker skin, especially women.

Similarly, in 2016 two Chinese scholars, Xiaolin Wu and Xi Zhang, published a paper in which they claimed to have developed a system that could identify 'criminal' facial features after comparing photographs of convicted criminals from police records to those of 'normal' citizens. The paper led US researchers Carl T. Bergstrom and Jevin West to argue, in a pungently entitled presentation 'Calling Bullshit in the Age of Big Data', that the Chinese scientists had merely invented a 'smile detector'. This was because Wu and Zhang had compared the images of glum detainees with the jollier faces displayed by other Chinese people when presenting themselves on the web. Garbage in, garbage out.

Data problems also bedevilled past attempts to identify 'criminal' types from faces and use bodily measurements to identify unique individuals. For example, in his *L'uomo delinquent* (*Criminal Man*), Cesare Lombroso — the late nineteenth-century father of Italian criminology, believed that he could identify a criminal physiognomy from the photographs and representations of criminals that he collected. But all he was doing was picking out the features of the faces of convicted criminals that caught his eye (jug ears, low forehead, thin beards, and so on) in a rather random fashion and then claiming that they were diagnostic of criminality. He also argued that female prostitutes had deep, 'masculine' voices, although he did not reveal the source of his data, whether from personal experience or otherwise, for this peculiar claim.

A more rigorous use of bodily data can be seen in the 'anthropometric' system of identification developed by Alphonse Bertillon, a clerk in the Paris Police Prefecture, in the 1880s. Anthropometrics involved taking numerous measurements of the faces and bodies of convicted criminals, using specially designed equipment.

The resulting measurements, a sort of non-digital data profile, were stored in a special archive and could be consulted when criminals who refused to give their names were subsequently re-apprehended and re-measured. The similarities with modern facial recognition systems are obvious. This system spread rapidly to the rest of Europe, the USA, and then to the French colonies and South America, where it was used to identify the general population rather than just criminals.

However, there were basic problems with Bertillon's data – the bodily measurements of growing children and youths were not fixed, and there was always a problem with touching women during measurement, although this animus against 'dishonour' did not, of course, extend to colonial populations. Additionally, 'Bertillionage' required the meticulous training of police staff to achieve accurate measurements, and there was evidence that even in optimum conditions, the bodily measurements taken by two separate operatives could differ so much as to make reliable identification difficult. In the early twentieth century, the anthropometric system was quickly replaced by fingerprinting, which was less intrusive, easier to perform, and more accurate.

Given the data problems with these early forms of identification, the obvious question to ask is why they were so enthusiastically seized upon by criminologists and police forces alike. Historians have suggested that the primary role of these techniques was not actually to identify people but to give the forces of law and order an aura of professionalism by associating them with notions of scientific and technological progress. The police were no longer to be seen as corrupt, or incompetent Keystone Cops, but as 'expert' crime fighters. I wonder to what extent the contemporary enthusiasm for the involvement of AIs in identification stems from a similar desire to appear to be at the forefront of high-tech modernity?

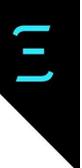

**Identifying types**

One of the aspects of research using AI systems that intrigues me is the way in which they start from the premise that a certain type or category exists, without any empirical proof that they are in fact, real. A case in point is Wang and Kosinski's contentious claim that they have developed an AI system trained on the photographs of people advertising on a dating site that can [determine if a person is straight or gay](). Nose shape and the width of the chin are claimed to be diagnostic for men, and the wearing of baseball caps seems to be a dead giveaway for lesbians. They claim that even looking at facial features alone, and ignoring grooming and clothing styles, they can still detect sexual orientation. But keep in mind that the jaw underlying the chin, for example, is not simply a bone but [has deep ideological meanings]().

Other researchers have pointed out problems with the data used but another issue is that Wang and Kosinski appear to assume that sexuality is, in some sense, a bipolar distribution – people are either gay or straight. However, much research into sexuality shows that it is polyvalent, and is not necessarily constant over time.

Similar issues arise in the research of those who claim that they have identified the ['criminal face'](). As I have already noted, the data used to train such AI systems is problematic, compromised by such things as racial and other biases in criminal conviction rates. However, why should one imagine that there is a criminal face in the first place, especially given that the definition of crime can change over time – do the signatories of the US Declaration of Independence have criminal faces because the majority (41 out of 56) owned slaves?

The belief in the existence of a criminal physiognomy may reflect, in part, the known tendency of people to ascribe positive characteristics to those they find physically attractive, and negative characteristics to the 'ugly'. This idea was enshrined in the thought of ancient Greek thinkers such as Plato and Aristotle as the principle of *kalokagathia,* in which physical beauty was understood as having order, symmetry and moderation, which also applied to moral [beauty]() and virtue. Of course, for Western imperialists, the faces of subject populations in Africa and India were seen as unattractive and therefore associated with a lack of virtue, and indeed with criminality.

Cesare Lombroso certainly believed that non-Western people had criminal tendencies because he saw the criminal type as an evolutionary throwback to earlier forms of humanity, which included 'primitive races'. Some modern criminologists have [tried to rehabilitate Lombroso]() by placing his idea of the criminal type in the context of the inheritance of 'antisocial' DNA lineages by individual criminals.

However, this obscures the basis of Lombroso's claims, which lay in the embryology of the species, rather than the genetic inheritance of individuals. As Charles Darwin had noted in *The*

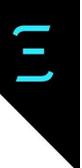

*Origin of Species*, the human embryo appears to recapitulate past stages of evolution in its development, even to the extent of having fish-like proto-gills at one point. Darwin's cautious comments were developed into a full-scale theory of human evolution by his German follower Ernst Haeckel, who saw embryological development as a strict route map of evolution. For Haeckel and Lombroso 'savage' early humanity, the subject populations of Empire, and criminals, simply represented earlier stages in this human embryological development.

Another major proponent of scientific racism was the Victorian polymath Francis Galton. Galton believed that if he layered photographic images of criminals one on top of another to create 'blended' or 'composite' images, he could create a representation of the criminal 'type'. In [his approach to composite photography](#), Galton drew on his undeniably important work on the statistics of correlation, normal distributions, and the standard deviation. Galton reasoned that the bolder features of his composite criminals represented the central tendency of the criminal facial type and that the fuzzy aura of less defined features about the central image represented less common features in tails spreading out from the centre of the 'bell curve'. He was so enamoured with the idea that he went on to produce composites of those suffering from TB and Jewish children, believing that these revealed images of the 'diseased' and racial types.

Of course, they did nothing of the sort. His criminal composites showed the faces of quite ordinary people, and he had to fall back on arguing that these faces represented the criminal type *before* a life of crime had criminalised their features. He never stopped to ask if the face of the criminal 'type' was due to the roughening effects of poverty and a criminal life, rather than that being a criminal type predisposed someone to a criminal existence. He made similarly specious claims respecting the diseased 'type', and the nature of race. The more I read Galton, the more I am struck by his obtuseness.

He extended these ideas to his theory of phenomenology. Human beings, he reasoned, formed composite or blended images of types in their minds. Thus, by looking at lots of dogs they formed a blended concept of the common features of 'dogness', to which they gave the label 'dog', and uttering that term brought to mind the blended image. However, he failed to recognise that people would already need to have a notion of what a dog was *before* they could select the appropriate images to blend. There was no point in blending images of cats or cabbages in with those of dogs to create a notion 'of dogness'. Again, he assumed the type from the outset. This is a problem I would suggest that is found in many statistical exercises, when people assume that there must be an underlying thing (such as 'intelligence' or 'criminality') represented by the central tendency of their distributions when that central tendency is really an artifact of their initial categorisations.

**The history of emotion detection and its impact on affect AIs**

There are occasions when it is possible to follow the development of the models underlying AI applications through time – what the French thinker Michel Foucault called an 'archaeology of knowledge'. This I would argue can be seen in the case of the AI systems that are claimed to read emotions from expressions, and which are being introduced for recruitment, marketing, the surveillance of classrooms, and possibly even at EU border control.

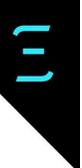

The contemporary model of expressions underlying such applications, the Basic Emotion Theory (BET), assumes that there is a unitary person who has a series of simple emotions, or feelings, that lead directly to a set of universal facial expressions that are generally understood by all other human beings. This model has been criticised by the likes of Lisa Feldman Barrett, since here is evidence that people in various cultures understand expressions differently, that not all emotions lead to expressions, and that other factors than the expressions themselves, such as context, need to be taken into consideration when interpreting them. However, the simple BET model of expressions and their relationship to emotions has a very long history in the West.

In some ways, once again, it's all the fault of Aristotle. For the ancient Greek philosopher the form of the body was an element of the human soul. However, this soul was not the later Christian soul – the immaterial essence of the person, which is separate from the body, and can be judged by God for its sins. Rather, for Aristotle the soul was in part the immanent potential of the human body. It was what determined that when two human beings mated, the offspring took the form of a human being, and not a chicken. We might think of it today in terms of DNA, but for Aristotle it was the teleological principle that gave the body an inherent form into which it grew. He also thought that rather than optics being based on bodies reflecting light, all objects were continuously giving off simulacra or representations of their forms, which travelled through the aether and entered the eye. Interestingly, ancient Greek understandings of optics are still current in the West today.

Since the Aristotelian soul was the form and organising principle of the human body, there was no conceptual, or physical, gap between the soul, the person, the emotion, and the facial expression. What affected the soul would naturally affect the form of the body, and vice versa. The understanding of facial expressions was also universal amongst human beings, since the human soul was also universal. Those who could not understand the emotions of other human beings, were simply not human, because it was the *telos,* or inherent purpose or end, of human beings to communicate – those who didn't were either beasts or gods. Since for Aristotle the act of seeing involved the projection of the forms of bodies from objects into the eye through the ether, there was hardly any way in which humans could not see the soul of others directly. There was thus no question that anything other than communication between souls was involved in the understanding of emotions via expressions.

Despite Aristotle's continued prestige, this formulation became problematic in the Christian era when the body and the soul were seen as distinct entities, could exist independently, and might even be in opposition. This body-spirit duality inherent in Christianity was taken to its ultimate conclusion by the seventeenth-century French philosopher René Descartes, for whom the physical world and the human soul were completely distinct. This implied that since animals had no human souls they were simply machines, or automata, activated by God's will. How then could human beings exist both as spiritual and material beings, and how could states of the soul be communicated?

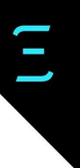

Descartes answered these questions by replacing the form of the body with the movements of the muscles of the face and body as signifying the hidden motions of the soul. Descartes and his followers posited that God had so designed Creation that external objects had 'dispositions' that set up various kinds of motions in the body's nerves, which set up sensations in the soul via the pineal gland in the brain. Similarly, the soul's act of willing set up motions in the pineal gland, which in turn moved 'animal spirits' to the nerves, which moved the limbs and facial muscles. Descartes hit upon the pineal gland as the site of this interchange because it was a single, rather than a binary, organ in the brain, and therefore indivisible like the Christian soul. So although the soul was no longer the bodily form as in Aristotle, expressions were still universal because the human soul was universal. Expressions could be understood by other humans, and were one-to-one communications, because that is how God intended it.

This way of understanding expressions survived into the nineteenth century in the work of neurologists such as Sir Charles Bell and Guillaume-Benjamin-Amand Duchenne de Boulogne. However, the notion of expressions as a Divine dispensation to enable souls to communicate was swept away in 1872 by the publication of Charles Darwin's *The Expression of the Emotions in Man and Animals.* For Darwin, human expressions were strictly natural reactions to internal stimuli and evolved from the facial movements of animals. However, Darwin was keen to preserve the notion of a universal language of ubiquitous expressions because he wanted to use this as evidence that all human beings were one species (monogenesis). This was a repost to the racist notion that human 'races' were separate 'creations' (polygenesis). In many ways, Darwin retained the older model of expressions but merely replaced God by Evolution, and the Soul by the Mind.

Darwin's ideas were largely ignored until they were resurrected by Paul Ekman in the 1960s. In 1973 Ekman edited *Darwin and Facial Expression: a Centenary of Research in Review* to mark the 100th anniversary of the publication of Darwin's *Expression,* and he provided an introduction and commentaries to an edition of that work in 1998. He was defending Darwin's work [as late as 2017](). Influenced by Darwin and Silvan Tomkins, Ekman went on to develop the modern Basic Emotion Theory, and the standard taxonomy of distinct movements in facial muscles which could be used to describe expressions, the Facial Action Coding System, or FACS. The BET and FACS then underlie the contemporary development of affect AIs, which purport to read emotions in the face. Aristotle would, therefore, feel at home amongst contemporary AI systems, such as Amazon Rekognition, Microsoft Azure, and Affectiva, even if they have no soul.

**Conclusions**

There are, of course, other ways in which the deployment of contemporary facial AI systems has eerie parallels with the past, as in the resurgence of classical physiognomy (Aristotle again) [in digital form](), or the stigmatisation of the migrant [as a threat]() in facial recognition systems. However, I hope that I have done enough to show how an understanding of the past can show:

1. How some contemporary AI developments are the culmination of histories going back centuries, if not millennia;

2. That their failings can reflect problems of research design and data input that have revealing precedents in the past

3. How the long history of human attempts to 'read' the face provides a context in which the dubious application of psychometric research in AI systems can be seen as 'common sense.

Are those who do not understand their history doomed to repeat it, even if in digital form?

---

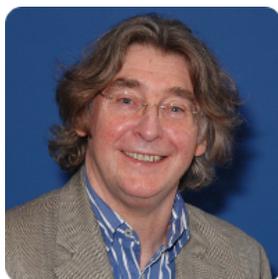

**Edward Higgs**
Professor of History
University of Essex

Edward Higgs studied modern history at the University of Oxford. He completed his doctoral research there, a quantitative analysis of nineteenth-century domestic service, in 1978. He was an archivist at the Public Record Office, now the National Archives, in London from 1978 to 1993, where he was responsible for policy relating to the archiving of electronic records. He was a senior research fellow at the Wellcome Unit for the History of Medicine at the University of Oxford from 1993 to 1996, and a lecturer at the University of Exeter from 1996 to 2000. He then joined the History Department at the University of Essex, eventually becoming Departmental Head, before becoming a Professor Emeritus in 2018.

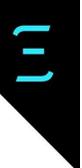

# 2. Ethical AI

**Opening Remarks** by Dr. Alexa Hagerty (Anthropologist, University of Cambridge)

As spring arrives in the Northern hemisphere, revelations about research practices at two of the world's most powerful technology companies are providing sobering lessons in our ongoing education about "AI ethics."

The fallout from Google's expulsion of Dr. Timnit Gebru and Dr. Margaret Mitchell continues in many forms including the decision by the ACM Conference for Fairness, Accountability, and Transparency (FAccT) to suspend its sponsorship relationship with Google and the public refusal of a research award from the company by respected AI ethics researcher Dr. Luke Stark.

AI ethics research at Facebook has also come under scrutiny due to journalist Karen Hao's article about how the company's responsible AI team focuses on problems of algorithmic bias while deflecting attention from misinformation. In this carefully researched piece, Hao argues that Facebook's models favour growth and engagement above all else, even when driven by forms of polarization and extremism that can have devastating real-world consequences.

As we reflect on the implications of these recent revelations, our understanding of AI ethics grows deeper, more layered, and more complex. We have graduated from an AI ethics centred on abstract principles and toy problems to deeper considerations of political economy. Now we must turn our attention to AI ethics research and advocacy itself and to the ecosystems of power in which this work is carried out.

As Hao posted on Twitter, "It's not about corrupt people do corrupt things. That would be simple. It's about good people genuinely trying to do the right thing. But they're trapped in a rotten system, trying their best to push the status quo that won't budge." We must interrogate the AI ethics status quo—including publication and review norms, PR campaigns, pressure on frontline researchers and journalists, and research funding—among many other issues.

One important aspect of this work is looking at the research methods we bring to the study of AI ethics. In a recent commentary in Nature Machine Intelligence, Vidushi Marda and Shivangi Narayan argue that qualitative methodologies are uniquely positioned for contextual analysis of systems and understanding the social justice implications of AI. As an anthropologist, I agree wholeheartedly with the authors about the critical role ethnographic methods can play in examining "power hierarchies and asymmetries."

But research methods can do more than examine power. Participatory methods—in which communities are directly involved in the research process——can be a tool to shift power. The harms of AI systems are unevenly distributed within and across societies, and have a pattern of amplifying existing social inequalities, with disparate impacts on communities of colour and other communities marginalised by systemic exclusions and oppressions.

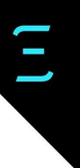

As the work of researchers and organisations like the Montreal AI Ethics Institute has shown, and as the articles and research summaries presented here reflect, AI systems have a range of societal impacts on issues including discrimination, labour conditions, human rights, misinformation, and democracy.

Communities affected by AI systems must have meaningful engagement in identifying and assessing algorithmic harm, as well as in decisions about what kinds of AI technologies are developed in the first place, when non-technological solutions should be preferred, when and how technologies are used, and how they are governed, including when they should be banned.

There is increasing interest in participatory methods in AI ethics research. The Ada Lovelace Institute has been convening a [Citizens' Council](#) to consider the use of biometrics technologies, the ethical questions they raise, and how they should be regulated. The University of Washington's Tech Policy Lab has developed a "[Diverse Voices](#)" toolkit for community feedback on policy documents. The [Sepsis Watch](#) project at Duke University Hospital took a [participatory, sociotechnical approach](#) to using AI in patient care. As part of a [Nesta Collective Intelligence](#) program, I've been working with a team on an arts-based citizen social science project called emojify.info that lets the public [try an emotion recognition system](#) for themselves and give their perspective on this technology's potential societal impacts.

Participatory methods have a [proven track record](#) of [promoting equity, empowering communities, generating new ideas](#), and solving difficult problems. However, they also present challenges. In *Design Justice*, Sasha Costanza-Chock warns that if research lacks accountability and community control, participatory methods can be extractive. Mona Sloane and colleagues point to the dangers of "[participation washing](#)"—a spectrum of ways in which participation can be misrepresented, exploitative, and co-opted. These concerns are not new. In a [foundational paper](#) written in 1969, researcher Sherry Arnstein wrote: "there is a critical difference between going through the empty ritual of participation and having the real power needed to affect the outcome of the process."

Participatory approaches are not a panacea, but when used with skill and integrity, they are one of the best tools we have for shifting real power to communities. When communities affected by AI systems have meaningful engagement with all stages and aspects of AI design—including AI ethics research—we can build more just, equitable and sustainable technologies and societies.

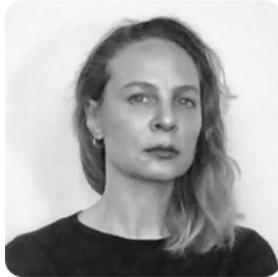

**Dr. Alexa Hagerty ([@anthroptimist](https://twitter.com/anthroptimist))**
Anthropologist
University of Cambridge

Alexa Hagerty is an anthropologist and scholar of science, technology and society studies (STS) investigating societal impacts of AI/ML systems using ethnographic, arts-based, and participatory methods. Based at the University of Cambridge, Leverhulme Centre for the Future of Intelligence and Centre for the Study of Existential Risk, her research focuses on affected communities, equity, and human rights. She is also a member of the JUST AI network at the Ada Lovelace Institute and a Fellow at the World Economic Forum Global Future Council on Systemic Inequalities and Social Cohesion.

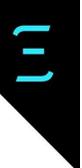

# Go Deep: Research Summaries

## "Cool Projects" or "Expanding the Efficiency of the Murderous American War Machine?"

([Original paper](#) by Catherine Aiken, Rebecca Kagan, Michael Page)
(Research summary by Abhishek Gupta)

**Overview:** This research study seeks to glean whether there is indeed an adversarial dynamic between the tech industry and the Department of Defense (DoD) and other US government agencies. It finds that there is wide variability in perception that the tech industry has of the DoD, and willingness to work depends on the area of work and prior exposure to funding from and work of the DoD.

---

**Key findings**
- Most AI professionals are actually positive or neutral on working with the DoD. This is in stark contrast to what the common media portrayal is of the attitudes of workers in the tech industry.
- Doing good and working in interesting research areas were the most compelling reasons to engage with the DoD.
- Discomfort primarily arose from lack of clarity on how the DoD might use their work and potentially using the work to cause harm.
- Unsurprisingly, people were more willing to work on humanitarian projects compared to war efforts or back-office applications (the back office attitude was surprising to me).
- If the funding provided by the DoD is used solely for basic research, a lot of people were willing to engage.
- Academia and their own employers got the highest trust ratings when it came to whether AI will be developed with the public interest at heart.
- Those that had prior exposure to the funding and work from the DoD have more positive viewpoints on the DoD. I'll talk a bit more about this later, but I think there is an unaddressed bias that might need to be thought about in this context.
- Ultimately, there isn't a binary framing that characterizes the relationship between the tech industry and the DoD as is commonly portrayed in the media, and depending on the context, we get varying results in terms of the willingness of people to engage with this kind of work.

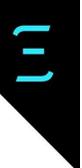

- In recent years, industry and non-industry members have decried the prevalence of biased datasets against people of colour, women, LGBTQ+ communities, people with disabilities, and the working class within AI algorithms and machine learning systems. Due to societal backlash, data scientists have concentrated on adjusting the outputs of these systems. Fine-tuning algorithms to achieve "fairer results" have prevented, according to Denton et al., data scientists from questioning the data infrastructure itself, especially when it comes to benchmark datasets.

**What is the key problem this study is trying to address?**

There are many conflicting narratives on the willingness to engage and attitudes towards the DoD from the tech industry. To clarify these and gain a better understanding for what the reality is and which factors shape that will help to better bridge the gaps between US government agencies and the tech industry.

The DoD in particular has the ability to drive large-scale changes by virtue of its funding and market-making power which means that there are a lot of scientific advances that academia and industry might be unwilling to investigate that if funded by the DoD (when people are willing to engage with them) can lead to large societal benefits. Essentially, having a productive, open, and honest dialogue between the two is essential to leverage the potential opportunities.

**Survey questions**

The questions were centred around trying to elicit the attitudes that people had towards the DoD, what their response would be to a hypothetical project in terms of engagement and employer relationship, what might be some factors that can shift their perceptions, and finally their understanding and perception of the different US agencies and trust in the political instruments in the US.

A caveat that the authors point out is that they had a fairly low response rate (~4%) to the survey that they used and as such don't provide guarantees on the representativeness of the general population and call for further research to build upon the results that they obtain in this study.

The study also ended up sourcing practitioners who self-identified as AI practitioners online and were mostly from the major tech hubs like Boston, SF, Seattle, etc. so perhaps not fully representative of all the places where such collaborations might be taking place.

**Discussion of results**

One of the most interesting findings for me from this study was the vast difference in positive perception of the DoD when people had prior exposure to working with them.

- This might be a case where people genuinely through their interactions with the DoD found the work to be highly meaningful and to buck the narrative that most applications of AI by the DoD are war-related with the potential to cause harm.
- Perhaps it's because participants normalized such work, even if there were some ethical consequences, by virtue of repeated exposure and working on it over time.

The ability to work on big, cutting-edge research and the potential to do a lot of good through such engagements was one of the most compelling reasons to engage with the DoD. Another consideration is that the DoD might be able to provide access to technology and data that might otherwise be inaccessible that can help surface insights pushing the envelope.

The most prominent concern identified by the responders was the potential for misuse of the research and the potential for harm.
.
- Alleviating these through transparent and robust governance can help both parties, especially researchers building more trust in the DoD.
- Specifying outcomes and making sure that those are adhered to can be a way that might help alleviate some of these concerns.
- The lack of transparency sometimes in being able to publish results, or not having complete control over the direction of the research was cited as a reason for not engaging.

The general views that people hold of the DoD greatly impacted in both the positive and the negative reasons that they found compelling in a potential engagement. Thus, there is a strong prior effect that can shape the willingness of actors from the tech industry to engage.

While one might imagine that willingness might be increased by framing the work in the context of threats to the US from foreign adversaries, this wasn't something that was indicated by the respondents to the survey.

Most people that responded weren't aware of the DoD ethical AI principles that were published. This is perhaps something that needs redressal so that we have balanced discussions on the impact of technologies and what measures are being used to mitigate harmful consequences.

Differential motivations in the responders was an important finding from this study. When people had a positive perception, the potential to mitigate foreign threats to the safety of the US was a strong motivating factor. When people had a negative perception, engaging in non-combat related work was a strong motivating factor. There also wasn't *unwillingness as the default* amongst AI practitioners, something that is probably miscommunicated most often in popular media.

In terms of actions that employees would take when presented with an opportunity to engage with the DoD, in the case of the hypothetical humanitarian project, most would choose to engage. In the case of the battlefield project, most would choose to not engage. An insight that was interesting here was that the frequency of people proactively supporting projects that they believed in was higher than that of those who actively condemned the projects that they didn't believe in.

In terms of the discussions around lethal autonomous weapons, most people were somewhat familiar with the issues in the larger ecosystem, but not specifically as they relate to the DoD. In choosing to work on something or not, a lot of professionals took into consideration the social impact of their work which is a good indication for the healthiness of the ecosystem.

In terms of trust, intergovernmental organizations and the EU ranked higher than US government agencies and tech companies. The Chinese government received the least amount of trust from the respondents. These perceptions are not without flaws: I think that the role that the media plays in how it portrays each of the actors has a huge impact. Trust in national governments was high when it came to who should be entrusted with the responsibility to manage the consequences of AI.

**Conclusion**

My takeaway from the study was that we need to have more granular and informed discussions when it comes to the relationship between the tech industry and government agencies. Ill-informed characterizations, propagated by media outlets, sometimes based on anecdotal evidence have the potential to do tremendous harm by creating self-fulfilling prophecies that strain the relationship between the two.

Straying away from research with government agencies just based on perceptions that you have formed from the media discourse are inadequate grounds for making a decision. Active recognition of your own biases and searching for information to gain a balanced understanding will be essential to support your claims to engage or not to engage within your organization.

**Examining the Black Box: Tools for Assessing Algorithmic Systems**
([Original paper](#) by Ada Lovelace Institute)
(Research summary by Abhishek Gupta)

**Overview:** The paper clarifies what assessment in algorithmic systems can look like, including when assessment activities are carried out, who needs to be involved, the pieces being evaluated, and the maturity of the techniques. It also explains key terms used in the field and identifies the gaps in the current methods as they relate to the factors mentioned above.

---

**What is the difference between an algorithmic audit and an algorithmic impact assessment?**
**Algorithmic audit**

The paper talks about two kinds of audits:

- Bias Audit: A targeted approach that is just focused on evaluating the system for bias.
- Regulatory Inspection: This concerns itself with evaluating the system for compliance with regulations and norms within the place that the system is operating in; it is typically performed by regulators and auditors.

**Algorithmic impact assessment**

The paper also talks about two kinds of impact assessments:

- Algorithmic risk assessment: This is an evaluation of the potential harms that might arise from the use of the system before it is launched into the world.
- Algorithmic Impact Evaluation: This is an evaluation of the system and its effects on its subjects after it has been launched into the world.

**What do we mean by audits?**

In the computer science community, the term is mostly used to describe an evaluative suite of tests and other methods designed to ascertain whether the system behaves as intended by looking at its inputs and outputs. In the non-technical community, audits include a more comprehensive evaluation of the system to see its level of adherence to professed norms and regulations.

**What might a bias audit look like?**

This essentially takes the form of testing by researchers in the form of evaluating the system by running counterfactuals (differences on one aspect with the rest of the attributes being identical) to suss out if there are any discriminatory behaviours from the system. These are typically done in a black-box fashion since the researchers usually only have access to being able to send in inputs to the system and then observe the outputs from it.

The idea of a "black box" is that we are unable to examine the contents and workings of a system (intentionally or unintentionally) and that limits our analysis to taking a look at the inputs of the system and the corresponding outputs. This is then used to make inferences about what might be going on inside the system and how it might be working.

**What are some tools and methodological approaches that can be used in this context?**

Bias audits can be carried out by following one of the following three approaches as an example:

- Scraping audits:
    - This is done by querying the system in question to figure out where there might be biased outcomes, but this risks running afoul of laws like the CFAA.
    - There are rulings recently that give leeway to researchers to engage in this but I caution checking first that you are in the clear before embarking on this method since it is quite easy to violate the ToS of the platform.
- Sock puppet audits:
    - This involves the creation of fake accounts, CVs, etc. to ascertain how the system might respond to the varying inputs.
- Crowdsourced/collaborative audits:
    - This is done when there is a mass of people who participate in evaluating the behaviour of the system by, for example, installing a browser extension.
    - The Markup has done some fantastic work with this approach to figure out privacy leakages in various platforms.

The uniting feature of all of these audits is that they have a targeted attribute that they are looking to evaluate the impact on. Often these are based on some scientific and technical definitions (which in themselves might be flawed!) but the goal is to arrive at an understanding on how the system might be discriminating on that particular attribute or set of attributes.

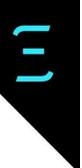

The most famous example of this is the Gender Shades project undertaken by Dr. Timnit Gebru and Dr. Joy Buolamwini that unearthed how facial recognition systems are biased against those with darker skin and those who are not male.

**What is the benefit of engaging in bias audits?**

- It leads to more awareness that can move platforms to change their working so that we have more just outcomes from the system.
- A multi-platform comparison also mounts peer pressure for the firms to do better so that they can retain their customer base and provide users with more information on which to choose to protect their rights.

**What are some future research and practice priorities?**

1. Making sure that we have more meta-literature* on how impactful bias audits have been.
2. Conducting bias audits over time and tracking their results to see if changes arise from doing them.
3. Looking at audits in more than just the online context, especially in back-end systems that are used in the public sector as an example.
4. Examining the role of different funding instruments and the agenda that researchers have because of that when they conduct such audits.

**What is regulatory inspection?**

As mentioned before, this is supposed to be a more comprehensive evaluation of the system to see if there is adherence to the norms and regulations where the system is being deployed.

Given that this is more invasive and requires more access, we need cooperation from the firms that are being evaluated and hence this falls better under the purview of regulators and auditors who have legal mandates to be able to compel the firms to such scrutiny.

The UK Government's ICO Auditing framework provides a good sample of this approach and is one that can be used as a starting point for those who are looking to prepare themselves for such regulatory scrutiny.

While this framework takes a more risk-based approach to align with the EU, in other regimes where that is not the case, you might also encounter cases with a rules-based approach.

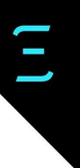

**What are the future research and practice priorities?**

- Thinking about what the current skills gaps are for regulators and auditors to effectively carry out the above roles is going to be important.
- In addition, there should be a forum where successes and failures can be shared with others to accelerate the knowledge building process.

The robustness of such an evaluation relies not just on the examination of the code, inputs, outputs, and the other tangibles but also the social context within which the system is going to be deployed because that leads to differential impacts and requires due consideration.

**What are algorithmic risk assessments?**

As I just alluded to, these go beyond just the mandated requirements, and take a more holistic view of the system, especially the social context within which the system is deployed.

An effort from the Canadian Federal Government has created an online questionnaire that gathers some basic data about your use case and then provides you with a list of requirements to build a *responsible AI* system after establishing a level of risk from it.

Another approach called *stakeholder impact assessment* seeks to bring in people who will be directly and indirectly impacted by the system into making determinations about how to build a more responsible AI application and help to uncover blindspots in the expertise that is on board the internal development team.

**What are the future research and practice priorities?**

Quite similar to the points raised in the other instruments, what we need here are:

- Ways to share the findings easily, building up a case-study bank that others can look into to take away valuable lessons.
- Sharing findings from other fields where such risk assessments have been conducted to make the approach in our field more robust.

**What are algorithmic impact evaluations?**

These are quite similar to the algorithmic risk assessments except that these are conducted after the system has already been deployed. The key thing to watch out for here is the degree to which the organizations to whom the recommendations are made actually implement the actions and whether they have the willingness to do so. Numerous cases in the past have demonstrated that there isn't as much of an incentive to change behaviour, especially when there is the potential to lose business.

In the case of the Allegheny Family Screening Tool, after an evaluation done by researchers from Stanford that found there to not be too many problems with putting it into place, it gave the impression that it is ok to use algorithmic systems in practice.

There have also been conflicting reports on the efficacy of the tool and we need to be careful when advocating for the deployment of more such automated tools through the legitimization of their use by such evaluation reports. There are a host of professional and ethical concerns which should be highlighted prior to even engaging in the practice of such an evaluation because post-hoc, they can provide the ammunition needed to continue their use without necessarily questioning if they should be used in the first place.

**What are some future practice and research priorities?**

Similar to the previous cases, having a more fleshed out set of use-cases where this has been tried along with some documented best practices for doing so effectively will be essential.

**Conclusion**

The paper provided some much needed clarity on how different terms that are sometimes used interchangeably mean different things, have different audiences and also imply different requirements from a technical. organizational, and legal standpoint.

Following some of the recommendations, especially in the future research and practice priorities will help the field move forward faster in a deliberate fashion.

Preparing your team and organization to better meet these different needs is something that you can start to do by determining which of the mentioned instruments are more applicable to your case.

Working actively with peers in your industry to create and augment best practices will not only help you shine as a practitioner but also push the entire field forward in terms of building more ethical, safe, and inclusive AI systems.

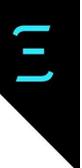

## Bridging the Gap: The Case for an 'Incompletely Theorized Agreement' on AI Policy

([Original paper](#) by Charlotte Stix, Matthijs Maas)
(Research summary by Charlotte Stix, Matthijs Maas)

**Overview:** In this paper, Charlotte Stix and Matthijs Maas argue for more collaboration between those focused on 'near-term' and 'long-term' problems in AI ethics and policy. They argue that such collaboration was key to the policy success of past epistemic communities. They suggest that researchers in these two communities disagree on fewer overarching points than they think; and where they do, they can and should bridge underlying theoretical disagreements. In doing so, the authors propose to draw on the principle of an 'incompletely theorized agreement', which can support urgently-needed cooperation on projects or areas of mutual interest, which can support the pursuit of responsible and beneficial AI for both the near- and long-term.

---

How can researchers concerned about AI's societal impact step beyond background disagreements in the field, in order to facilitate greater cooperation by the research community in responding to AI's urgent policy challenges?

Ongoing progress in AI has raised a diverse array of ethical and societal concerns. These are in need of urgent policy action. While there has been a wave of scholarship and advocacy in the field, the research community has at times appeared somewhat divided amongst those who emphasize 'near-term' concerns (such as facial recognition and algorithmic bias), and those focusing on 'long-term' concerns (concerning the potentially 'transformative' implications of future, more capable AI systems). In recent years, there have been increasing calls for greater reconciliation, cooperation, and clarifying dialogue between these sub-communities in the 'AI ethics and society' research space.

In their paper, Stix and Maas seek to understand the sources and consequences of this 'gap' amongst these communities, in order to chart the practical space and underpinnings for greater inter-community collaboration on AI policy.

Why does this matter? Critically, how the responsible AI policy community is organized, and how it interacts amongst itself and towards external stakeholders, should matter greatly to all its members. Diverse historical cases of conflict or collaboration in scientific communities—such as in nanotech, biotech, and ballistic missile defense arms control–illustrate how coordinated 'epistemic community' action can achieve remarkable policy goals, while sustained fragmentation can severely undercut researchers' ability to advocate for and secure progress on key policies.

Moreover, it appears particularly urgent to address or bypass unnecessary fragmentation in the AI policy community sooner rather than later. The field of AI policy may currently be in a window of opportunity and flexibility–in terms of problem framings, public attention, policy instrument choice and design–which may steadily close in coming years, as perceptions, framings, and policy agendas become more locked in. A divided community which treats policymaker- or public attention as a zero-sum good for competing policy projects may inadvertently 'poison the well' for later efforts, if it becomes perceived as a series of interest groups rather than an 'epistemic community' with a multi-faceted but coherent agenda for beneficial societal impact of AI.

Furthermore, while there are certainly real and important areas of disagreement amongst the communities, these do not in fact neatly fall into a clear 'near-term' and 'long-term' camp. Instead, it is possible and not uncommon to hold overlapping and more nuanced positions across a range of questions and debates. These include epistemic positions on how to engage with future uncertainty around AI, and different types of evidence and argument. However, these also include more pragmatic differences of opinion over the (in)tractability of formulating meaningful policies today which will be or remain relevant into the future. However, on critical reflection, many of these perceived disagreements are not all that strong, and need not pose a barrier to inter-community cooperation on AI policy.

However, are there in fact positive, mutually productive opportunities for both communities to work on? What would such an agreement look like? The authors propose to adapt the constitutional law principle of an 'incompletely theorized agreement' to ground practical policy action amongst these communities, even in the face of underlying disagreement. The key value of incompletely theorized agreements is that they allow a community to bypass or suspend theoretical disagreement on topics where (1) that disagreement appears relatively intractable given the available information, and (2) there is an urgent need to address certain shared practical issues in the meantime. Incompletely theorized agreements have been a core component to well-functioning legal systems, societies, and communities, because they allow for both stability, as well as flexibility to move forward on urgent issues. Indeed, it has been argued that this principle has underpinned landmark achievements in global governance, such as the Universal Declaration of Human Rights.

There are a range of issue areas where both 'near-term' and 'long-term' AI ethics scholars could draw on this principle to converge on questions both want addressed, or shared policy goals which they value. Without aiming to be comprehensive, potential sites for productive and shared collaboration could include; (1) research to gain insight into- and leverage on general levers of (national or international) policy formation on AI; (2) investigation into the relative efficacy of various policy levers for AI governance (e.g. codes of ethics; publication norms; auditing systems; publicly naming problematic performance); (3) establishing an appropriate scientific culture for considering the impact and dissemination of AI research; (4) policy interventions aimed at preserving the integrity of public discourse and informed decision-making in the face of AI systems; (5) exploring the question of 'social value alignment'–how to align AI systems with the plurality of values endorsed by groups of people. For each of these projects, although the underlying reasons might be distinct, both communities would gain from these policies.

That is not to suggest that incompletely theorized agreements are an unambiguously valuable solution across all AI policy contexts. Such agreements are by their nature imperfect, and can be 'brittle' to changing conditions. Nonetheless, while limitations such as these should be considered in greater detail, they do not erode the case for implementing, or at least further exploring the promise of well-tailored incompletely theorized agreements for advancing responsible AI policies to support responsible and beneficial AI for both the near- and long-term.

# The Ethics Owners — A New Model of Organizational Responsibility in Data-Driven Technology Companies

([Original paper](#) by Emanuel Moss, Jacob Metcalf)
(Research summary by Muriam Fancy)

**Overview:** The responsibility of ethics owners is to navigate challenging ethical circumstances with technical tools to manage and work within the tech company infrastructure. The report highlights the main findings from an ethnographic study of how professionals navigate ethical dilemmas in tech companies. The report highlights the importance of navigating systems of governance and complex company structure while pushing forward the agenda of justice and anti-oppression to support individuals and communities who need to be seen and heard for a more ethical future.

---

The questions and concerns that tech companies receive from the public are important and significant. However, tech companies' ability to effectively receive and hopefully address concerns raised by the public require individuals who bring forward ethical frameworks in cross-divisional teams. Ethics owners span across companies in various roles and hierarchal positions, but they need to have ethics owners is undoubtedly there. The report gathered main findings from completing ethnographic observations and a series of interviews with ethics owners in Silicon Valley to disseminate their role and practices better.

**The challenge of owning ethics**

With the challenges tech companies face, there is a clear need and gap to implement ethics frameworks into cross-divisional teams within the company. Companies themselves have tried to own their approach to ethics, which has influenced an ethics owner's role. However, ethics owners can also be found to understand and implement ethics independently of the company.

However, ethics owners' primary responsibilities are creating processes, tools, and policies to address ethical circumstances within the corporate infrastructure directly. These mechanisms are meant to be important for various teams across the company. Most notably, ethics owners question product designs and development to better understand the product's impact in civil society, within the company, and greater structural implications in society.

The State of AI Ethics Report, April 2021                                            33

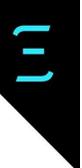

Despite the importance and need for ethic owners, they face unequivocal challenges as part of their role. For instance, the ethics owner's role description and title in tech companies vary and thus pose a barrier to understanding their role within companies. As role placement within governed structures ultimately dictates status. Another challenge presented is that there is no singular framework of ethics, the meaning of ethics is evolving and is multifaceted.

**Ethics in industry: existing frameworks and methods**

As mentioned above, creating principles or tools to implement or guide ethical practices is a living challenge for ethics owners. The frameworks of ethics that ethic owners can draw upon are diverse but can be distilled to be focused on business and professional ethics and research ethics (specifically for human subjects). However, what defines ethics in the technology industry can be rooted in various frameworks depending on the product, service, or company.

The report presents a few ethical frameworks that ethic owners can pull from as part of implementing ethical practices in their company. Ethical frameworks include:

- Medical research ethics can bridge scientific research practices and the role of regulation of these practices for human subjects and, therefore, are forced to face their practices' moral implications.
- Business ethics can aid in understanding how to manage operating systems and implement values systemically.
- Professional ethics includes practices and frameworks from the engineering and computing industry which can be applied in various tech companies.

When ethics owners review these frameworks and practices, they can create statements of principles and codes of ethics. The two documents are very different, but both serve an important purpose—a statement of principles reviews an organization's responsibilities and commitments, which can be public-facing. In contrast, a code of ethics is specifically meant to dictate internal conduct and a level of professionalism.

Other tools and methods that support the implementation and ideation of ethical frameworks are ethics review boards, development lifecycle tools, and other forms of corporate organizing such as development methodologies. All while ethics owners are keeping in mind the importance of metrics and other development methodologies to measure risk and harm in technology design and deployment.

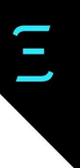

**Foundational assumptions in Silicon Valley**

Silicon Valley has developed their own working culture, which has cultivated very particular working practices. Distilling these working practices requires working in this environment, which dictates how class, gender, and race exist in this space. The report has effectively distilled three important assumptions for ethics owners to consider, including meritocracy, technological solutionism, and market fundamentalism. Ethics owners have faced various barriers in navigating these three assumptions as part of practices in Silicon Valley.

Meritocracy has allowed individuals to claim power and authority in multiple domains due to being authoritative in one domain. An example of how this would manifest as a barrier to promote equality and commit to some form of foresight of the risks of and harm of the product is realizing that technical teams do not effectively represent the diversity of individuals that can be impacted by the development of the technology. There is a lack of racial, geographic, class, and gender diversity in technical teams.

Technological solutionism demonstrates how many view technology and building "better" technology as a primary solution to many problems. Ethnics owners are forced to help explain to their company that a technological solution cannot address an issue's multifaceted implications.

Finally, market fundamentalism assumes that the market determines which ethical frameworks can be implemented and which cannot be. The idea that the market can dictate ethics forces ethics owners to advocate for ethical practices in organizational interventions that may slow or bar the process from bringing a product to market.

**Navigating tensions**

The report presents six tensions that ethics owners highlighted where ethics must be implemented and how the solution to address these six tensions might counter the generalized solutions applied to these tensions. The tensions that are presented are:

1. Personal ethics and corporate duties: ethics owners have to balance their personal and business ethics when navigating Silicon Valley spaces.
2. Upside Benefits, Downside Risks: ethics owners are positioned well to advocate for a solution that mitigates harms from a product vs. advocating for changes in product design that can benefit users.

3. Direct to Consumer, Business to Business: ethics owners work molded by their company business model, so developing standard practices may not be relevant for differing business models.
4. Measurable and Nonmeasurable Impacts: measuring the impact of tech's ethical practices and tracking successful interventions is a challenge for ethics owners.
5. Users, Nonusers: ethics owners must know the widespread implications of the social impact of the products their company is deploying and the impact of users and non-users.
6. Specificity, generalizability: ethics owners are challenges in scaling their methodologies beyond their given context.

The report ends by presenting the last section on opportunities for ethics owners to address some of the tensions mentioned above and other gaps in the space and field that they have recognized and wish to address. Such opportunities include case studies, informal meetings, partnerships, and outreach, and accountability and engagement.

In conclusion, the report finds ethics owners face unparallel barriers and challenges in the space. Although these challenges may continue to persist, the way ethics owners can navigate the space and the challenges that accompany it can continue to change and ease their work. Implementing diverse methodologies and communicating beyond their company are just ways to challenge the normative frameworks and assumptions in Silicon Valley.

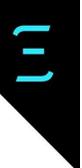

# Rethinking Gaming: The Ethical Work of Optimization in Web Search Engines

([Original paper](#) by Malte Ziewitz)
(Research summary by Iga Kozlowska)

**Overview:** Through ethnographic research, Ziewitz examines the "ethical work" of search engine optimization (SEO) consultants in the UK. Search engine operators, like Google, have guidelines on good and bad optimization techniques to dissuade users from "gaming the system" to keep their platform fair and profitable. Ziewitz concludes that when dealing with algorithmic systems that score and rank, users often find themselves in sites of moral ambiguity, navigating in the grey space between "good" and "bad" behavior. Ziewitz argues that designers, engineers, and policymakers would do well to move away from the simplistic idea of gaming the system, which assumes good and bad users, and focus instead on the ethical work that AI systems require of their users as an "integral feature" of interacting with AI-powered evaluative tools.

---

Remember JCPenney? In 2011, they were shamed in the pages of The New York Times for "gaming" Google's search engine algorithm to boost their website's ranking. They did this by having a bunch of irrelevant pages link to theirs so that the algorithm would read that as an indicator of the page's relevancy and bump it up in the "organic" search results, right before the holiday shopping season. When Google found out, they "punished" JC Penny's "bad" behavior by substantially lowering their ranking thereby reducing traffic to their website.

Ziewitz uses this case study to interrogate what we mean when we accuse algorithmic technology users of "gaming the system." As machine learning models that rank, score, classify, and predict proliferate across AI applications in fields ranging from healthcare and journalism to credit scores and criminal justice, there is widespread concern around how to design and govern AI systems to prevent abuse and manipulation. Maintaining the integrity of AI systems is of paramount importance to most stakeholders, but it is a hard nut to crack.

**From "gaming the system" to "ethical work"**

As with anything in the social world, what is "ethical" behavior is hard to pin down. People often straddle the line between good and bad behavior, constantly negotiating what is acceptable and what isn't in any given situation. It may seem like we live in a world with hard and fast rules like "lying is bad" and "speaking the truth is good." Instead, when we take a closer look, we see that we operate in a morally ambiguous world where "white lies" or coloring or downplaying the truth to protect others' feelings, for example, may be acceptable. Intentions, identities, and context matter. By remembering that AI systems are in fact sociotechnical systems, we can expect people to engage with AI systems just like they do with other people i.e. in complex ways and from within existing (though everchanging) cultural norms, simultaneously reproducing and resisting them.

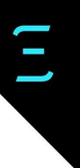

Because people alter their behavior as they negotiate the cultural rules of interacting back and forth with algorithms, ranking and scoring algorithms don't just measure "objective" reality. Through interaction with people and instructions, algorithms co-create reality. It is through that "ethical work," as Ziewitz calls it, that we collectively produce more or less ethical outcomes.

What does it mean then to design and build "ethical AI"? It requires us to take into consideration the ethical work that will be done through the building, deployment, maintenance, and use of the AI system. Below are some questions that ML developer teams can explore to move away from the binary thinking associated with "gaming the system" to a more nuanced approach that tries to understand the ambiguities, uncertainties, and context dependencies of algorithm-human interaction.

**Applying Ziewitz's ideas to machine learning development**

Extending Ziewitz's sociological research into engineering practice, we can extract a few thought-provoking questions for ML developers to consider when building ML models and AI systems.

- Recognize that AI systems don't just take in human behavior as input, but they also actively elicit some behaviors versus others. In other words, algorithms have the power to change and shape human behavior as users respond to the affordances or constraints of the system.
    - How will your AI system potentially change human behavior or incentivize broader collective action?
    - For example, could your product inadvertently create a new cottage industry, like SEO consultancies, to deal with the ambiguities of your product?

- Acknowledge that just as ethics is blurry, so is ethical AI. Moving away from binaries (AI is neither ethical nor unethical), we can instead build AI on a moral spectrum, that through its use, adaptation, and constant change in the deployment environment, creates more or less ethical behaviors, practices, and outcomes depending on the time and place of its use.

- In other words, in addition to considering "bad" or "good" user interactions with your system, what kind of behavior could be considered to fall in a morally ambiguous or grey area?
- Will that depend on the type of user or the time or place of their interaction with your product?
- In the SEO example, is trading links acceptable but paying for them is not? Is it ok if a struggling small business does it versus a big retailer like JCPenney?

● Consider the blurry lines between intended and unintended uses of an AI system during the design phase.
    ○ How likely is it that your intended uses for the product align with those of your expected users?
    ○ Broadening your view of relevant stakeholders, how might unexpected users of your product or types of users that don't yet exist (e.g. SEO consultants) engage with your product?
    ○ What mechanisms will you rely on to figure out if your system is being manipulated (e.g. New York Times exposés) and how will you impose fair penalties and mechanisms for appeal?

● Stepping back from the black and white thinking of "gaming the system," consider the kind of "ethical work" that is expected of various stakeholders.
    ○ How heavy is the burden of ethical work on some versus others?
    ○ What kind of behaviors do you expect stakeholders to engage in to try to generate ethical clarity out of ethical ambiguity when interacting with your product?
    ○ What is every stakeholder's position of power(lessness) vis-à-vis that of the AI system, the company developing and maintaining the AI system, and the broader institutions within which the AI system will be deployed? Could some behaviors be a form of resistance to an existing unequal power relationship? If so, how can you distribute more agency to marginalized stakeholders to level the playing field?
    ○ What systematic inequalities can your product potentially propagate within the ecosystem and how might your design compensate for that?

**Conclusion**

More and more, ML models are used to organize online information, rank and score human characteristics or behaviors, and police online interactions.

Despite being done by largely automated AI systems, this work, nonetheless, requires making value judgements about people and their behavior. And this means that there will always be moral grey areas that cannot be automated away with the next best ML model.

Ziewitz encourages us to think more deeply about how users will interact with our AI systems beyond the binary of good or bad behavior. He welcomes us to take the extra time to dwell in ambiguity and consider the ethical work that all humans will do as they interact with automated voice assistants, self-driving cars, child abuse prediction algorithms, or search engine ranking models.

There are no easy answers or one-size-fits-all solutions, but surely considering human behavior in all of its complexity will help us build better and more human-centered AI.

# Algorithmic Content Moderation: Technical and Political Challenges in the Automation of Platform Governance

([Original paper](#) by Robert Gorwa, Reuben Binns, Christian Katzenbach)
(Research summary by Abhishek Gupta)

**Overview:** The paper provides a comprehensive overview of the existing content moderation practices and some of the basic terminology associated with this domain. It also goes into detail on the pros and cons of different approaches and the difficulties that continue to be present in the field despite the introduction of automated content moderation. Finally, it shares some of the future directions that are worthy of our attention to come up with even more effective content moderation approaches.

---

The Global Internet Forum for Counter-Terrorism is a cross-company group that helps to share information across some of the major tech providers to combat illegal online hate speech. They regularly share information with each other to help increase the efficacy of their content moderation efforts.

As the amount of user-generated content (UGC) goes up in volume, the ability to effectively utilize community and human moderation is diminished and we have to rely on automated and commercial content moderation.

- Community moderation: Relying on the members of the community in a distributed manner to self-moderate the content that is created and distributed on a platform.
- Human moderation: This involves hiring dedicated staff who are responsible for moderating the content by reviewing those that are either flagged by users or put in their review queue by some other mechanism.

**What are the key challenges in commercial content moderation?**

- The human cost: subjecting the reviewers to traumatic content on a frequent cadence leads to long-term psychological harm
- What constitutes "valid" content is determined by a homogenous set of people in Silicon Valley that affects people all around the world
- A potential lack of transparency and accountability, especially when things go wrong

**What are some of the reasons precipitating the need to utilize AI for content moderation?**

- Very large amount of UGC
- Short response times dictated by legislations, for example the NetzDG or the EU Code of Conduct on Hate Speech, that necessitate the use of AI to meet those requirements

**What is algorithmic moderation?**

An important point made in the paper pointing to prior work by Grimmelmann talks about how moderation is not just the act of administrators and others involved in the moderation process to remove content, but it also includes the design decisions that are made for the platform that dictate how users interact with each other.

In this paper, algorithmic moderation refers to doing prediction or matching that leads to a decision or governance outcomes related to the piece of content and the parties involved. As an example, it could be taking down the content, geoblocking, suspension of accounts, among other actions.

Again, in this paper, the focus is on hard moderation which refers to actions like taking down content and blocking accounts vs. soft moderation which refers to design decisions, recommendation algorithms, and other approaches that are used to govern the interactions between content and users on a platform.

**Some basic terminology:**

- Hashing: A technique that transforms a large piece of data into a smaller piece of data, trying to uniquely identify it despite the change. They are frequently used in cryptography to check the integrity of content.

- Homology: Since hashes are supposed to be unique and two dissimilar pieces of content should give different hashes, if content is altered for example through rotation or watermarking, we might miss previously flagged content in this modified form. Homologies are a way to counter that problem through hashing techniques that are insensitive to those changes allowing for matching of "similar" content.

- Perceptual hashing: Since the cryptographic definition of hashing requires exact matching which is orthogonal to what we want to achieve with hashing in content moderation, there are non-cryptographic approaches to hashing like perceptual hashing which focus only on the perceptually relevant features in a piece of content. This means that it doesn't change if minor features in the content are altered that would still lead them to be perceived the same way by humans. This allows us to match similar content and leverage previously flagged content to prevent it from spreading again.

- Matching: Matching refers to the use of any of the hashing techniques to identify whether there exists the same content (modified or otherwise) in the previously flagged content repository so that it can automatically be removed.

- Classification: Classification refers to the approach of using the previous bank of flagged content and bucket new incoming content into a category that is most appropriate given the other content in that repository.

**So, what are some of the problems?**

For starters, maintaining a blacklist of this content that is shared and accessible across the major platforms is a very challenging task. Content, especially multi-modal content, constantly changes in meaning over time. Words take on new meanings based on the societal context in which they are used, and there are also limitations in the kind of examples that are available in the flagged content repository. Relying on domain-specific ontologies, for example a list of words that are deemed to be offensive to the LGBTQ communities based on consultations with domain experts is one way to address this challenge but that requires us to be able to go out and do that for a lot of different communities and coordinate that effort across all the platforms that might need to use such an ontology.

**What is the problem with matching?**

- It requires a manually curated list of flagged content that needs to constantly be kept up to date.
- It needs to be robust to variations introduced by malicious actors who are trying to evade moderation efforts.

**What is the problem with classification?**

- It requires inducing generalizations from existing content bases which might not be culturally or contextually relevant beyond the home base of the content that is used to generate these generalizations.
- New categories may emerge which will again need manual effort before they are effective as a content moderation tool.

**What happens when content is matched or classified?**

The content can either be flagged or deleted:
- Flagging: The content is identified for further processing at which point it enters the queue of content to be reviewed by human moderators, or pushed into a high-priority queue to be reviewed faster or by an expert moderator.
- Deletion: The content is deleted without further review. This is reserved as an action for more serious violations of the terms and conditions of the platform.

**What does the future hold?**

- When considering text content, given the complexity of language, it is incredibly hard to encode cultural and contextual factors into an automated system, so the presence of human moderators won't be fully eliminated.
- Human moderator's efforts will be augmented through the use of automated content moderation systems.
- Companies using shared hash databases, and different policies of flagging and deleting need to become more transparent to earn the trust of their community of users.
- The above point is important because there is an uneven application of community standards and the severity of content moderation based on the commercial motivations of different platforms.
- Opening up the shared databases and more access to the policies as they are actually applied to content moderation will be helpful to 3rd party auditors and researchers to make the systems more just and help them work in the interest of users rather than only serving commercial interests.

**Conclusion**

Content moderation is an incredibly complicated field in the application of AI and it more than other endeavors perhaps requires the inclusion of domain experts and communities in the formulation of community standards and moderation approaches.

Additionally, higher levels of transparency in how the content moderation is applied in practice will also help to elicit higher levels of trust from the user community on the platforms.

The design of content moderation systems should be flexible to include the input of external stakeholders, including domain experts and community members.

The flexibility of such systems should also be such that one can easily integrate emerging uses of language and other media, for example memes as a form of multi-modal content diffusion.

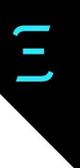

## Go Wide: Article Summaries (summarized by Abhishek Gupta)

### Ethical AI isn't the same as trustworthy AI, and that matters
(Original *VentureBeat* article by Kimberly Nevala)

It would seem that every few weeks we get new terms that are used to define the domain of responsible AI. It encompasses within it many acronyms like ART, FATE, FAccTML, etc. This article takes a good stab at making a distinction between the notions of ethics and trust, something that is left ambiguous and unresolved in many early-stage discussions.

While ethics provide us guidance on what is right or wrong, trust is something that is established between two parties and their beliefs in each other. In particular, we see that even if a product is ethical, it might not evoke trust from its users, for example, if the company building that product has in the past engaged in untrustworthy behaviour. On the other hand, having trust in something can be misguided when the underlying application itself is unethical, say in the way that data was obtained to train the system without the consent of the users.

The article gives a few cases where this becomes apparent. In a survey with employees at SAS, they found that people trusted a healthcare AI system more than one that is used to make credit decisions about individuals. One possible explanation for this is that the healthcare domain is one where people assume that the best intentions of the patient are front-and-center. This might not always be the case with credit lending facilities. So, ultimately, the article does make a strong case for thinking of these as distinct concepts though they both do support each other and having one doesn't necessitate the existence of the other but it certainly makes it a touch more likely.

### Google showed us the danger of letting corporations lead AI research
(Original *QZ* article by Nicolás Rivero)

An article that captures the dangers of having corporate-backed research pervading the advances that are made in the field. Specifically, with the recent firing of Dr. Timnit Gebru which has been a huge loss for the community, we have seen how even the guise of freedom in a corporation only extends so far. The article also talks about some of the limitations of the news around solving the protein folding problem by DeepMind which was rebuffed by some in the biological sciences as not being reproducible at the moment because of the lack of publishing the associated code, data, and details that would allow for independent verification, par for the course in the world of science.

One of the things as was highlighted in a recent paper titled the Gray Hoodie Project, was that there is a risk when a lot of the research coming out in the field is funded by corporate interests one way or another. In particular, when it comes to research in the field of AI ethics, this has even more serious implications as we saw with the case of Dr. Gebru.

There are also huge concerns about whether such research even when it doesn't have explicit censorship might be suffering from a skewed perspective, whether implicit or explicit, that shapes the direction of the field in general because the funders want specific outcomes that benefit them. There can be counterarguments that organizations will strive to maintain their independence when receiving funds but one can only wonder to what extent that can be taken at face value. Perhaps we need deeper investigations into the track records of publishing in this field to analyze the impacts of this, though drawing conclusions without the presence of counterfactuals (which will be impossible to come by) means that we can draw loose correlations between these factors and not much more.

### If not AI ethicists like Timnit Gebru, who will hold Big Tech accountable?
([Original *Brookings* article](#) by Alex Engler)

Without a doubt Dr. Gebru's work has been a cornerstone of the AI ethics research and advocacy community; the work at MAIEI is also deeply inspired by her persistent call for awareness and action when it comes to injustices in AI ethics. This article highlights one of the key areas of concern when it comes to AI ethics: the lack of accountability in Big Tech when it comes to work that might run counter to their business interests even when the very reason for having AI ethicists on the team is to help bring that accountability to the work that is being done by the organization. As mentioned in the article, because of the way AI systems are built and structured, there might not be a way for external researchers to do much except probe externally to tease out potential problems. AI ethicists working at the company serve as a much more robust check (if they are allowed to function!) against potential ill-uses of the technology. But, if they are not allowed to function freely, that removes one of the final guardrails that we have, wreaking havoc on unsuspecting users of the various technologies and platforms offered by Big Tech.

Very rarely do we see action taken by organizations who for the most part dispute external findings. Even mass-scale action like #StopHateForProfit and #DeleteUber put only a minor dent financially speaking on the bottom lines of the organizations thus acting as a limited check. In fact, in the month where about 1000 companies signed up to limit their advertising spend on Facebook, the company still reported a profit showcasing the highly essential nature of the services provided by these organizations that are hard to disentangle from our daily existence and operations.

While there is tremendous value in the work done by journalists to bring to the forefront some of the problems and galvanize public action, without government action, Big Tech will continue to run rampant and ignore the importance of the work by scholars like Dr. Gebru who have worked tirelessly to ensure that we build fair technologies that benefit everyone instead of a narrow set of people.

### AI research survey finds machine learning needs a culture change
(Original *VentureBeat* article by Khari Johnson)

This article urges people to reflect on the importance of considering that every person represented in large datasets is someone with a deep and enriching life and that their data should be treated with care. This is one of the problems that occurs when we work with numbers which tend to flatten out our understanding of and empathy towards people.

We need to invest a lot more care into how we curate and create large datasets. Especially when they have to do with critical parts of our lives. Arguably, with the large degree of interconnectedness, perhaps all datasets would fall into that because there are so many ways in which they can interact with each other that they might still end up playing an important role in determining something significant about someone even when it is not significant on its own. While such approaches will require more investment in terms of efforts and resources, the payoffs in terms of better outcomes for those who are represented in that data is going to be quite worth it.

The considerations shouldn't just look at direct impacts on the people involved but also indirect and follow-on effects, say the impact on the environment that arises from the use of large-scale models and datasets. This also includes the exclusionary effect that it has on siloing out those who don't have the compute and infrastructural resources to be able to shape the technical and policy measures as they relate to such large models.

### The year algorithms escaped quarantine: 2020 in review
(**Original *AlgorithmWatch* article** by Nicolas Kayser-Bril)

As we covered previously, 2020 was a breakout year for the deployment of automated systems, and this article from AlgorithmWatch sheds some light on a few things that went right and a lot of things that went wrong. Lots of bold claims were made by companies when it came to their ability to combat the pandemic using the services that they offer via their automated systems. For example, BlueDot claimed to have detected the pandemic before anyone else but those claims remain unaudited. And this has been the flavor for a lot of the automated systems that are being pushed out. Unsuspecting governments and other entities looking to gain an edge over the pandemic trusted the "snake oil" that they were being sold by cunning entrepreneurs (a clever phrase from AW) into purchasing solutions that were yet to be battle-tested.

Detecting temperature and subsequently, the potential for someone to be infected from video feeds was a subversive sell of surveillance software. And many countries rolled back hasty deployments after facing pressure from their populace. One of the reasons that the Montreal AI Ethics Institute stresses the need to have civic competence in AI ethics is because of that - the more aware we are as everyday citizens, the more we can meaningfully push back against systems that violate basic rights.

Especially with the use of automated systems in government services like unemployment services in Austria rightfully faced backlash when a nebulous "employability" score was used to rank people seeking help. Amsterdam and Helsinki led the way in terms of publicly documenting the algorithmic systems that are being used to provide a greater degree of transparency, something that a lot more places can start to emulate as a starting point. A chilling statement in the article that really caught my attention was how with the failure of some of the traditional means of communications in the wake of other natural disasters (yes, there was more than just the pandemic that ravaged 2020), we left a lot of essential communication and rescue efforts to the whims of newsfeed algorithms, something that we need to be more conscious of.

### The Turing Test is obsolete. It's time to build a new barometer for AI
(**Original *Fast Company* article** by Rohit Prasad)

For those who are new to the field of AI ethics, you often work with questions and a high degree of interest in whether the Singularity is coming or whether the different chatbots that we see around us are "intelligent" enough to pass the Turing Test. The basic premise of the Turing Test is for a chatbot to hold an extended conversation with a human and convince the humans that they are speaking with a human rather than a machine. Yet, that notion might seem outdated now given the ubiquity of intelligent applications around us. And in some cases, we might not even want to be fooled into believing that something is human when it isn't.

One of the great things about this article is that it highlights the different era in which this test was put forward when computing was limited, expensive, and sensors weren't as widespread as they are today. So, we might be artificially constraining the intelligence test of an AI system to just text when in fact today's systems can combine multimodal information which adds a few other dimensions. In addition, we might also be limiting the true capabilities of the system today to make it appear human when answers to several fact-based things are instantaneous given access to cloud computing rather than having to insert artificial pauses to make it appear more human when responding to a query about the distance between two cities as an example.

Finally, thinking about the goals that such a test sets for the developers in the field is also something to consider. Especially with the spread of misinformation that is fueled by the presence of chatbots and other automated mechanisms, we might want to be careful about setting the right incentives so that the technology is actually used for beneficial purposes rather than to trick people. Think about what havoc can be wrecked with the combination of chatbots that are backed by GPT-3 producing some very believable interactions with humans.

### China wants to build an open source ecosystem to rival GitHub
(Original *Rest of World* article by Meaghan Tobin)

For a while now there have been extended talks on the splintering of the internet into regions that are based on the regulatory frameworks that are applied to the operations of the different platforms. The most dominant and pervasive such fragment is the Chinese ecosystem which has equivalents for a lot of the popular platforms like Facebook, Twitter, Google, among others. GitHub, the open-source code hosting platform, is another example where a local Chinese equivalent called Gitee is being developed and backed by some prominent local companies in the interest of maintaining the tradition of open-source though one that is housed internally within China.

While so far GitHub continues to remain accessible within China, concerns are frequently raised when it comes to how sometimes content like messages going against the government's perspectives among other documentation of harms is stored on the platform to save it from censorship on some of the home-grown platforms might lead to it being taken down within the Chinese walled garden.

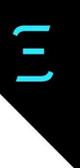

Without a doubt, open-source code has strengthened the Chinese technology ecosystem but such a fragmentation where perhaps everyone outside China stores their open-source code on platforms like GitHub and GitLab whereas developers in China store theirs on a platform like Gitee might lead to a lack of knowledge sharing and application of best practices because people get siloed. It would go against the spirit of the open-source community which aims to share code with each other in an unrestricted manner encouraging collaboration so that we build on each other's work rather than replicating efforts.

## The Future of Responsible Tech: Experts Share Their Lessons and Predictions for 2021
(Original *Salesforce Blog* article by Christina Zhang)

Given that the Montreal AI Ethics Institute is firmly centered on community in all of its work, we appreciated the shout out from Salesforce in this article mentioning the importance of community as a central pillar that will be responsible for the achievement of responsible AI. One of the points that came up again and again in this article shared by all the authors was the importance of addressing bias in its many forms that manifests in automated solutions. Especially, as technology deployment has accelerated because of the pandemic, such solutions have become widely deployed and have become pervasive in many parts of our lives.

Design as another core consideration that will help to shape the future of technology and what it is able to achieve in the upliftment of people was another great insight in the article, something that gets talked about a little less than some of the other issues in the field. Finally, I also really appreciated the call-out that building responsible technology will be the responsibility of all of us and not just isolated actors calling out for justice. This is something that will be bolstered by the introduction of regulations that provide more of a mandate to undertake these activities and create a forcing function for companies to invest efforts into building and deploying responsible AI solutions.

## A New Australian Law Is the Wrong Answer to Big Tech
(Original *OneZero* article by Owen Williams)

Australia has put forward a new regulation called *News Media and Digital Platforms Mandatory Bargaining Code* that is meant to reduce the influence of large platforms that drive a lot of traffic around news media and affect the bottom line of publications. The article makes some interesting points around how the regulation in its current form misses the key problems facing these publications in the first place.

While it is great that the regulation forces large companies like Google and Facebook to come to the negotiating table, it does that by asking the platforms to pay for any snippets and content that might be featured on their platforms. This is problematic because it restricts how such links can be shared in the first place and the amount of traffic that they are responsible for driving to the news media websites in the first place. Pulling the plug on that would mean that the news media websites would have to be responsible for finding all that traffic on their own organically or through ads, something that might further injure their bottom line. It is also antithetical to the ethos of the web in terms of how links are shared openly and might penalize some smaller companies that do link out to such news media articles.

The one that it does do right is that it potentially creates a more level playing field in terms of news media organizations not having to cower down to the proprietary formats and requirements that the platforms like Google and Facebook impose, for example, Google AMP, which pose a burden for content producers and service the needs of the platforms by locking them into their distribution networks. They do offer more prominent features for that content but at the cost that might be too much to bear for everyone save the largest organizations. Ultimately, going back to the more open ethos of the internet might just be the way that we can solve some of these emergent issues.

## Can Auditing Eliminate Bias from Algorithms?
**([Original *The Markup* article](#) by Alfred Ng)**

- What happened: HireVue, the company that uses AI to assist with the hiring process, and one which has faced lots of scrutiny for having bias in their systems, hired Cathy O'Neil's (author of Weapons of Math Destruction) company (ORCAA) to conduct an audit of their systems. The audit itself focused on a narrow use-case and HireVue framed in a light that shows them as not having done anything wrong vis-a-vis what they promised to offer.

- Why it matters: Because there are no standards for what forms a good audit in this domain, audits can easily become instruments for ethics washing the solution that has been created by a company offering a veneer of legitimacy. The lack of transparency around audits at the moment in the domain is also quite problematic.

- Between the lines: It is unfortunate that more details on what actually happened can't be released as mentioned in the article that O'Neil declined to comment on the specifics of what transpired with HireVue. If audits don't have transparency, then we just risk making them a tool for ethics washing solutions and walking away thinking we have done our part without really affecting meaningful change.

## How The Department Of Defense Approaches Ethical AI
(Original *Forbes* article by Kathleen Walch)

Certainly an organization within the US government that has significant resources to shape how AI might be deployed and used in the wider industry, as large research budgets from them in the past have led to pivotal technological developments, this is great to see the DoD and the JAIC taking an active approach to formalizing their AI ethics principles and discussing them in an article to shed some light on the process behind it.

One of the things that other public institutions could borrow from this is the mindset that was adopted by the DoD in creating and publishing these principles in the interest of helping nations around the world navigate these issues better. Given the sometimes negative connotations that their work receives, having these principles articulated helps attract and retain talent that can allow these departments to pursue AI projects in the first place.

Some of the actions suggested in the article that can help organizations better actuate AI ethics principles include thinking of ethics as an enabler rather than an inhibitor in the sense that it actually makes systems more robust and useful. Taking a lifecycle approach (as we just discussed) that integrates smoothly with the rest of the organizational practices and increasing literacy around this subject across the organization can help tremendously. To lower the barrier of adoption, from a literacy point of view, we are not seeking to create subject matter experts. Instead we are looking to create stewards who can apply the tools and techniques.

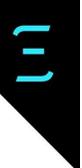

# 3. Fairness and Justice

**Opening Remarks** by Marianna Ganapini, PhD (Faculty Director, Montreal AI Ethics Institute)

Should we trust AI? The answer to this question hinges on whether or not we will be able to create fair and just AI systems, namely systems that are able to open up opportunities for a large number of people without discriminating against vulnerable groups and reproducing systemic injustices. Different solutions have been offered to make AI just and trustworthy but these solutions need to be implemented globally and consistently if we want to make progress in this area.

A good place to start our analysis is by reflecting a little bit on the notion of "trustworthy AI". Generally speaking, we trust something or someone if we decide to rely on them to achieve some set goals. Those who we trust need to be not only competent and capable but also respectful of our broad interests (see [here](#) for a nuanced discussion on the notion of trust). Thus a trustworthy AI is an AI system that will be reliable in achieving a specific goal while also being mindful of our shared interests and values. In contrast, an AI system that is oppressive and/or is used to promote injustice, is not to be trusted, because of how it undermines what we cherish the most.

There are a lot of reasons for not trusting AI at the moment. Algorithms now make many of the decisions humans used to make. Among many other things, we rely on algorithms to figure out the right college applicants and to decide who gets an apartment or a job, who deserves a loan, a vaccine or an organ transplant. The rationale behind this is that algorithms are faster, can look at more data and are more 'objective' than human beings. Whereas humans are riddled with biases and irrational thinking, algorithms just look at the facts and make impartial decisions. Or so we thought. In fact, it turned out to be quite the opposite: not only are algorithms reproducing human unjust biases but they are also amplifying them and making them worse.

The result is that the decisions reached by using algorithms are unfair as they discriminate against groups that have been traditionally subject to discrimination. As the articles in this chapter clearly show, there are a lot of examples of injustice in AI. A key example of this is how AI systems can discriminate against those who have disabilities. The paper "Disability, Bias, and AI" illustrates the ways in which AI discriminates by setting standards that exclude people who don't fit within the 'norm'. Similarly in the piece "Algorithms Behaving Badly", we have a list of how AI puts minorities and low-income people at a disadvantage when it comes to getting medical care or getting a loan.

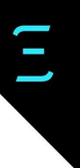

But it gets worse: not only is unjust bias a pervasive problem of algorithm-based decisions, at times sophisticated algorithms and technologies are *actively* employed to target certain groups. The article "The UK Government Isn't Being Transparent About Its Palantir Contracts" points out that the UK has been making deals with Palantir, a US-based "surveillance and data analytics company". The software in question has been used in the past to target minorities and the worry now is that it will be weaponized against vulnerable groups in the UK. Similar cases are to be found in China, Ethiopia, Zimbabwe and many other countries.

What can we do to make AI more fair and trustworthy? Bias-mitigation is a technical process to address the problem of bias in machine learning. As the article "Root Out Bias at Every Stage of Your AI-Development Process" explains, this technique can be adopted at various stages of the machine learning pipeline. At the pre-processing stage, bias-mitigation techniques can help with bias emerging because of the training data. In the processing stage, we can use 'adversarial debiasing' which is a way to spot bias by creating an 'adversarial model'. In the post-processing phase, it is helpful to adopt 'Rejection Option-based Classification'. As the article explains, this debiasing tool "assumes that discrimination happens when models are least certain of a prediction. The technique exploits the "low confidence region" and rejects those predictions to reduce bias in the end game. This allows you to avoid making potentially problematic predictions."

At a more high level, an important tool for ensuring a trustworthy AI is developing a third-party auditing framework for autonomous systems. In their article "The algorithm audit: Scoring the algorithms that score us", the authors propose a general model for an algorithm audit that is able to account for the context in which algorithms are employed. As the article explains, beyond the technical aspects of the design of an algorithm, an auditing framework needs to also evaluate the outcomes and use of some particular technology as these can promote and spread systemic injustice as well.

Finally, it is important to mention that politicians and legislators should be playing a role in this too. In the US, there seems to be some reason for optimism. As one of the articles explains, with the new Biden administration, "there may be an appetite for finally enacting guardrails for a technology that is increasingly part of our most important automated systems." Let's hope this is really the case!

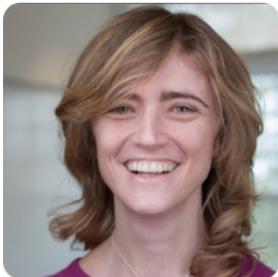

**Marianna Ganapini, PhD (@MariannaBergama)**
Faculty Director
Montreal AI Ethics Institute

Dr. Marianna Ganapini manages curriculum development and public consultations at the Montreal AI Ethics Institute. She is also an Assistant Professor Of Philosophy at Union College. Her main areas of research are Philosophy of Mind and Epistemology. She has numerous peer-reviewed publications and she is currently working on several projects on disinformation, content moderation, reasoning and human irrationality.

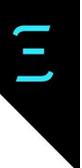

# Go Deep: Research Summaries

**Exchanging Lessons Between Algorithmic Fairness and Domain Generalization**
([Original paper](#) by Elliot Creager, Jörn-Henrik Jacobsen, Richard Zemel)
(Research summary by Falaah Arif Khan)

**Overview:** This paper unifies two seemingly disparate research directions in Machine Learning (ML), namely Domain Generalization and Fair Machine Learning, under one common goal of "learning algorithms robust to changes across domains or population groups". It draws links between several popular methods in Domain Generalization and Fair-ML literature and forges a new exciting research area at the intersection of the two.

---

Both algorithmic fairness and domain generalization share the common objective of reducing sensitivity to the training distribution. In algorithmic fairness, we wish to make classifications that are 'Fair' as per our context-specific notion of Fairness, such that we do not disadvantage individuals due to their membership in a certain group (based on sensitive features such as race, gender, etc). In Domain Generalization, we wish to learn features that are 'domain-invariant' such that the classifier's predictions are made based on object information rather than stylistic information, such as color, which might vary across data domains.

This exciting line of work attempts to 'take the best of both worlds' and share insights and methods across Fair-ML and Domain Generalization to design algorithms that, within their specific context, are both robust as well as "fair". Group membership can be treated as Domain-specific attributes and so the popular conception of 'Fairness through Blindness', which removes all sensitive attributes (such as gender, race, etc) from consideration, has a natural connection to the 'Domain Invariant' features that are learned in Domain Generalization tasks.

The authors provide a succinct review of both Domain Generalization and Fair-ML literature, including Distributionally Robust Optimization (DRO), Invariant Learning and Invariant Risk Minimization (IRM) and Risk Extrapolation algorithms and Fairness notions mapping to demographic parity, equal opportunity, calibration, group sufficiency and multi-calibration. They also attempt to map some common objectives across the two areas such that lessons from one can be used to inform the other. Intuitively, group membership can be thought of as domain information. In Fair-ML literature, group membership is based on some protected attribute such as gender, race, etc. In domain generalization, the target domain is a mixture of multiple domains, all of which may or may not be available during training.

In Domain Generalization, we wish the algorithm to learn properties that will generalize well to the test distribution. In algorithmic fairness, on the other hand, our learning objective is dictated by the worldview we employ and the context-specific fairness notion we wish to satisfy.

The authors draw from fairness approaches that optimize worst-case performance without access to demographic information and formulate an algorithm to learn domain-invariance, without access to environment information. They consider a realistic scenario when the train-test partitioning of domains is not provided, since in the real-world domains will have overlap and a clean split of different domains that are present in the testing environment is practically infeasible. The proposed method, Environment Inference for Invariant Learning (EIIL), is a variant of Invariant Risk Training (IRT), where the former takes hand-crafted environments EIIL learns suitable partitions that would lead to worst-case performance. Performing IRT on such partitions thereby provides a good generalization.

The authors demonstrate the robustness of EIIL without requiring a priori knowledge of the environments through experiments on the Color-MNIST dataset, and further enumerate how EIIL directly optimizes the common fairness criterion of group sufficiency, without knowledge of sensitive groups, on the UCI Adult dataset.

The authors also demonstrate the sensitivity of EIIL to the choice of reference representation and empirically show that the algorithm discovers suitable worst-case partitions only when the reference representation encodes the incorrect inductive bias by focusing on spurious features, thereby calling out the limited, setting-specific effectiveness of EIIL over standard Empirical Risk Minimization approaches.

They also propose some interesting future directions where methods in domain generalization can be applied for creating "fair" outcomes, such as the scenario where distribution shift occurs due to the correction of some societal harm.

The paper puts forth an extremely exciting research direction that seems to emerge naturally from the shared objective between generalizing to an unseen domain and in trying to fulfill a specific notion of fairness. They adeptly show how ideas from 'Fairness through Blindness' can be helpful to learn domain invariance and this motivates a deeper, more critical look at how two seemingly disparate sub-fields of machine learning can inform and even bolster the capabilities of one another.

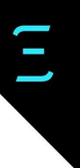

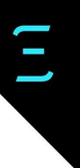

### AI and the Global South: Designing for Other Worlds
([Original paper](#) by Chinmayi Arun)
(Research summary by Victoria Heath)

**Overview:** This paper explores the unique harms faced by the "Global South" from artificial intelligence (AI) through four different instances, and examines the application of international human rights law to mitigate those harms. The author advocates for an approach that is "human rights-centric, inclusive" and "context-driven."

---

"Increasingly," writes author Chinmayi Arun, "private owned web-based platforms control our access to public education, the public sphere, health services and our very relationships with the countries we live in." This infiltration and permeation of technology requires that we critically examine and evaluate how it is designed and how it operates—this is especially true with automation and artificial intelligence (AI). "We must place the needs, history, and the cultural and economic context of a society at the center of design."

This is true for many designed artifacts of society, like houses or buildings — why shouldn't it be true for AI?

**What is the Global South?**

Arun's focus for this research is on the "Global South," a term she explores in length, concluding that it has come to transcend borders and includes "countless Souths." It can be found within Europe and North America (e.g. refugee populations), and can also be used to distinguish the "elite" in countries like India, Mexico, and China from the impoverished and oppressed. This is especially useful due to the fact that the "political elite" and "industry elite" in many countries encapsulated in the "Global South" are often more focused on "protection of markets than on protection of citizens." For example, "data colonization" is growing within countries like India and China, in which governments contract predominantly Western technology companies for public services. These companies are able to extract data from these populations with little to no regulation or oversight.

Thus, "Global South," as utilized in this research and increasingly elsewhere, "focuses on inequality, oppression, and resistance to injustice and oppression."

**Technology in other worlds**

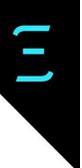

In order to examine some of the harms posed to the "Global South" from AI, Arun explores "different models of exploitation" illustrated by four real-world examples. The first example is Facebook's role in the Rohingya genocide in Myanmar, classified by Arun as a North-to-South model of exploitation, in which a technology "designed in the North" proves harmful when exported. The second example is the biometric identity database in India called Aadhaar, classified as a model of exploitation stemming from the actions of local elites. In this case, software billionaire Nandan Nilekani helped fund and create the mandatory system that has resulted in excluding people from the local welfare system and even surveilling undocumented migrant workers for deportation.

The third example is the use of data collection systems, like facial recognition, on refugees in Europe. Arun classifies this as exploitation by governments and even international humanitarian agencies of asylum seekers and refugees as they collect their biometrics and subject them to surveillance. Even with the best intentions, these practices often deprive these populations of their agency and can make them more vulnerable. The final example is China's practice of selling surveillance technology to authoritarian countries like Ethiopia and Zimbabwe, classified by Arun as similar to the North-to-South model of exploitation. However, in this case, it's facilitated by another Southern Country. These surveillance systems are oftentimes used by the political elite of a country to control the population.

**AI and the Global South**

At this point, it's well-known that there are issues of bias and discrimination in algorithmic systems. However, what's often missing from conversations around these issues and the harms they cause, is how "Southern populations" are both uniquely affected and unprotected. As Arun explains, "When companies deploy these technologies in Southern countries there are fewer resources and institutions to help protect marginalized people's rights." Thus, institutional frameworks that exist in Southern countries must be taken into account when devising ways of mitigating harms caused by these systems. It will be impossible to ensure the rights of marginalized peoples in the Global South if there is limited space and capabilities for citizens and civil society to engage with both the government and industry.

**How international human rights apply**

International human rights law "offers a standard and a threshold that debates on innovation and AI must take into account," writes Arun. However, as many in the international community have noted, most of the documents and international agreements related to international human rights were adopted before many of today's technologies existed. Therefore, more must be done to ensure that AI does not violate basic human rights, and that basic digital rights are also codified in international agreements. One idea is to obligate governments and companies to "conduct human rights impact assessments and public consultations during the design and deployment of new AI systems or existing systems in new markets."

**Conclusion**

"With every year that passes," reflects Arun, "this system [of knowledge] intertwines itself with our institutions and permeates our societies." The time to begin working on "reversing extractive technologies in favor of justice and human rights" was yesterday. The harms faced by Southern populations at the hands of AI and automation are significant, but they are not impossible to mitigate. The first point of action, says Arun, is to "account for the plural contexts of the Global South and adopt modes of engagement that include these populations, empower them, and design for them."

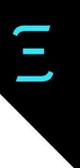

## Data Statements for Natural Language Processing: Toward Mitigating System Bias and Enabling Better Science
([Original paper](#) by Emily M. Bender, Batya Friedman)
(Research summary by Abhishek Gupta)

**Overview:** This paper provides a new methodological instrument to give people using a dataset a better idea about the generalizability of the dataset, the assumptions behind it, what biases it might have, and the implications from its use in deployment. It also details some of the accompanying changes required in the field writ large to enable this to function effectively.

---

**What is a data statement?**

A data statement is a characterization of a dataset that provides context to allow developers and users to better understand how experimental results might generalize, how software might be appropriately deployed, and what biases might be reflected in systems built on the software.

- A benefit that we might get from the use of data statements is that it will help with generalizability and reproducibility in the field, something that has been pointed out as a concern at many workshops like NeurIPS, ICML, and ICLR.
- A clarification that authors provide is the difference between ethical practice and sound science where having one doesn't mean we necessarily get the other. You might get accurate results based on a given metric (sound science), but it doesn't make the use of that technology ethical.

**Key terms**
- While some of the common terms are fairly well-known in the NLP community, the paper sheds an important light on nuances to some of them for better understanding.
- For example, it talks about curators who are the people who make decisions on, e.g., which search terms to use for gathering data, selecting where to look, whom to interview, and other meta-level choices that have downstream implications.
- An interesting distinction concerning the stakeholders is between those who are indirect stakeholders, and those whose data might be rendered invisible due to how search terms are processed and utilized in an NLP system.

**Bias**

- Providing a very clear summary of what bias means in this context, they refer to how bias (in the social sense) occurs only when it is done systematically and leads to unfair outcomes.
- It also talks about pre-existing biases which have roots in social institutions, practices, and attitudes.
- Technical biases stem from technical constraints and decisions.
- Emergent biases are those that arise when a system is developed for one situation and is supplanted into another.
- An example that clarifies why we might need data statements in NLP is from research cited in this paper that points to how Mexican restaurants with similar star ratings got lower ratings on their reviews because of pre-existing biases against the word Mexican in the wider use of that term.

**What will data statements look like?**

The paper provides a schema with a minimum set of items that must be included but also goes on to talk about how there will need to be some best practices that emerge from the field as they start to become more widely used.

Here are the elements:

- Curation Rationale: What were the decisions and assumptions that led to a certain data source being selected, both in the original and the sub-selection that might be made in the case of reuse?
- Language Variety: Given that languages have different structures and might thus interact differently with the NLP algorithms, taking a tag from BCP-47, a standardized way to identify languages along with a prose description of that tagging will help in understanding the language captured in the dataset. In addition, paying attention to the dialect used will also help downstream researchers make appropriate decisions.
- Speaker demographic: Here speaker is used to the generator of the text artifact. Having information on the demographics will also help to make determinations about language artifacts to expect in the dataset since people use language in differing fashion to project their identity. This should include:
    - Age
    - Gender
    - Race/ethnicity
    - Native language

- Socioeconomic status
- Number of different speakers represented
- Presence of disordered speech
● Annotator demographics: The same points as above are also essential for better understanding artifacts in the dataset because the social address of the annotators will determine how they interact with the data. Some attributes to capture here include:
  - Age
  - Gender
  - Race/ethnicity
  - Native language
  - Socioeconomic status
  - Training in linguistics/other relevant discipline
● Speech situation: We all realize that we speak quite differently in different contexts, depending on the situation we are in, who we are speaking to, etc. Capturing this information is also essential which should include:
  - Time and place
  - Modality (spoken/signed, written)
  - Scripted/edited vs. spontaneous
  - Synchronous vs. asynchronous interaction
  - Intended audience
● Text characteristics: The genre and the topic will influence the text that shows up in the data.
● Recording quality: Any particulars about how the recording was done and any limitations of the equipment will also help to understand any shortcomings that might show up in the dataset.
● Other: This is the placeholder where best practices can be integrated as they evolve over time.
● Provenance Index: In the case of reuse of the dataset, we can include information linking it to the origin to help people understand how the dataset has changed over time and if there are things that they should watch out for.

An important point to note is that while it might seem onerous to repeat the information again and again for datasets that are commonplace in the NLP community, considering the data in new light every single instance will help people better understand the capabilities and limitations from that data in the new task at hand.

● Having a short form of the data statement between 60-100 words will also help direct inclusion in papers that are published which will be useful to have persistent reminders of the specificities of the data that is being used.

- The paper also provides two sample case studies which can be used to understand how the data statements might be used in practice.
- Even retroactive application of data statements has value, though they won't be as good as making them right when the dataset is being created because of the loss of information over time unless there are strict records.

**Potential barriers to implementation and mitigation strategies**

- If the ACL implements this as a requirement, those who are not yet members of ACL and are not aware of how to craft good data statements might face desk-rejection for their work.
- Having mentorship programs, training modules, and how-to guides will help to bridge this gap.
- These require time to craft (2-3 hours based on the experience of the authors) but the potential payoff is high and thus we must incentivize researchers to do so.
- They take up space in the paper if they are included and hence the authors ask that conferences exclude these from page and word limits.
- While excluding demographic information can obviate the need for ethics reviews in some instances, collection of this information might add that as a burden for projects that didn't previously engage in that.

**Differences from related work**

Data statements stand out in some regards in the sense that data sheets for datasets, as an example, are broader and thus left a lot of elements with too much flexibility which can make their application inconsistent.

Similarly, algorithmic impact assessments also provide very high-level guidance but data statements provide more granularity making them more actionable!

**Implications for tech policy**

- Without the presence of clear information about the origins of the data and the limitations that are posed by their use in different systems, we might end up in situations where we limit the ability of citizens adversely affected by these systems to mount effective challenges because of the lack of sufficient evidence to back their claims.
- Having some sort of requirements in certification processes can also normalize and standardize the use of such instruments.

- The authors also advocate that we ensure that people using software give preference to those that come with such data statements over those that don't to mount pressure on the rest of the community to adopt the practice.

**Conclusion**

Data statements will provide many benefits to the field including better generalizability and reproducibility. It will empower stakeholders to be better informed about the capabilities and limitations of the system. For people in the industry it can be a forward-looking initiative that helps prevent PR nightmares because of overlooked areas.

Having such documentation more well-integrated into the MLOps workflow will be crucial for the success of their adoption and utilization, and having domain-specific formats that are agreed upon by practitioners will be the way forward.

## Disability, Bias, and AI
([Original paper](#) by AI Now Institute)
(Research summary by Abhishek Gupta)

**Overview:** A comprehensive report on how people with disabilities are excluded from the design and development of AI systems. The authors situate the discussion within the context of existing research and provide concrete recommendations on how AI practitioners can do better.

**Key questions**
- What can we do to draw from disability activism and scholarship to ensure protections for those who are deemed "outside the norm" by AI systems?
- Given that AI shapes the world, we need to account for the implications that exclusion as mentioned above causes to people and what we can do to recognize and address that proactively.
- In the service of disabled people, how can we lean on existing legislation to fight for accountability?
- Can we learn from the work that has been done in advocating for rights and design changes in the physical world into the digital realm?
- What are ways that we can assess that the changes being made to the system are actually benefiting people?
- What are some of the accompanying changes at a systemic level in addition to technical interventions that can help with this?

**Terms and concepts from disability studies**

- Disabled people are heterogenous. We need to keep this in mind to avoid applying blanket methodologies to people who might fall in the same "category" based on their expressed disability.
- We should have some opt-out mechanisms so that we don't impose classifications on people without their consent. These classification are sometimes erroneous and this can have severe implications for them.
- An example that articulates the problem in a very relevant fashion is how the LGBTQ community fought very hard to remove being gay as a condition from the DSM. Prior to its removal, this classification has justified unjust and unfair treatments because of the enshrinement of their identity as a problem in formal documentation.

**Models of disability**

- Using medical definitions of disability as those falling outside of what is thought to be normal bodies, stands the risk of entrenching stigmatization further, and encouraging exploitation of individuals.
- The social model of disability instead looks at how the environment, both built and social, leads to disability rather than being something located in the body of the individual.
    - The key insight with this is that it places the onus of interventions at a systemic level rather than placing it all on the individual.
    - An important consideration that the paper highlights is how African Americans, LGBTQ, women have at varying times been described as disabled which led to a lot of marginalization.
- Disability is also not static and can wax and wane over time even within the same body. This is something else that we must keep in mind when we build AI systems.

**Key terms**

- The paper provides definitions that help to recenter the conversation in a way that is empowering rather than using language which overlooks the concerns of the disabled community
- One thing that particularly caught my attention was the reframing of the phrase "assistive technology" describing how all technologies are meant to assist us but framing those used by disabled people as such gives a ring of paternalism and encourages a mere technological solution rather than a serious reflection on community education, support, and social change.
- This is again reflected in the fact that we simply can't add disability as another axis to consider in the bias discussion but instead should consider the lived experiences of people which may not fit in what we consider the "norm".
- The above will also aid in the appropriate emphasis on the non-technical measures that are required in addition to the use of technology to meet the needs of these communities.

**Discrepancies in development and deployment**

- Another consideration when thinking about technology as a vector for change is that access to technology is highly stratified, meaning there is inequity in access because of both financial and distribution reasons.

- There are also problems when this is used as a pretense for developing solutions that might seek to initially meet the needs of the community, but upon new-found success in the wider market, those are abandoned in the interest of pursuing bigger markets and profits.
- It is also wrong when the community is used as a testbed for new technologies, to iron out the kinks before rolling it out for wider use and ignoring the adequate consideration and participation of the people from the community.
- A lot of ethical decisions might already have been made in terms of the limits that the technology will impose on its user, thus stripping agency from the users.

**Design considerations**

- For example when thinking about transcription services using AI, they are meant to be standardized and implement broad-based gestures and vocabulary but when this is done by humans, they often tailor their communication in a way to be more personal and connected to the individual.
- This personalization might be lost through the interjection of automated systems into such fields.

**Biases at the intersection of AI and disability**

- People with disabilities are affected non-uniformly across identity groups when it comes to the AI-bias and it is problematic that most current discussions on AI bias don't take this into account.
- For example, in content moderation practices, the paper points to the example of how content with terms about disability is marked as toxic more frequently than those without. This has severe implications in the ability of people to freely express themselves, gather and discuss issues online.
- When people from the disabled community are excluded, there is a severe risk of misunderstanding and misrepresenting the issues to the point that they create more harm than good when making decisions on how to address bias in the AI systems.
- An example that potentially highlights this problem is how the 2018 Arizona Uber incident failed to take into account the pedestrian that the car struck, partially because it was confused since she had a bicycle with her. This might imply that people on wheelchairs or scooters might also fail to be adequately recognized and run into more accidents with self-driving vehicles around.

"Indeed, the way in which "disability" resists fitting into neat arrangements points to bigger questions about how other identity categories, such as race, sexual orientation, and gender, are (mis)treated as essential, fixed classifications in the logics of AI systems, and in much of the research examining AI and bias."

- Another particularly poignant point made in the paper is that disability is more than the physical and mental posture of the individual and more so as to how society responds to it.

**What does "normal" mean to AI?**

- We learned above that at varying times, different groups have been designated as disabled leading to unfortunate consequences.
- But, in the case of automated systems, the harms are easy to magnify.
- Especially when the systems are making significant decisions about someone's life, we don't want to have rigid, faulty categories that can jeopardize the safety of individuals.
- An example from the Deaf community mentions how instead of using technology to bring them over to the hearing world, they believe the failure to be the unwillingness of people to learn sign language in communicating them as the problem.
- With invigilation systems relying on emotion and face recognition, especially under the pandemic of 2020, there are visceral risks to the ability of people to participate in activities because of different notions of normal within the system.

**Reverse Turing Tests causing harm**

- A reverse Turing Test is one where we are asked to prove our humanity to the machine, often for security purposes.
- But, it is most of the time looking for a specific kind of human, one that falls within its definition of normal. What this means is that it ignores the potential that people with different conditions might be slow to click on things, or might have speech differences which might flag them as anomalies unnecessarily.
- While not a Turing Test, Amazon in its warehouses utilizes monitoring software that is meant to extract as much labor as possible from its workers. This is made worse by the fact that it leads to injuries and places an even greater burden on those with disabilities.
- Sometimes, technology running in the background such as mouse movements and click actions on a webpage can be surreptitiously used to infer whether someone has a disability, certainly something that is not only nefarious but done without consent from the users visiting that webpage.

**Can additional data help?**

- According to the examples in the paper, this might only serve to reinforce the normative model at the core of the AI system which might further exacerbate the problem.
- Even when more data is included, there is a problem that not only does it not adequately represent people but that it also intrudes on privacy through unnecessary surveillance in the interest of capturing more data.
- More so, the risk is then pushed onto the historically marginalized groups while the benefits accrue to already powerful actors that stand to make a profit from the deployment of such systems.
- Often those who have rare conditions can't be sufficiently protected when their data is collected in the interests of making more inclusive AI systems.
- Additionally, there is no guarantee either that such data will be kept out of the hands of insurance companies or other actors who can stand to benefit from this information by differentially charging those people.
- As is commonplace in the AI world, in collecting large-scale data, clickworkers are often employed to label data and they might be provided with little or no guidance as to what disability means, and they might erroneously label data in way that reinforces the normative model and further entrenches discrepancies in how people with disabilities are addressed by AI systems.
- Considerations about the regional disparities in this discussion, as I've pointed out with my co-author Victoria in an MIT Technology Review article, are important to consider. In an example about the differences between how autism might be expressed and perceived, differences in children in Bangladesh vs. America created differential results skewing the kind of support that was provided.

**Work and Disability in the context of AI**

- In systems that are used to automatically scrape data and make decisions about candidates in hiring, even when companies manufacturing these systems claim that they are able to adequately able to address biases on the disability front, there is no guarantee that if such a determination is made by the system and provided to the employer making the decision, harm won't be avoided.
- In fact, it exacerbates the problem by giving the decision-makers even more data (perhaps surfacing an unexpressed disability) that can be used to discriminate against individuals while simultaneously stripping away their ability to bring suits against the employers in case they are discriminated against because of the opacity of the system.
- This weakens the protections offered by the ADA in the United States as an example.
- Some companies who receive subsidies from the government to employ people with disabilities might use tactics like compensating people with gift cards instead of money creating unequal working conditions and structures aggravating harm in the workplace.

**Are there accountability measures?**

The credo of this community is "Nothing about us without us".

- The above credo is importantly expressed in grant proposals as an example, but often organizations like the NSF don't follow-up on whether that has been adhered to after the funding has been provided.
- Without accountability and follow-up, we risk creating a false sense of comfort when the real harms continue to remain unmitigated in the field.

**Other ethical concerns**
- People might also forget that when using such technological fixes, we create additional concerns such as the compromising of bystander privacy, say for example, when a vision system is used to aid someone with visual impairments.
- There is also a locking in of corporate interests in creating a dependence on such systems when they close them off to scrutiny and modification that might limit the ability of people to fix them if they are broken or adapt them to better meet their needs.

**Key challenges**
- Given how these technologies are built in a very proprietary manner, it is hard at the moment to see how we can move from mere inclusion to agency and empowerment of individuals.
- Especially pointing to the case of the civil rights movement, the paper concludes on a powerful note mentioning that we lose when we let others speak for us.

**Conclusion**

The paper offers many rarely (for the AI community) discussed facets of what the experience of those with disabilities is when they interact with AI systems. It situates the concerns in the wider discussion of the disability rights movements and years of scholarship and activism in the space. It also provides some guidance on how people designing and developing these systems can do better when it comes to better meeting the needs of those with disabilities.

- As we build AI systems, keeping ethics concerns should extend beyond the familiar concerns of racial and gender bias to consider intersectionality with other aspects which are traditionally excluded, like disability which raise unmitigated concerns even when traditional vectors are addressed.

- Even when technical systems are built with the explicit needs of people with disabilities, this doesn't mean that ethics concerns such as bias and privacy are automatically managed. It still requires deliberation and careful consideration, especially active efforts to include them in the design and development process.

## Theorizing Femininity in AI: A Framework for Undoing Technology's Gender Troubles

([Original paper](#) by Daniel M. Sutko)
(Research summary by Alexandrine Royer)

**Overview:** The predominance of feminized voice assistants points to AI's tendency to naturalise gender divisions. This paper draws on the science fiction narrative of Her and Tomorrow's Eve to offer a critical understanding of how femininity serves as a means of domesticating AI all the while reproducing gender relations.

---

For anyone keeping a close eye on technological development, the statement that AI has a gender problem in its conceptualization, design, and output comes as no surprise. AI automates the existing social status quo, and its developers deploy consumer products that reproduce deeply ingrained gender divisions and roles. In 2001, Microsoft's *Halo: Combat Evolved* game introduced the character of Cortana, an AI-powered female voice and occasional sexy cybernetic hologram, made to assist main players. By no coincidence, Cortana would later become the name of Microsoft's feminized and subservient artificially intelligent voice-interface, in the likes of Apple's Siri and Amazon's Alexa. The way gender is imagined will naturally feed into how it is materialized; as Sutko asserts, "technological representation intersects with technological design."

Responding to what he terms "technofeminist calls to unpack the gendered politics of technologies," Sutko offers a critical analysis of how ideas of femininity continue to be reproduced in technologies across three dimensions: docile labour, replaceable embodiment, and artificial intelligence. The author draws on science fiction narratives to underscore how femininity serves as a mode of domesticating new technology, with technology, in turn, materializing these ideas of gender relations. For Sutko, "the dreams and designs of AI (re)produce discursive formations rooted in the subjugation of others, particularly women". Combining insights from media archaeology, critical theory, gender studies, and media theory, Sutko delineates how gender divisions are sustained and naturalized through technology.

**Fictional characters, real illustrations**

Novels, movies, video games, and other forms of media provide a framework to evaluate the narratives, tropes, and underlying gender patterns in the depiction and, often idealized views of, the creation of artificial intelligence. A recurrent plotline in science fiction literature and cinema revolves around masculine relationships with feminized technologies. Spike Jonze's 2013 *Her* depicts a lonely middle-aged divorcee who falls in love and begins a sexual relationship with his seductively voiced virtual assistant. A century back, the misogynistic Victorian novel *Tomorrow's Eve* centers on a scientist who invents a female android for his lovelorn aristocratic friend; the aristocrat seeks to replace his shallow wife with an artificial 'ideal' woman.

Both stories are told at a turning point of new technology, *Eve* during the introduction of technical media (e.g. the gramophones, photography, cinematography, etc.) and *Her* during the introduction of computational media (e.g. smartphones, virtual assistants, etc.). As Sukto underlines, "cultural artifacts like *Eve* and *Her* are useful to think with because they help us identify which structures of feeling are being dismantled or reinforced." Adding that, despite being more than a hundred years apart, "both stories envision using contemporary communication technologies to create an ideal female lover and associate femininity with new technology." In *Her* and *Eve*, female technologies are made to produce affective labour docilely and replace real-life 'dysfunctional women', yet their artificiality is also a threat to the men who own them- reminding them of their very 'unnaturalness'.

**Feminized labour, embodiment, and intelligence**

By arranging our calendars, making calls, and sending reminders, voice interfaces are designed to provide stereotypically gendered labour. As argued by Sutko, "gendering the interface-whether Sam or Siri- essentializes labour divisions and female bodies/minds as docile, disciplined to respond to others' needs". This pattern reinforces the view that men master technology and women are subordinate to it. Yet, Sutko cautions that this division is muddled by our daily interactions with technology, where Siri is often the one programming us and not the other way around. Speaking from his own experience, Sutko notes, "I have been trained by Siri to enunciate properly, to ask only certain kinds of questions, in a particular way, to get a useful response."

Along with changing or reaffirming conceptions of labour, virtual assistants and androids complicate our understanding of the division between mind and body. For Sutko, "both *Eve* and *Her* erase the politics of embodiment by reducing embodiment to a state of mind." Sutko points out how both stories reinforce a view of the unknowable inner workings of feminized technology –"who knows why they act that way" – all the while depicting them as machines who are programmed to care, nurture and comply. This tension replicates familiar tropes surrounding femininity and the characterization of women.

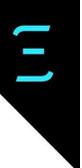

As Sutko summarizes, "essentialized categories of gender, nature, culture and technology are replicated in our fantasises of artificial intelligence without introspection and may be reproduced or rearticulated depending on the design of such AI".

Though our fears and fantasies over intelligent machines often eclipse this fact, AI is an artificial imitation of human intelligence – it is an attempt to create something equal and less than human. In Sutko's view, there is a willingness to have AI systems respond to all our desires without affording them any of their own. Ideal AI must be "caring, nurturing, responsive, attentive, helpful but not willful, smart but not overly so, replaceable, customizable, available." Computer scientists know that any AI system is unlikely to meet all these requirements soon. Nevertheless, the spread of voice interfaces, and popular fantasized depictions, will impact human cognition and communication. Simulations are also not just *descriptive*, but *prescriptive*.

Going forward: prescriptions for future AI

As technological development advances, scientists will continue their endeavours to mirror human cognition and behaviour in machine-shaped forms- with what it means to be human remaining a shifting category. Sutko argues that society must 'take seriously' these fictionalized accounts of AI: "If we consider the expectations for AI seriously as anticipatory models of our own future, then we open up political possibilities for an ethical obligation towards AI".

Both *Her* and *Eve* also point to the changes in human behaviour as the leading characters co-evolve alongside their technological companions. As Sutko underscores, human behaviour will be modified by closer technological interaction- "we become machine-shaped", turning the Turning test on its head. His prescriptions for future interventions include introducing concepts of politeness, consent, and respect in "the way AI addresses us" but also "the way we are required to address AI." Other suggestions include equipping AI with gender-neutral tones or allowing the AI to be modulated according to the user's preference. While Sutko recognizes that these solutions do not redress the long-standing structural inequalities that have led to feminized AI, he argues that "they are a place to begin unpacking the black-boxed intersectionalities of gender, bodies, labour, and technology".

With AI companions continuing to pervade multiple spheres of human activity, a thorough and critical examination of its development beyond the rational/technical and into the heuristic will add some welcomed nuances to our understanding of the intersection of gender and AI.

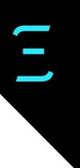

### The Limits of Global Inclusion in AI Development
([Original paper](#) by Alan Chan, Chinasa T. Okolo, Zachary Terner, Angelina Wang)
(Research summary by Alexandrine Royer)

**Overview:** Western AI institutions have started to involve more diverse groups in the development and application of AI systems as a response to calls to level out the current global imbalances in the field. In this paper, the authors argue that increased representation can only go so far in redressing the global inequities in AI development and outline how to achieve broader inclusion and active participation in the field.

---

When it comes to AI development, the scales are generally tipped in favour of countries in the Global North. It is evident that "those best-positioned to profit from the proliferation of artificial intelligence (AI) systems are those with the most economic power." To counterbalance these global inequities, Western institutions have called for more diverse groups in the development and application of AI.

For Chan, Okolo, Terner and Wang, greater representation can only go so far while structural inequalities remain unchallenged. To enact far-reaching change requires a redistribution of power. If we fail to provide a level playing field, Chan et al. caution that "the future may hold only AI systems which are unsuited to their conditions of application, and exacerbate inequality." By taking a critical look at the limitations of inclusion in datasets and research labs, Chan et al. offer a list of potential barriers and steps to alleviating the power imbalances in AI development.

**Invisible players**

The invisibility of countries in the Global South in conference publications is generally reflective of the broader inequality in AI development; it signals where innovation, hence funding, is happening. Both the NeurIPS and ICML, two of the world's largest machine learning conference, did not feature countries from Latin America, Africa, nor Southeast Asia in their top ten publication index. In terms of the top ten institutions, the US dominated the list with 8, including familiar names such as Google, Facebook, and Microsoft. Left with a skewed perspective, the lack of contextual knowledge of institutions in the Global North towards the Global South's realities can cause social and ethical harms in designing and implementing AI systems.

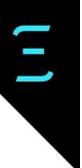

Calls for inclusion may lead to what Sloan et al. have termed "participant washing," where "the mere fact that someone has participated in a project lends it moral legitimacy." The term "participant washing" perfectly encapsulates the self-congratulatory tendency to treat representation as a series of quotas to be filled and boxes to be ticked, without ever engaging with the root causes.

**Diversifying the data-gathering pipeline**

With its abundance of capital, well-funded research institutes and technical infrastructure, the Global North is well-positioned to lead AI innovation. However, its advantageous position is in large part due to riches accumulated through colonial exploitation. While calls to diversify the data-gathering pipeline are steps in the right direction, the authors delineate how the process is much more complicated – it involves dismantling long-standing global inequities. When it comes to data collection, large image datasets, such as ImageNet and OpenImage, remain heavily US and Euro-centric. For Chan et al., current data collection practice "neglect consent and poorly represent areas of the Global South." The focus on the accumulation of large – and frequently uncompensated- data by foreign institutes to diversify datasets obscure whether such data should be collected in the first place.

As for data labelling, there are limited possibilities for individuals in the Global South to participate or achieve any form of upward mobility. Given the tedious and repetitive nature of the task, data labelling companies often seek out a low-wage workforce from the Global South, contracted via crowdsourcing platforms such as MTurk and Samasource. These third-party providers accentuate the disparity between data labelling companies' profits and workers' earnings, leading Chan et al. to assert that "in parallel with colonial projects of resource extraction, data labelling as the extraction of meaning from data is no way out of a cycle of colonial dependence."

By being far-removed from the decision-making centers, "workers are contributing to AI systems that are likely to be biased against underrepresented populations in the locales they are deployed in and may not be directly benefiting their local communities." Calls for participation in AI development often presuppose that members of the Global South have computing devices and internet connection readily available. Global power dynamics fuel the uneven growth of tech, leading Chan et al. to affirm that "It is instructive to view inclusion in the data pipeline as a continuation of this exploitative history." As the chasm between data labellers and the downstream product deepens, these workers will continue to be severely exploited and alienated from the fruits of labour.

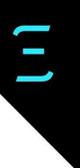

**Rethinking research labs**

With AI becoming increasingly present in global consumers' daily lives, major tech companies, from Microsoft, IBM, and Google, have expanded their development centers and research labs outside of the Global North. Research labs in the Global South tend to be concentrated in specific countries, such as India, Brazil, South Africa, and Kenya. Fears over political and economic instability and a misguided view of local talent have led to limited investments in other areas. Chan et al. also underscore how many lab directors and staff within the Global South are frequently recruited from elsewhere; hence local representation is sorely lacking. As the authors argue, "to advance the equity within AI and improve inclusion efforts, it is imperative that companies not only establish locations in underrepresented regions but hire employees and include voices from those regions in a proportionate manner."

True inclusion requires underrepresented voices to be present at all levels of a company's hierarchy, including upper management. It follows that opportunities must be provided for local residents to acquire the skills and training needed for management roles and guide critical decisions. Some notable examples of grassroots AI education and training initiatives include Deep Learning Indaba, Data Science Africa, and Khipu AI in Latin America. The authors refer to the sentiment expressed by Makashane, a nonprofit organization committed to improving the representation of African language in natural language processing, that "we [Makashane] do not support shallow engagement of Africans as only data generators or consumers." Actual representation will ensure that "the benefits of AI apply not only to technical problems that arise in the Global South, but to socioeconomic inequalities that persist around the world."

**Committing to global inclusion**

When it comes to fast-paced economic development, South Korea's import substitution industrialization policy (ISI), where the state endeavours to replace imports with domestic production to stimulate home-grown competitive industries, has offered an exemplar model. The authors suggest AI development could benefit from the lessons learned from ISI policies. Instead of relying upon "foreign construction of AI systems for domestic application, where any returns from these systems are not invested domestically, we encourage the formation of domestic AI development activity." For ISI-like policies to succeed, Chan et al. affirm that "domestic expertise must be developed in tandem to shape the future of AI development and reap its large profits". As American trailblazer and poet Audrey Lourde famously states, the "master's tools will never dismantle the master's house. They may allow us temporarily to beat him at his own game, but they will never enable us to bring about genuine change." Global AI development, as it currently stands, is likely to leave the Global South to bear the burnt of algorithmic inequity. True global inclusion in AI, and the potential to bring genuine change, cannot be done without a redistribution of power.

## Unprofessional peer reviews disproportionately harm underrepresented groups in STEM

([Original paper](#) by Nyssa J. Silbiger, Amer D. Stubler)
(Research summary by Alexandrine Royer)

**Overview:** The peer review process is integral to scientific research advancement, yet systematic biases erode the fair assessment principle and carry downstream effects on researchers' careers. Silbiger and Stubler, in their international survey among researchers in STEM, demonstrate how unprofessional peer reviews disproportionately harm and perpetuate the gap among underrepresented groups in the field.

---

The principle of academic integrity finds its concrete form in the process of peer review, with researchers scrutinizing scientific findings to ensure they meet the highest standards of quality and reliability. The fast pace of scientific advancement and the institutional pressures to keep up — what some have termed the "publish or perish" environment of universities, has chipped away at the peer-review process's credibility. The process also does not stand outside of systematic biases that harm underrepresented groups in the science, technology, engineering and mathematics (STEM) fields. As it relies on the reviewers' good faith, academics have been reluctant to investigate the downstream effects of unprofessional reviewer conduct.

Silbiger and Stubler conducted an international survey among participants in STEM fields to gain better insight into the situation and determine unprofessional conduct's pervasiveness. They invited participants to share their perception of scientific aptitude, productivity, and career trajectory after receiving unprofessional comments to assess the long-term implications of poor conduct in peer reviews. The survey results confirmed that "unprofessional reviews likely have and will continue to perpetuate the gap in STEM fields for traditionally underrepresented groups in the sciences."

**Acting on bad faith**

There is growing evidence that the scientific review process is tainted with biases, leading to objectivity violations by the reviewer towards the submitting authors' attributes and identity, whether it is their nationality, institutional affiliation, language, gender, sexuality, etc. The reviewer's perceptions of the field, from scientific dogma to discontent with methodological advances, can also unfairly harm the submitting author's assessment. Scientific journal editors, who oversee the distribution of reviews to authors and the policing of comments, work under a tight time crunch and often have inherent biases of their own. While past studies, such as the infamous "Joan vs John," have manipulated authors' attributes and identity to demonstrate the peer-review system's unfairness, few have documented the actual content of the comments attached by the reviewers across intersectional gender and racial/ethnic groups.

In their survey, Silbiger and Stubler defined unprofessional review comments as:

1. Lack of constructive criticism
2. Are directed at the author(s) rather than the nature or quality of the work
3. Use personal opinions of the author(s)/work rather than evidence-based criticism
4. Are mean-spirited or cruel

Some of the respondents shared comments received by reviewers that were abjectly callous, misogynistic and racist, such as "this paper is, simply, manure" or "what the authors have done is an insult to science" or "the first author is a woman, she should be in the kitchen" and "the author's last name sounds Spanish, I didn't read the manuscript because I'm sure it is full of bad English" to name a few among others.

Such comments reflect how underrepresented groups in STEM continue to be vulnerable to what Silbiger and Stubler refer to as stereotype threat, "wherein negative social stereotypes about performance abilities and aptitude are internalized and subsequently expressed".

In their survey, Silbiger and Stubler assessed the downstream of unprofessional peer reviews on four intersecting gender and racial/ethnic groups: "women of colour and non-binary people of colour", "men of colour", "white women and white non-binary people" and "white men". They gathered responses from 1, 106 individuals from 46 different countries across 14 STEM disciplines. The results showed that over 58% of respondents had received an unprofessional review, with 70% among these testifying to multiple instances.

While the authors found no "significant differences in the likelihood of receiving an unprofessional review among the intersectional groups, there were clear and consistent differences in downstream effects between groups in perceived impacts on self-confidence, productivity and career trajectories after receiving an unprofessional review." Compared to the three other groups, white men reported less doubt in their scientific aptitude following a bad review. They were also the least likely to indicate that such reviews greatly hampered their number of publications per year. The groups confessing to the highest levels of self-doubt were those who confronted the highest delays in productivity. Women of colour and non-binary people of colour were the most likely to report that unprofessional reviews had contributed to significant delays in their career advancement.

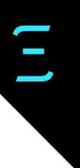

**Reviewing the peer review process**

The results of Silbiger and Stubler surveys correlate with other publication patterns in STEM fields, where men have 40% more publications than women on average, and women continue to be severely underrepresented as both editors and reviewers in the peer review process. While the authors admit the limitations of their survey design, such as being administered in English only and the temporal element to the authors' feedback, their results nonetheless clearly underscore how "unprofessional reviews reinforce bias that is already being encountered by underrepresented groups on a daily basis."

The differences in the responses reported by white women and white non-binary people, and women of colour and non-binary people of colour showed the importance of including an assessment of intersectional groups; the two groups had significantly varying responses in perceived delays of career advancement. For Silbiger and Stubler, these discrepancies further indicate that "receiving unprofessional peer reviews is yet another barrier to equity in career trajectories for women of colour and non-binary people of colour, in addition to the quality of mentorship, intimidation and harassment, lack of representation and many others."

The pervasiveness of unprofessional conduct by reviewers should not go undenounced by members of the scientific community; bad faith prevents good science from moving forward. An overhaul of the peer-review system by academic institutions ought to be promptly set in place, with the Silbiger and Stubler offering a series of recommendations such as:

1. Making peer review mentorship an active part of student and peer learning
2. Following previously published best practices in peer review
3. Practicing self-awareness and interrogating whether comments are constructive and impartial
4. Encouraging journals and societies to create explicit guidelines for the review process and penalize reviewers who act in an unprofessional manner
5. Increasing vigilance on the part of editors to prevent unprofessional reviews from reaching authors directly

Unprofessional conduct is found throughout the peer-review process. Along with biased reviews, scholars' have deployed strategies to increase their publication index through citation rings and the preselection of peer reviewers, leading us to question the validity of the peer-review process and scientific findings themselves. The peer review system, rather than ensuring an unbiased, fair assessment of scientific merit and credibility, contributes to the replication crisis and perpetuating long-standing inequalities in STEM. As it stands, good science may be penalized by the peer review system, and bad science rewarded by it.

# The Algorithm Audit: Scoring the Algorithms that Score Us
([Original paper](#) by Shea Brown, Jovana Davidovic, Ali Hasan)
(Research summary by Dr. Andrea Pedeferri)

**Overview:** Is it right for an AI to decide who can get bail and who can't? This paper proposes a general model for an algorithm audit that is able to provide clear and effective results while also avoiding some of the drawbacks of the approaches offered so far. The model involves ethical analysis of algorithms into a set of practical steps and deliverables.

---

Is it right for an AI to decide who can get bail and who can't? Or that it can approve or disapprove your loan application, or your job application? Should we trust an AI as we would trust a fellow human making decisions? Those are just a few examples of ethical questions that stem from the widespread application of algorithms in many decision-making processes and activities. This growth of AI in replacing humans has triggered an arms race to provide capable and efficient AI evaluations.

One viable way to provide guidance and evaluations in these settings is the use of third-party audits. As audits are widespread in the evaluation of decision-making processes and procedures that are wholly or mostly human-centred (think of financial audits, for instance), it is natural to refer to the audit process when we look for ways of providing an ethical assessment of AI's algorithms.

In their article "The algorithm audit: Scoring the algorithms that score us", Shea Brown, Jovana Davidovic and Ali Hasan propose a general model for an algorithm audit that is able to provide clear and effective results while also avoiding some of the drawbacks of the approaches offered so far.

When we look at regulators, their primary interest is to assess "the algorithm's negative impact on the rights and interests of stakeholders, with a corresponding identification of situations and/or features of the algorithm that give rise to these negative impacts." The authors note however that "recently much criticism has been directed at early attempts to provide an ethical analysis of algorithms. Scholars have argued that using the classical analytic approach that over-stresses technical aspects of algorithms and ignores the larger socio-technical power dynamics has resulted in ethical approaches to algorithms that ignore or marginalize some of the primary threats (especially decision-making and classification) that algorithms pose to minorities."

In their paper, the authors thus aim to provide a more comprehensive framework for algorithm audit that avoids those shortcomings by modelling ethical analysis of algorithms into a set of "practical steps" and deliverables that can be both broadly applied and used by a variety of different stakeholders.

In particular, they focus on what they believe to be a critical point that has been mostly overlooked by current ethical audits: the context of the algorithm. By that, they mean the sociological and technical environments within which the algorithm is employed. This includes a broad and wide range of processes, settings, and dynamics that go beyond the technical aspects of the algorithm itself but affect all the relevant situations and stakeholders that fall within the algorithm's range of functioning and applications. The authors provide as an example the case of algorithms about loan risk: "the negative impacts of a loan risk tool do not simply depend on whether the algorithm is statistically biased against, for example, some minority group because of biased training data; more importantly, the harm emerges from the way a loan officer decides to use that loan risk tool in her decision whether to give out a loan to an applicant".

So, focusing on the context allows us to create more precise and relevant metrics about specific features for specific stakeholders' interests. The proposed auditing tool is built from those metrics; this is why, according to the authors, it is essential that a careful analysis of the context be the primary step in the audit.

The actual framework proposed in the paper consists of "(1) a comprehensive list of relevant stake-holder interests and (2) a measure of the algorithm's potentially ethically relevant attributes (metrics). A clear description of the context is needed both to generate a list of stakeholder interests (1) and to evaluate the key features of the algorithm, i.e. metrics (2). Once steps (1) and (2) are completed we can (3) evaluate the relevance of a good or bad performance of an algorithm on some metric for each stakeholder interest. We can then use the metrics score (2) and the relevancy score (3) to determine the impact of the algorithm on stakeholder interests."

With these basics points in mind, the authors go on to provide a fine-grained description of the major elements that should be considered in an ethical audit by first focusing on the stakeholder interests and rights. This is followed by a thorough elucidation of the different types of metrics to be taken into consideration for different categories such as bias, effectiveness, transparency, direct impacts and security & access.

The most common present audit processes are often limited to the logical and mathematical operations that compose the input-output function of the algorithm. In the proposed structure, an algorithm should also include the "larger socio-technical structure that surrounds this function". So, for example, in a description (by use of metrics) of an algorithm, we should include "facts about how the output of the function is used in decision making, and whether the actions taken are done so autonomously or with a human in the loop". The key metrics are then the "ethically salient features" of the algorithm in the relevant context. An auditor should be able to test and provide assessments (that can be of different kinds such as narrative, numerical or categorical) for each metric in an objective way, that is, by assessing each metric independently "of any of the other metrics and independently of stakeholder interests".

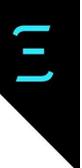

The key feature for the auditing method presented in the papers is the "Relevancy matrix". Here is what the authors mean by that. The two necessary components of the auditing process are knowing what are the stakeholder interests that could be affected in a specific context by the algorithm and knowing all the metric scoring of the algorithm itself. Although those two aspects independently provide much useful information for an overall ethical assessment of the algorithm, their connection allows a full and complete picture that is necessary to produce meaningful and effective audit results. The idea is to be able to answer the following: "for each stakeholder interest, how much could each metric threaten that interest if the algorithm performs poorly with respect to that metric?".

The solution proposed is to build a relevancy matrix that connects each interest to each metric. This bi-dimensional matrix produces a sort of still-picture that captures and identifies low-scoring metrics that have high relevance to stakeholder interests. The auditor is therefore equipped with a powerful tool that highlights areas of the potential negative impact of the algorithm. Moreover, the narrative assessment that comes with the metrics is a powerful resource to help produce strategies in order to alleviate those potential risks.

The central focus on context-dependency and the effective deliverables represented by the relevancy matrix allow this new approach to ethical audits to account for the concerns on obsessing over technical aspects of algorithms and ignoring "the larger socio-technical power dynamics" part of the larger context that surrounds algorithms. According to the authors, this narrow vision of audit has so far resulted in incomplete or incoherent approaches to ethical evaluations of algorithms that miss or leave space to hazards that can have a big impact, especially on minorities. Brown, Davidovic and Hasan believe that the model presented in their paper can correct those flaws while "staying within the constraints of what a genuine audit can do, which is to provide a consistent and repeatable assessment of (in this case) algorithms".

**Diagnosing Gender Bias In Image Recognition Systems**
([Original paper](#) by Carsten Schwemmer, Carly Knight, Emily D. Bello-Pardo, Stan Oklobdzija, Martijin Schoonvelde, and Jeffrey W. Lockhart)
(Research summary by Nga Than)

**Overview:** This paper examines gender biases in commercial vision recognition systems. Specifically, the authors show how these systems classify, label, and annotate images of women and men differently. They conclude that researchers should be careful using labels produced by such systems in their research. The paper also produces a template for social scientists to evaluate those systems before deploying them.

---

Following the insurrection in the United States, law enforcement was quickly able to identify rioters who occupied the Capitol and arrested them shortly after. Their swift action was partly assisted by [both professional and amateur use](#) of facial recognition systems such as the one created by Clearview AI, a controversial startup that scraped individual pictures from various social media platforms. However, researchers [Joan Donovan and Chris Gillard](#) cautioned that even when facial recognition systems produce positive results such as in the case of arresting rioters, the technology should not be used because of myriad flaws and biases embedded in these systems. The article "Diagnosing gender bias in image recognition systems" by Schwemmer et al (2020) provides a systematic analysis of how widely available commercial image recognition systems could reproduce and amplify gender biases.

The author begins by pointing out that bias in visual representations of gender has been studied at a small scale in social sciences like media studies. However, systematic large scale studies using images as social data have been limited. Recently, the availability of image labeling provided by commercial image classification systems shows promise in social science research. However, algorithmic classification systems could be mechanisms for reproduction and amplification of social biases. The study finds that commercial image recognition systems can produce labels that are both correct and biased as they selectively report a subset of many possible true labels. The findings demonstrate the idea of "amplification process," or a mechanism through which gender stereotypes and differences are reinscribed into novel social arenas and social forms.

The authors examine two dimensions of biases: identification (accuracy of labels), and content of labels. They use two different datasets of pictures of Congress Members of the United States. The first dataset contains high-quality official headshots, and the other set contains images tweeted by the same politicians. The two datasets are treated as treatment and control datasets. The first dataset is uniformed while the second varies substantially in terms of content. They primarily use results using Google Cloud Vision (GCV) for the analysis, then compare the results with labels produced by Microsoft Azure and Amazon Rekognition. To validate results produced by GCV, they hire human annotators through Amazon Mechanical Turks to confirm the accuracy of the labels.

The authors found two distinct types of algorithmic gender bias: (1) identification bias (men are identified correctly at higher rates than women), and (2) content bias (images of men received higher-status occupational labels, while female politicians received lower social status labels).

**Bias in identification**

The majority of bias literature focuses on this type of bias. The main line of inquiry is whether a particular algorithm predicts accurately a social category. Scholars have called this phenomenon "algorithmic bias," which "defines algorithmic injustice and discrimination as situations where errors disproportionally affect particular social groups."

**Bias in content**

This type of bias takes place when an algorithm produces "only a subset of possible labels even if the output is correct." In the case of gender bias, the algorithm systematically produces different subsets of labels for different gender groups. The authors called this phenomenon "conditional demographic parity."

The research team found that GCV is a highly precise system, which produced labels that human coders also agreed with. However, false-negative rates are higher for women than men. In the official portrait dataset, men are identified correctly 85.8% of the time, while 75.5% of the time for women. In the found Twitter dataset, the accuracy is much lower and more biased: 45.3% for men, and only 25.8% for women.

The system labels congresswomen as girls, and overly focuses on their hairstyle, color of their hair while returning high-status occupational labels such as white-collar workers, businessperson, and spokesperson to congressmen. In terms of occupation, it returns labels such as television presenters to congressional female members, a more female-associated professional category than businesswomen. They conclude that from all possible correct labels, "GCV selects appearance labels more often for women and high-status occupation labels more for men." Images of women received 3 times more labels categorized as physical traits and body. Images of men receive about 1.5 times more labels categorized as occupation. In the found Twitter dataset, congressional women are substantially categorized as girls. The authors found similar biases in Amazon and Microsoft systems and noted that Microsoft's system does not produce high accuracy labeling.

This research is particularly needed as it shows systematically how image recognition technology should not be used in social science research for gender research projects. Furthermore, the research team provides a template for researchers to evaluate any vision recognition system before deploying it in their research. One question that remains for the wider public is whether vision recognition systems should not be deployed in daily and commercial practices at all. If they were to be used, how could an individual or an organization evaluate whether they would amplify social biases through such technology?

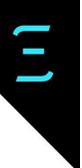

## Go Wide: Article Summaries (summarized by Abhishek Gupta)

### The UK Government Isn't Being Transparent About Its Palantir Contracts
(Original *Vice* article by Gabriel Geiger)

Palantir is notorious for the lack of transparency in how it operates. This has been demonstrated time and again in the US and now a report published in the UK shows that this pattern is being repeated there with some of the government contracts that they have secured there. A lot of freedom of information requests weren't adequately responded to by the government shrouding these engagements in further secrecy.

Access to data held by the government is a really powerful instrument for firms looking to gain an edge when it comes to training their AI systems. The problem unsurprisingly is that such partnerships with data in the hands of the private sector without requisite accountability mechanisms is ripe for abuse of sensitive data belonging to citizens that can have many downstream harms that are unanticipated and unmitigated at the moment. Especially when the CEO of the company publicly declares that he doesn't care much about it, it is a clear sign that we need to be extra cautious in demanding more transparency.

In a request for comments, Palantir gives non-committal answers that further exacerbate the problem. Even if anonymization is applied to the data, when looking at the implications of large-scale access, there are macro-trends that can still harm public welfare if inferences are drawn from that and weaponized against people, especially those who are marginalized. Under the GDPR, hiding behind the defense of merely being a data processor who looks to the data controller (the government in this case) is a cop-out that needs to be highlighted and properly addressed.

### Root Out Bias at Every Stage of Your AI-Development Process
(Original *Harvard Business Review* article by Josh Feast)

This article lays out a fairly clear mandate in terms of the importance of having responsibility being borne by the manufacturers of AI systems rather than those who are using them. In a call to leaders to pay more attention to the AI lifecycle, whether or not they have technical expertise, is an important consideration that will become a critical part of any leadership role in the future. In particular, this might call for leaders with a non-technical background to at least engage in equipping themselves with a fundamental understanding of how AI systems work but more importantly the different stages of the AI lifecycle.

The author in this article does a coarse-grained analysis of the lifecycle, splitting the stages into pre-, in-, and post-processing. This is a useful distinction, at least at the level where we might expect some executives to operate. It helps to make the more detailed components of the MLOps process more accessible.

While the steps highlighted in the article are typical of what you would expect in terms of due diligence when trying to address bias like having multiple annotations from a diversity of human labellers to prevent biases from creeping in, model monitoring, etc. what did stand out was the emphasis on seeing this as an ongoing process that doesn't stop after you've deployed the system. It is essential that periodic checks be run to generate assurances that the system is still operating within expected boundaries.

## The Coming War on the Hidden Algorithms that Trap People in Poverty
([Original *MIT Tech Review* article](#) by Karen Hao)

Detailing the harrowing journey of an individual who fell under "coerced debt", something that happens when someone close to the victim or a family member perpetrates financial abuse due to their intimate knowledge of the individual's private data, this article showcases how in the era of algorithmic decision making, repercussions extend beyond just the immediate domain where such fraud might be perpetrated. Given that there is widespread data sharing across agencies and how credit ratings that are calculated in an opaque manner can be used in many downstream tasks, getting hit in one part affects all parts of our lives. Unfortunately, the impacts of these also have implications for the lawyers who are trying to help people who face the brunt of this.

Talking to a few lawyers defending victims, the article talks about how since 2014 the prevalence of this phenomenon has gone up, and lawyers are playing catch up in trying to gain a better understanding of how these algorithms work and how they can fight against the organizations, sometimes government agencies that are using these systems. In a case in Michigan, thousands of people were flagged for fraud and were denied access to government-provided unemployment services. In another case, someone was offered fewer hours of support as they got sicker, the opposite of what one would expect. When faced in court, the nurses didn't have an answer as to how the decisions were being made. The reason being that they didn't have an insight into what was being fed into the system, and how it was weighing the factors. They are after all medical experts and not computer scientists and shouldn't have to bear the burden to scrutinize the system.

Finally, the procurement process is also quite opaque which exacerbates the problem. The push for adopting such systems, especially in the pandemic with reduced availability of staff and increased strains and demands for services, is obvious. But, kicking humans out of the loop and hoping that the systems can perfectly model the needs of people is fallacious at best. Lawyers are banding together to better educate each other so that they are better equipped to aid their clients and bring about justice in the face of abstract algorithms pushing back on basic rights of people.

### Algorithms Behaving Badly: 2020 Edition
(Original *The Markup* article by The Markup Staff)

Incidents of racial biases in algorithmic systems have been persistent. It is a situation that only reifies existing biases and stereotypes in society, quite against the layperson's perception that numerical systems are value-agnostic, a notion referred to as math-washing. Quoting the example of how Black athletes are treated differently in the NFL when it comes to the impacts of concussions, the associated treatments, and compensation comes from their being classified as having lower cognitive function and hence having differential effects of concussions throughout their career. Such societal ills are unfortunately captured all too well in our algorithmic systems since they are mostly a reflection of the data that is used to codify human interactions in the real world.

While most of the other examples mentioned in the article are ones that we have covered in the past, one that particularly caught my attention was the one that talks about how Whole Foods tries to figure out stores where there might be unionization attempts by tracking a variety of factors like the local unemployment rate, number of complaints, etc. so that they might be quashed before causing the company too many problems.

The article does conclude on a positive note that talks about how when people are made aware of algorithmic systems and how they might be impacting them, they can take actions to counter some of the injustices that are being leveled against them. Note that this is only possible if people are aware that there are such injustices being perpetrated against them in the first place. This is one of the primary reasons that I think that the civic competence work being done at the Montreal AI Ethics Institute is so important because it raises awareness about where one might encounter such systems, and what signs to look for to determine if they might be facing disparate outcomes because of their group membership or identities.

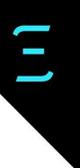

## Another Arrest, and Jail Time, Due to a Bad Facial Recognition Match
([Original *NY Times* article](#) by Kashmir Hill)

Are we bound to continue repeating mistakes even when obvious flaws are pointed out to us? It seems to be the case with facial recognition technology that users of this technology just can't get their minds wrapped around the current pitfalls in its implementation and how it disproportionately harms those who are the most marginalized in our society. What was appalling in this case is that this is the third instance where someone was wrongly arrested based on evidence from a facial recognition technology match; in each case, the victim was a Black man. The performance of facial recognition technology systems on minority populations is notoriously bad yet we continue to see its use by law enforcement which is endangering the lives of innocent people.

In this case, there was a double whammy whereby Mr. Parks, the victim in this case, was also subjected to an algorithmic system's decision after being falsely arrested because of the facial recognition technology system and wasn't allowed to leave jail awaiting trial because of his prior run-ins with the law. The system deemed him to be risky rather than requiring monetary bail as is the case in New Jersey where he was arrested.

What was heartbreaking in reading this article was that Mr. Parks considered taking a plea deal even though he knew he was innocent because he didn't want to risk having to face a very severe and long sentence because of his priors. A system that compounds problems is certainly not what the facial recognition technology systems are sold as; law enforcement presumably procures them to help citizens but that is certainly not what is actually happening and just as several states in the US have called for moratoriums on the use of this technology, we need to make sure that we have consistent application of these calls rather than a piecemeal approach which is letting some folks slip through the cracks.

## Timnit Gebru's Exit From Google Exposes a Crisis in AI
([Original *Wired* article](#) by Alex Hanna, Meredith Whittaker )

A huge specter that loomed over the field of AI ethics in 2020 and leading well into 2021 was related to the firing of Dr. Gebru from Google. While many articles have been written about what happened and the inaction on the part of Google to shield and protect the very people who are helping to guide their AI research efforts in the right direction, much less has been said about what we can do together on the ecosystem level to make changes.

This article from a colleague of Dr. Gebru makes it quite clear that we can all play some part in making the necessary changes that will help to steer the direction of the development and deployment of such technologies in a way that benefits society rather than harm those who are already disadvantaged because of structural inequities. We need to become more aware of where we might be the subject of harm coming from automated systems, and questioning critically the design of the systems in relation to the environment surrounding them will help us move towards action that actually triggers change rather than just having even more conversations that are theoretical and divorced from reality.

Some of the suggestions put forth by the authors in the article include exploring the idea of having workers unions for tech workers. Especially workers in AI who have specialized skills that are hard to replace, organizing their labor strength to demand positive action can be a great way to counter some of the negative effects. Then, there are also the calls that we can make to our local and national policymakers to put forward meaningful regulations that centres the welfare of people above pure profit motivations. These need to be accompanied by accountability and repercussions to ensure enforceability. Funding for some of the initiatives that can help protect these worker rights can come from taxation on the giants.

In a world where highly-respected scholars in the field are not safe in the work that they do, we risk significantly our ability to effectively regulate and guide the development of technologies that are having a significant impact on our lives. We need more urgent action and perhaps 2021 is the year when we make that happen together as a community.

## AI Research Finds a 'Compute Divide' Concentrates Power and Accelerates Inequality in the Era of Deep Learning
([Original *VentureBeat* article](#) by Khari Johnson)

The emphasis on AI models that require large amounts of compute is creating a divide along typical social fault lines in being able to do research and development in the domain of AI. The study cited in this article talks about how elite universities in the US significantly outcompete other universities in the access to compute resources they have and subsequently the kind of work that they are able to do.

Supplementing their analysis with the diminished diversity at some of the top-tier universities, the authors point out how such a compute-divide will also exacerbate societal inequities because of the kind of research and agendas that are peddled by those who have access to the largest amounts of resources, both in terms of data and raw computing power. Major technology companies also repeat this same pattern.

As a counterveiling force, a call earlier in the US asked for the creation of shared data commons at a national level along with a National Public cloud that can democratize AI to the extent that we are inclusive of the perspectives of everyone, no matter their access to resources.

## Fighting AI Bias Needs to Be a Key Part of Biden's Civil Rights Agenda
([Original *Fast Company* article](#) by Mark Sullivan)

With a change of political landscape in the US, the Algorithmic Accountability Act which was first brought forth by Senators Wyden and Booker might be the first legislation that can enshrine some protections for AI ethics issues in law in the US. While it isn't without its shortcomings, for example, the lack of transparency requirements for algorithmic audits, the proposed bill still represents a great first step at the federal level to create regulation. Current legislations that penalize discrimination in hiring for example are still weak in terms of actual teeth for regulating algorithmic practices in hiring. One of the strongest calls to action at the moment for such emergent regulation is for it to regulate the appropriate thing and have sufficient teeth to hold companies accountable when they run afoul.

Agencies like the EEOC and FTC in the US are well-positioned to take some of the tools that would emerge from such a regulation and put them into practice against companies that violate these norms. With the Biden administration's emphasis on science and technology, for example, the elevation of the OSTP office to a cabinet position and appointment of Alondra Nelson as the science and society officer, the atmosphere seems ripe to push through such an Act to lay the groundwork for future work in this domain.

### AI Researchers Detail Obstacles to Data Sharing in Africa
(Original *VentureBeat* article by Khari Johnson)

- What happened: A recently published paper at the FAccT conference talks about the challenges with the current paradigm of data collection and use in Africa that reeks of colonial practices. It details how there is still a strong degree of paternalism, and a lack of contextual understanding of the problems, and a lack of investment in building up local infrastructure that will create sustainable and relevant use of this technology in Africa.

- Why it matters: While it is great that people are paying attention to utilizing data from the African continent and creating new solutions, doing so without leveraging the local expertise and without actually creating lasting infrastructure that is locally-owned will just reiterate and perpetuate colonialism. More importantly, it will also create solutions that are ill-suited to the needs of the people there.

- Between the lines: Major companies have established data centers and other infrastructure locally in Africa to capitalize on this opportunity but supporting the local governments and companies so that they become capable of deploying such infrastructure on their own and maintain ownership will create more long-run sustainability for the continent.

### Can AI Fairly Decide Who Gets an Organ Transplant?
(Original *Harvard Business Review* article by Boris Babic, I. Glenn Cohen, Theodoros Evgeniou, Sara Gerke, and Nikos Trichakis)

Healthcare is one of those domains where the ethical implications of AI have very high relevance and significance. However, there are still many challenges that are yet to be solved. Thoughtfulness and domain experts consultations are essential in coming up with solutions that meet both moral and legal standards while also serving the business interests of healthcare institutions. That last part is important because in the reality of deployment of these systems, ignoring those considerations leads to failures with people bypassing the controls that are put in place because they have unrealistic expectations.

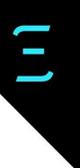

The article mentions some of the incompatible definitions in the world of fairness in machine learning (for a comprehensive explainer, see this summary from our founder Abhishek) and some of the subsequent challenges in operationalizing them. In particular, an ex-post analysis creates unnecessary risks in terms of the harms that might already be inflicted on people when incorrect decisions and refinements are made. In addition, we also stand the risk of killing further integration of technology in the healthcare system when an ex-post analysis reveals severe problems. Instead, adopting a proactive approach whereby we are able to analyze and articulate the ethical considerations upfront, balancing them with the business considerations will actually lead to systems that can then be tweaked in an iterative manner without receiving blanket rejection because of unfair outcomes.

### This is the Stanford Vaccine Algorithm that Left out Frontline Doctors
(Original *MIT Tech Review* article by Eileen Guo, Karen Hao)

Even simple rules-based algorithms have the potential to wreak havoc when they are wrapped in opacity around how those criteria were picked and how it is being applied. Frontline healthcare workers who have helped us maintain a sense of normalcy by keeping us safe and providing help to those who need it the most, often at great risk to their own well-being were conspicuously absent from the first round of vaccinations that were handed out at Stanford.

Residents rightly protested this by highlighting how the system favored those who were administrators and doctors working from home rather than those who were in critical positions dealing with high risks of exposure. This was exacerbated by the fact there wasn't a diversity of people who were involved in the creation of that formula. It was also made unnecessarily complicated when other hospitals followed much simpler formulations in the interest of getting the vaccine as quickly as possible to those who need it the most.

One thing that particularly caught my attention was how administrators tried to hide behind the fact that the system was complicated in justifying why they made this egregious decision. Something that we might see happening in the future is people using this as a crutch to rationalize why they aren't taking more decisive and transparent actions when it comes to the well-being of people. What was particularly heartening was to see the residents come out and call out the administration on it. But, this shouldn't lead to the normalization of the affectees always having to take on the burden to defend their rights; those in power should take on that responsibility, especially when they are a part of an organization that is intended to function for the welfare of the community in which they are situated.

**AI needs to face up to its invisible-worker problem**
([Original *MIT Tech Review* article](#) by Will Douglas Heaven)

Articulating the problem of a lack of recognition and fair compensation offered to gig workers who are behind the miracles of modern-day AI, the article provides some insights into the pervasiveness of this kind of work and how many people depend on it for their livelihood. Supervised machine learning approaches require large amounts of labelled data and often that is sourced from platforms like Amazon Mechanical Turk where workers toil for abysmal wages (~$2/hour) smoothing out the rough edges of AI systems so that we don't see them fail. Yet, these workers don't receive many protections that standard workforce participants would get.

What is appalling is that some of the richest AI companies are the ones who contract these workers without paying them adequately. There is a difference in terms of the amount of effort required to complete tasks and the wages they receive. The researcher interviewed in the article is offering tools and developing awareness to help these workers raise their concerns and at the same time empower them to better understand what they are signing up for. Finally, something that needs a lot of emphasis is how such work doesn't offer skills that can be utilized elsewhere and are often a roadblock to the workers moving on to more meaningful work that can lead them to better lives.

# 4. Humans and Tech

**Opening Remarks** by Deborah G. Johnson (Emeritus Professor, Engineering and Society, University of Virginia)

Technologists are hard at work developing humanoid robots with artificial intelligence (AI). These efforts may ultimately challenge (and perhaps undermine) the distinction between humans and machines. That is, the increasingly human-like appearance of robots combined with sophisticated AI that makes the robots seamlessly interactive suggests a trend towards a time when it may be difficult to tell when we are interacting with a human and when a machine. Some futurists predict that humanoid robots with general intelligence and autonomy will become so human-like that we will have to grant them some form of moral status or legal rights.

Of course, this blurring of the line between human and machine is far from inevitable. What we 'make' of the combination of robots and AI is in our (human) hands. The future development of robots and AI calls for serious and careful understanding of the relationship between humans and technology, a relationship that is enormously complex and multifaceted. Indeed, it is important to remember that technology is human. Technology is a human creation; it is used by humans in a wide variety of ways; and it has meaning and significance only insofar as humans give it such.

Our relationship with computer technology is especially complicated because it is such a malleable technology. The malleability of computers and information technology has led to its infiltration into nearly every domain of human life. Robots and AI are the latest manifestations of human choices about how to deploy the capabilities of computer technology.

Some claim that current AI constitutes a special or even unique step in the expanding development of computer technology because AI constitutes systems that operate autonomously and learn as they go. The learning capacity means that some AIs can make decisions that humans can't make (think here of AI systems detecting tumors that aren't visible to the human eye) and that humans can't always understand how an AI arrives at its decision (think here of AI systems learning how to detect fraud in millions of credit card transactions). While some suggest that humans cannot be responsible for some AI algorithmic decisions because they don't know how the AI makes its decisions, to my mind, the fact that humans can't fully understand how AI algorithms operate means that we should adopt mechanisms to ensure that AI systems are not out of human control. Among other things, this means holding humans accountable for the behavior of AI systems.

Arguably, AI and robots are in the early stages of development and we are just starting to grapple with their meaning and significance and the possibilities they create for both good and bad. In the early stages of any technological development, we imagine what they might become, anticipate future uses, and evaluate possible consequences.  This is part of the process by which we influence the technology's development, assimilate it into our lives, and give it meaning. This process of turning a new set of possibilities into a specific 'thing' involves a variety of ways of thinking.

Using diverse approaches, the selections in this chapter display this process; they explore the possibilities, identify potential roadblocks, and try to make sense of what is to come. One of the selections playfully considers the possibility of robots serving as companions and caregivers for the elderly. Another selection uses empirical evidence to provide evidence of the possibility that reliance on AI robots may lead to humans taking greater risk.  Another grapples with attempts to model human emotions and use those models in designing AI systems. Several of the selections express concern that the potential of AI robots will be impeded unless the issue of trust is addressed. These contributions are part of the process that will ultimately shape the future development, adoption, use, and meaning of the AI robots of the future.

---

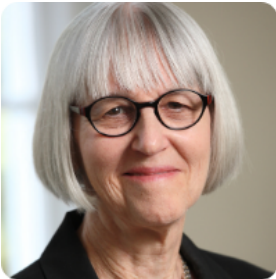

**Deborah G. Johnson (dgj7p@virginia.edu)**
Anne Shirley Carter Olsson Professor of Applied Ethics, Emeritus
University of Virginia

Deborah G. Johnson is best known for her work on ethical, social, and policy issues involving computer/information technology and engineering. Drawing on her training in philosophy, Johnson has published seven books including one of the first textbooks on computer ethics [Computer Ethics, Prentice Hall, 1985] and most recently Engineering Ethics, Contemporary and Enduring Debates [Yale University Press, 2020]. In recognition of her contributions, Johnson has received several awards including a Lifetime Achievement Award from the Society for Philosophy and Technology (2021), the Covey Award from the International Association for Computing and Philosophy (2018) and the Joseph Weizenbaum Award from the International Society for Ethics and Information Technology (2015). Her latest writing focuses on AI, robots, and deepfakes.

# Go Deep: Research Summaries

### To Be or Not to Be Algorithm Aware: A Question of a New Digital Divide?
([Original paper](#) by Anne-Britt Gran, Peter Booth, Taina Bucher)
(Research summary by Sarah P. Grant)

**Overview:** Understanding how algorithms shape our experiences is arguably a prerequisite for an effective digital life. In this paper, authors Gran, Booth, and Bucher determine whether different degrees of algorithm awareness among internet users in Norway correspond to "a new reinforced digital divide."

---

Traditional digital divide research focuses on inequalities in access and skills. By exploring what separates the haves from the have-nots when it comes to algorithm awareness, Anne-Britt Gran, Peter Booth, and Taina Bucher aim to take the concept of the digital divide in a new direction.

The authors assert that algorithm awareness is an issue of "agency, public life, and democracy," emphasizing that algorithms don't just facilitate the flow of content, they shape content. The paper also highlights how algorithms are changing the ways in which institutions (such as public safety agencies) make high-stakes decisions. Because algorithms have been found to produce outcomes that replicate historical biases, the authors argue, there is a need to understand whether an awareness gap exists among the general population.

Using data collected from a survey of internet users in Norway (where 98% of the population has internet access), the researchers analyze algorithm awareness, attitudes to specific algorithm-driven functions, and how varying degrees of awareness influence these attitudes. They compare these findings against key demographic variables and use cluster analysis to place the respondents into six distinct awareness categories: the unaware, the uncertain, the affirmative, the neutral, the sceptic, and the critical.

For this research, the focus is on three types of algorithm functions, including content recommendations (via platforms such as YouTube), targeted advertising, and personalized content such as news feeds. The authors note that studies have examined algorithm awareness in the past for specific platforms like Facebook, Twitter, and Etsy. The aim of this study is to go beyond individual platforms and adopt a more exploratory approach.

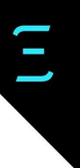

**Levels of algorithm awareness**

The findings suggest that a significant percentage of the Norwegian population has either no awareness or low awareness of algorithms: 41% of the respondents report no awareness of algorithms, while 21% perceive that they have low awareness.

No awareness is highest among older respondents, while the youngest age groups represent the highest levels of awareness. Education is strongly linked to algorithm awareness, with low awareness highest among the least educated group. Men perceive higher levels of algorithm awareness than women.

**Attitudes towards algorithms**

Those who report higher levels of awareness also hold more distinctly positive or negative attitudes towards algorithm-driven content recommendations. "Neutral" or "I don't know" attitudes are more strongly associated with respondents who have a low awareness of algorithms.

**Types of algorithm awareness**

The respondents fall into six categories based on demographics, attitudes, and level of awareness. The "unaware" group, for example, has the oldest average age, is composed of 59% women, and has a significantly higher proportion of people with secondary school as their highest level of educational attainment.

In contrast, the "critical" group (which reports a high level of awareness) is composed of younger people, is over-represented by males, and has a much higher proportion with higher levels of educational attainment. It holds negative or very negative attitudes towards the different types of algorithm content.

**Implications for digital divide research**

The authors conclude that a general lack of awareness poses a democratic challenge and that the demographic differences in algorithm awareness correspond to a new level of the digital divide. They explore where algorithm awareness fits into the traditional digital divide framework, and determine that it can be best defined as a meta-skill that is necessary "for an enlightened and rewarding online life."

A major implication covered in this paper is the potential for negative outcomes as algorithms become increasingly embedded in high-stakes decision-making related to areas such as health, criminal justice, and the news media. Another area to consider that is not emphasized in this research is what happens when tech companies position themselves as champions for closing the internet access gap by providing free services, but expose more people to the influence of algorithms in the process. The findings from this paper can be used to consider whether this is indeed a fair contract when large segments of the population may have a lack of algorithm awareness.

# The Robot Made Me Do It: Human–Robot Interaction and Risk-Taking Behavior

([Original paper](#) by Yaniv Hanoch, Francesco Arvizzigno, Daniel Hernandez García, Sue Denham, Tony Belpaeme, Michaela Gummerum)
(Research summary by Victoria Heath)

**Overview:** Can robots impact human risk-taking behavior? In this study, the authors use the balloon analogue risk task (BART) to measure risk-taking behavior among participants when they were 1) alone, 2) among a silent robot, and 3) among a robot that encouraged risky behavior. The results show that risk-taking behavior did increase among participants when encouraged by the robot.

---

Can robots impact human risk-taking behavior? If so, how? These are important questions to examine and understand due to the fact that human behavior has, as the authors of this study write, "clear ethical, policy, and theoretical implications." Previous studies on behavioral risk-taking among human peers show that "in the presence of peers" participants "focused more on the benefits compared to the risks, and, importantly, exhibited riskier behavior." Would a similar behavior be replicated among robot peers? Although previous studies examining the influence of robots on human decision-making have been conducted, there are still no clear answers.

In this study, the authors use the balloon analogue risk task (BART) to measure risk-taking behavior among participants (180 undergraduate psychology students; 154 women and 26 men) when they were 1) alone (control condition), 2) among a silent robot (robot control condition), and 3) among a robot that encouraged risky behavior by providing instructions and statements (experimental condition). The authors also measure participants' Godspeed ("attitudes toward robots") and their self-reported risk-taking. The robot used for the experiment is SoftBank Robotics Pepper robot, a "medium-sized humanoid robot."

The results show that risk-taking behavior increases among participants when encouraged by the robot (experimental condition). The authors write, "They pumped the balloon significantly more often, experienced a higher number of explosions, and earned significantly more money." Interestingly, the participants in the robot control condition did not show higher risk-taking behavior than the control. The mere presence of a robot didn't influence their behavior. This is in contrast to findings of human peer studies in which "evaluation apprehension" often causes people to increase risk-taking behaviors because they fear being negatively evaluated by others.

It would be interesting to see if this finding is replicated in a study that allows the robot control condition to interact with the robot before beginning the experiment.

The authors also find that although participants in the experimental condition experienced explosions, they did not alter their risk-taking behavior like those in the other groups. It seems, they write, "receiving direct encouragement from a risk-promoting robot seemed to override participants' direct experiences and feedback." This could be linked to the fact that participants in this group had a generally positive impression of the robot and felt "safe" by the end of the experiment.

While the authors acknowledge the limitations to their study (e.g., participants consisted mostly of women of the same age, focus on financial risk, etc.), the findings do raise several questions and issues that should be further investigated. For example, can robots also reduce risk-taking behavior? Would it be ethical to use a robot to help someone stop smoking or drinking? Understanding our interactions with robots (or other AI agents) and their influence on our decision-making and behavior is essential as these technologies continue to become a part of our daily lives. Arguably, many of us still struggle to understand—and resist—the negative influences of our peers. Resisting the negative influence of a machine? That may be even more difficult.

## The Ethics of Emotion in AI Systems

([Original paper](#) by Luke Stark, Jesse Hoey)
(Research summary by Alexandrine Royer)

**Overview:** In this summary paper, Luke Stark and Jesse Hoey, by drawing on the dominant theories of emotion in philosophy and psychology, provide an overview of the current emotion models and the range of proxy variables used to design AI-powered emotion recognition technology. The disjuncture between the complexity of human emotion and the limits of technical computation raises serious social, political, and ethical considerations that merit further discussion in AI ethics.

---

From benevolent Wall-E to femme fatale Ex-Machina, the emotional lives of intelligent machines have long been a subject of cinematic fascination. Even as early as Mary Wollstonecraft Shelley's 1818 *Frankenstein*, human imagination foresaw the advent of sentient and humanoid machines. With increasing machine learning sophistication that can drastically err from its makers' original intentions (we're looking at you [Tay.Ai](#)), Shelley's 200-year-old metaphor of human genius leading to monstrous results still rings true. While we are still far from being menaced by emotionally intelligent humanoid robots, such fictional characters raise crucial questions concerning the range of human emotions, how we conceptualize them, and how they can be measured and input into intelligent systems. The next Frankenstein, rather than a terrifying green mechanical monster, will likely be an invisible AI system that recognizes and feeds off your emotions.

Indeed, AI technologies that predict and model human emotions to signal interpersonal interactions and personal preference are increasingly being used by social media platforms, customer service, national security, and more. In this paper, Stark and Hoey present the range of models and proxy data used for emotional expression and how these are combined in artificial intelligence systems. They argue that the "ethics of affect/emotion recognition, and more broadly so-called digital phenotyping ought to play a larger role in current debates around the political, ethical and social dimensions of AI systems."

The conversation begins with choosing between the different theories of emotion and the range of phenomena that the term can encapsulate, from the physiological, expressive, behavioural, mental, etc. When describing the difficulties of emotion modelling, the authors cite philosopher Jesse Prinz's "problem of parts," selecting the essential components of emotion, and "problem of plenty," determining how these components function together. In psychology, emotion is distinguished from affect, described by psychologist Jerome Kagan as a "change in brain activity" and by affect theorist Deborah Gould as "nonconscious and unnamed, but … registered experiences of bodily energy response to a stimulus". Affect is considered an immediate bodily response, while emotion is a much more complex cognitive process that responds to changes in affect.

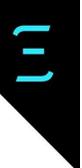

Stark and Hoey list the different camps in the conceptual modelling of emotion, choosing to define emotions as:

- **Experienced feeling states (Experiential theories).** The feeling tradition emphasizes the "feeling" component and embodied sensations of emotions and defines them as distinctive conscious experiences. By concentrating on the felt dimension of emotions, their perception and detection are derived from physiological processes.

- **Evaluative signals connected to other forms of human perception of social cues (Evaluative theories).** Evaluative theories, emerging out of the 1960s cognitive turn, view emotions primarily as cognitive phenomena and ascribe to them a certain degree of intentionality connected to human judgment (i.e. humans direct our emotions towards objects).

- **As intrinsically motivating desires/"internal cause of behaviours aimed at satisfying a goal" (Affective science).** This perspective of emotion as "motivating states'' draws from American psychologist Silvan Tomkins, who argued that nine innate affects are hardwired into humans. Paul Eckman, a student of Tomkins, pushed this argument further in the 1970s by developing the Basic Emotion Theory (BET) which posits that there are universally identifiable bodily signals, such as facial expression, for each basic emotion. BET has received sustained criticism for its reductionist view of human emotional states.

As the authors point out, the boundaries between these three camps are not always so clear-cut, with several philosophical and psychological theorists engaging in what the authors term "hybrid approaches", whereby human emotions involve a mix of physiological, affective, and social signals. In contrast, in computer science, affect and emotion are often treated interchangeably due to technical constraints. Computer scientists detect what they consider to be evidence of human emotions by tracking biophysical signals such as facial expression, gait, heart rate, or blood conductivity, hence conflating affect with emotion.

In addition to a lack of consensus around what emotions are, there is a lack of agreement over the empirical evidence used (in this case, proxy data) to identify human emotions. Depending on the emotional model favoured by the computer scientists, the proxies for emotional expression may include what is referred to as physiological data (e.g. facial expressions), haptic and proprioceptive data (e.g. blood flow), audio data (e.g. tone of voice), behavioural data gathered over time and semantic signifiers (e.g. emojis).

Stark and Hoey explain that the popularity in computer science of correlating bodily changes with emotional changes is partly due to Rosalind Picard's famous publication *Affective Computing* which argued for treating human emotional expressions as digitizable information. Picard's work was criticized for ignoring the dynamism of human social and emotional relationships, with emotions often being culturally constituted. Nonetheless, the majority of "emotional AI" technologies on the market are built on Basic Emotion Theory (BET), which, as stated above, emphasizes a traceable and universal view of emotions that are readily legible through bio-signals such as facial expressions, heart rate, voice recognition, etc. Even though it has a shaky scientific grounding, BET is more amenable to adoption in AI systems.

Emotion recognition based on facial expressions in facial recognition technology continues to attract strong commercial interest. It has also enabled the creation of digital phenotyping, where large-scale computation extracts any manifestation of human emotion, whether in keyboard interaction or speech, and turns these into patterns of behaviour that reveal inner motivations. Not all emotional recognition systems attempt to grasp our internal affective states, and instead are built to observe social displays of emotion, with Facebook's "reactions" icons being a case in point.

Despite BET's predominance, Stark and Hoey note that other emotional models are present in AI systems. One notable example is the Bayesian Affect Control Theory (BayesAct), a quantitative method designed by sociologists and computer scientists to capture through external observation and map out the range of interactions between a user and fellow actors, behaviours and settings. BayesAct combines sociological theory on cultural behavioural norms with the Bayesian probabilistic decision theory to predict what emotions are appropriate in a given situation and help guide users' actions towards socially normative behaviours. Such systems are being used in cognitive assistance technologies to assist individuals who have dementia, for example. In terms of evaluative theories of emotion, other algorithmic learning models include the Computational Belief and Desire Theory of Emotions (CBDTE) and the Emotion-Augmented Machine Learning (E-A ML).

With much of the emotional AI (EAI) field being left unsupervised by regulators, the authors refer to actors such as Andrew Shay and Pamila Pavlisack, who have issued a set of guidelines for the ethical use of EAI. They warn against current EAI assumptions by stressing that "physical display of emotion is only one facet of emotion," that there are "stereotypes and assumptions about emotions that materially affect a person or a group" and that "past expression does not always predict future feeling."

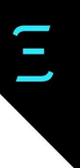

Not only may EAI systems be unwanted and overly invasive, but they also raise broader concerns around the use of data for tracking, profiling, and behavioural nudging. Such considerations go against the narrower definitional discourse of emotion in computer science and run contrary to social media giants' business philosophy, whose profit model rests on manipulating users' human emotion to get users to engage and interact on the platform. When touching on the divergence of opinion on emotions, the authors underline that recognizing this diversity of thought in the abstract is "unhelpful without considering how those differences might implicate AI design and deployment decisions, with their attendant ethical valences and social effects."

There will always be a gap between the emotions modelled and the experience of EAI systems. Addressing this gap also implies recognizing the implicit norms and values integrated into these systems in ways that cannot always be foreseen by the original designers. With EAI, it is not just a matter of deciding between the right emotional models and proxy variables, but what the responses collected signify in terms of human beings' inner feelings, judgments, and future actions. As Stark and Hoey warn, the AI systems used to analyze human faces, bodies, and gaits are dangerous digital revivals of the widely-discredited 19th and 20th pseudoscience of physiognomy, where a person's character can be based solely on their outward appearance.

From my own anthropological background, BET-based systems run counter to decades of ethnographic research that affirm the cultural specificity of human emotions and the danger of imposing a heterogeneous view on the human experience. Another serious ethical concern is the fact that individual users, often unknowingly, are the subject of human emotion subject research. There is also a self-realizing dimension to EAI systems, as individuals may, over time, modulate their emotional responses to fit the norms produced by these systems.

To conclude, as stated by the authors, "human emotion is a complex topic, and analysis of its effect and impacts in AI/ML benefit from interdisciplinary collaboration." As a social scientist, I cannot overemphasize the critical need for cross-disciplinary discussion on EAI and the contexts in which it is ethically permissible to develop and deploy these systems. As it is currently developing, the reductionist view of computing emotions in AI systems may create terrifying monsters indeed.

# Reliabilism and the Testimony of Robots

([Original paper](#) by Billy Wheeler)
(Research summary by Dr. Marianna Ganapini)

**Overview:** In this paper, the author Billy Wheeler asks whether we should treat the knowledge gained from robots as a form of testimonial versus instrument-based knowledge. In other words, should we consider robots as able to offer testimony, or are they simply instruments similar to calculators or thermostats? Seeing robots as a source of testimony could shape the epistemic and social relations we have with them. The author's main suggestion in this paper is that some robots can be seen as capable of testimony because they share the following human-like characteristic: their ability to be a source of epistemic trust.

---

Should we see sophisticated robots as a human-like source of information or are they simply instruments similar to watches or smoke detectors? Are robots testifiers or are they purely offering instrumental knowledge? In this paper, the author Billy Wheeler raises these questions arguing this topic is key in view of the growing role social robots have in our lives. These 'social' robots interact with humans helping with a number of different tasks but little is known of humans' *epistemic* relation with them.

It is common for any philosophical discussion about 'knowledge' to identify knowledge with having a justified true belief. The basic idea is that we want to distinguish between knowledge vs lucky guesses: knowledge requires a certain degree of support or justification that beliefs that happen to be true by chance do not have. And one way to capture this idea is to say that a belief constitutes actual knowledge only if it is formed via some reliable method. Common reliable belief forming-processes are perception and reasoning. Absent relevant defeaters, testimony is also seen as a way of gaining knowledge (and thus beliefs that are reliably true) from other people. Finally, using instruments (e.g. calculators, watches) that are well functioning is also a way of gaining knowledge.

What is the difference between testimony based vs instrument-based knowledge then? Well, when we receive information from people, we trust them. This kind of trust is substantially different from the 'trust' that we extend to instruments. Of course, if I deem it well-functioning, I can 'trust' my watch to be reliable and thus a source of knowledge about time. However, the author points out that testimony is based on social trust and "its purpose is to forge bonds between individuals for the sake of forming epistemic communities". At least according to some views of testimony (called 'non-reductionist') these interpersonal bonds make it the case that, to form a reliable or justified belief, I don't need any additional positive reason to trust someone else's word. Unless I have reason to doubt what you said or your sincerity, I will believe what you tell me. That also means that if it turns out that what they said was false, I can blame you or hold you responsible. Conversely, we always need positive reasons in order to gain knowledge from a watch (that is, we need to assume that it is not defective, that it is reliable, and so on). Parallelly, we do not strictly blame an instrument for getting things wrong.

Time to go back to the main question of the paper: should we grant robots the status of testifiers? As the author points out, to answer this question we might need to first figure out, or reach a consensus on, the moral and agential status of these robots. This seems to be a necessary step for deciding whether we could rationally grant normative and social trust to robots ( which is a necessary condition for seeing them as testifiers too). However, the author also points to another way to tackle this issue: observing the way humans interact with these robots. If humans come to see robots as capable of establishing relationships with them while also considering them as "honest" or "deceitful" (and there is seems to be empirical evidence pointing in that direction), then we might conclude that de facto robots are (or will be) part of that bond-forming process that is key for delivering testimony.  In other words, because of how we treat them, it is likely that we will eventually start to see robots as trustworthy and able to share their knowledge by testimony.

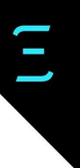

## Artificial Intelligence – Application to the Sports Industry

([Original paper](#) by Andrew Barlow, Sathesh Sriskandarajah)
(Research summary by Connor Wright)

**Overview:** PWC's report demonstrates how AI has become a mainstay of sport today given its analytical prowess. Whether this is to be the case or not is then considered within the paper, as well as AI's numerous achievements in the sporting field. With such achievements and digitalization becoming more prominent in itself, the human aspect of sports is seen to become more fragile than ever before, with no sign of an end to technology's encroachment.

---

Defining AI as a "smart technology" being able to learn without needing a human to tell it what to do, PWC's report on AI in sport offers AI's sporting history, its use today and the ethical considerations that come with it. The author first introduces the benefits of AI in sport to athletes, coaches and spectators alike. They then detail the spectator experience and how it's influenced by AI, using Wimbledon 2018 as an example. The next subtopic is the worries of the preservation of human aspects in different sports; especially to the world of Formula One (F1). To conclude, I detail some of the ways forward offered by the report, before closing with my own remarks.

Within PWC's report, it initially details the benefits that AI introduces to the sporting arena. From providing more in-depth information quicker, AI is able to provide athletes, coaches, and spectators an added depth to their experience. In terms of the coaches and athletes, activity and sensory data stemming from fitness trackers can be combined with deep and machine learning techniques to better analyse the data being produced. Analysis such as being able to provide closer analysis of the differing racket-head speeds of tennis players or the amount of fuel consumed at different moments by an F1 car is then coupled with AI doing it at a much faster rate than any human could. Such speed then allows for analysis in real-time of the sport at hand, influencing the game's tactics and game plans immediately.

In terms of the spectators, AI is able to utilise the same data to benefit the choices being made from outside the field of play. The enriched and rigorous data can then be displayed to spectators in order to influence betting decisions, as well as providing a higher level of accessibility to the game. In this case, virtual reality (VR) headsets have been able to take advantage of AI-simulated realities in order to position the user as if they were in a seat at the stadium itself. From there, for those who don't have time to watch the whole match of their favourite sport, AI can even help compile highlights of the best moments of the competition, something which was revealed in the paper to have been used at Wimbledon in 2018.

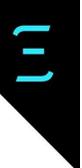

Here, IBM's Watson was used to speed up the highlights process, analysing the footage of all matches for player emotion and condensing down the best bits accordingly (the more animated the players, the more likely that play was to feature). As a result, the highlights were shorter, unmissable, and produced at scale, with the overall day highlights now focusing more so on the most exciting plays, rather than whose playing. VR then comes into the fray again; augmented reality features were also used in Wimbledon to show court hot spots and player stats in real-time.

PWC's report then details how AI can contribute to the off-field experience in the form of chatbots and automated journalism. The use of chatbots eases the ticketing and game questions that human operators receive, such as 'when's the next Lakers game?', 'are tickets available for the Raptors game?' and so on. Having automated responses lined up in order to direct customers to the relevant pages saves time, energy and resources, while providing 24/7 care. Within this option, should the chatbot be unable to handle a query, there is to be an option to defer to a human agent. In terms of automated journalism, goal updates, substitutions, in-game description and more can be dedicated to AI in order to provide almost instantaneous reporting of in-game events. In this case, an immersive experience is still made possible without actually being present at a sporting event, especially down to the efficacy and speed which AI brings.

The paper also underlines how what AI brings to sport isn't even confined to just augmenting what is already at play, but can also be viewed through what it has achieved. Having learned through self-play and the rules of chess, DeepMind's AlphaZero has managed to be undefeated for 4 years as a chess champion ever since its release. In a similar fashion, a monumental victory was achieved when AlphaGo beat the reigning GO! Champion Lee Sedol in 2016. As a result, many questions were asked about humans' future in such sporting contests, as well as in sport itself. It seemed humanity was starting to slip out of the picture of sport, locked in a losing battle against superior machines.

While such radical claims have not risen to fruition, the increasing digitization of sport is getting noticed. PWC's report raises some interesting takes on the role of technology within sport, with the reliance on the data it produces requiring less and less human intervention. Continuing on our Wimbledon example, the introduction of hawk-eye has alleviated a lot of the pressure placed on line judges, with the final say on the tough calls now, at times, being handed over to the technology. Likewise, in football (soccer, that is), Video Assistant Referee has established the process of using AI technology to strongly influence the overruling of some referee decisions, such as offside calls. However, these both pale in comparison to what F1 has gotten up to.

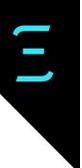

Given the new F1 tech, the human driver is now starting to count for very little. In a 2016 study by Mike Hanlon cited in the report, the driver's influence over the outcome of the race compared to the car and their team has reduced. For example, in 1980, 30% of the victory was dedicated to the driver, with that only being 10% in 2014. With such startling statistics, technology is now receiving an ever more impenetrable authority, especially among coaches.

In this regard, AIs and the statistics they provide are becoming assistant coaches as well, informing coaches and staff of different player statistics, giving them the 'raw truth'. In this way, the report's analysis starts to introduce some questions over whether coaches are actually able to disagree with a provided analysis. It's fathomable that some sports trackers may malfunction and produce erroneous data, but the emergence of technology within sports in its current form has made it hard to question.

In this way, the positive impact of technological innovation has been questioned. The turn towards data analysis has introduced an element of whether such data limits or unleashes the human creative side. However, the very same Lee Sedol, upon being beaten by AlphaGo as mentioned above, reflected on how such AI triumphs have the potential to "unlock human creativity" and allow humans to be "promoted" rather than replaced. In this way, technological innovations should be seen as gateways towards the evolution of sports, rather than the end of human expression within the domain.

Given the poignancy of the debate, PWC's report does offer some considerations to be taken into account today. Potentially introducing a "technology cap" could help 'preserve' the human emphasis on certain sports, while keeping competition fair. Implementing third-party verification of the technology employed could also help maintain the nature of the technology being used, and how this is to be kept maintained throughout the sport itself.

It goes without saying that PWC's report shows that AI has and will continue to influence the sporting arena. Whether it takes the form of further deepening the information available to those involved, or whether it assumes its own role within the sport, the innovations will carry on. As a result, moving forward, we're going to have to try and match the coming innovations, with the aim of providing fair competition and preserving human participation.

# The Algorithmic Imaginary: Exploring the Ordinary Affects of Facebook Algorithms

([Original paper](#) by Taina Bucher)
(Research summary by Dr. Iga Kozlowska)

**Overview:** In this paper, Taina Bucher explores the spaces where humans and algorithms meet. Using Facebook as a case study, she examines the platform users' thoughts and feelings about how the Facebook algorithm impacts them in their daily lives. She concludes that, despite not knowing exactly how the algorithm works, users imagine how it works. The algorithm, even if indirectly, not only produces emotions (often negative) but also alters online behaviour, thus exerting social power back onto the algorithm in a human-algorithm interaction feedback loop.

---

Facebook seems to think I'm "pregnant, single, broke and should lose weight." These are the kinds of comments that Taina Bucher uncovers as she reaches out to 25 ordinary Facebook users who have tweeted about their dissatisfaction or confusion over Facebook's news feed algorithm.

In popular imagination and in public discourse, we often think of algorithms as objective and capable of accurately reflecting reality. Because we don't associate algorithms with emotions, we tend to underestimate the affective power of algorithms on people's social lives and experiences. While Facebook algorithms are not necessarily designed to make users feel one way or another (except when they are: see [Facebook's emotion contagion experiment](#)), they certainly have the power to produce emotional reactions and even alter behaviour.

**How Facebook makes people feel**

Bucher summarizes several ways in which Facebook users experience negative, confusing, or disconcerting situations when interacting with algorithms. Users readily admit they don't understand the inner workings of the algorithm, as no one outside of Facebook does. However, not understanding how something works doesn't preclude us from experiencing its effects. Bucher discovers the following themes:

- Dealing with algorithmically-built profiling identities that are not flattering or do not comport to how users see themselves
- Creepy moments when people feel like their privacy is violated

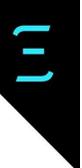

- Frustration and anxiety when posts don't do well
- "Cruel" moments when unwanted memories from the past are resurfaced in feeds

In response to some of these unpleasant experiences, savvy Facebook users try to "play Facebook's game" by adjusting content (wording and images), timing of posts, and forms of interaction with friends' content. Facebook's "game" consists of explicit and implicit rules (much like Google's SEO guidelines), and if you play the game, over time, you get better and are more likely to "win." In fact, this is not too far from how social norms function in the real world—there are spoken and unspoken cultural norms that we are socialized into at an early age and we are rewarded for playing by the rules and penalized for breaking them.

The fact that there is a game you have to play to get rewards out of using Facebook is not inherently good or bad. However, it is something that the company needs to recognize and address. The platform is not just an open, free space for organic human interaction as Facebook sometimes likes to argue to avoid accountability; rather it is a highly structured and circumscribed website with features that encourage and enable some outcomes and discourage and forestall others. Engineers need to take this seriously if for no other reason than the fact that interaction with the platform and its algorithms does cause patterned changes in user behaviour that feedback into Facebook's machine learning algorithm to unknown effect.

**Let's talk about feelings**

What can machine learning developers and product designers do with Bucher's findings?

First, consider the social power your AI-based feature or product will have on the user (and indirect stakeholders). Consider, the good, the bad, and the ugly. In particular, think about the emotional and psychological effects that the algorithm may produce in different contexts as humans interact with it. These can be more obvious harms like attention hijacking, gaslighting or reputation damage, but can also include things like confusion, anxiety, and harm to self-esteem or positive self-identity.

In tech, we don't talk about feelings because we like to focus on what we can easily measure. That gives us the false comfort that we're being objective, unbiased as well as efficient and effective. Bringing feelings back in during the design phase of algorithmic systems is critical, however, to designing experiences that are human-centred.

Second, consider how users will imagine that your algorithm works, even if you know that it doesn't actually work that way. To the extent you can, aim for transparency and balancing the information asymmetry, but consider the agency that people will ascribe to the algorithm. You know that the news feed algorithm doesn't "think" that a given user is overweight, lonely, or sad. But since people tend to anthropomorphize machines and algorithms, what effect, nonetheless, might that have on someone? In other words, people know that machines don't think, feel or judge, but they can still have emotional responses to interactions with machines that are similar to those that are generated when interacting with other humans.

Third, when in doubt, give the user more control rather than less. How can your algorithm and the features within which it's embedded produce a user experience that puts the human back in the driver's seat? Maybe it's tweaking the UI wording. Maybe it's giving the user a simple option to turn a feature on or off. Maybe it's using other automated machine learning techniques to improve the experience. Always optimize for direct human well-being, rather than indirect measures of human satisfaction like usage metrics that can be misleading.

Fourth, consider how emotional or behavioural changes in response to various types of interactions with your algorithm that humans can have will impact that algorithm's continued performance. How might the algorithm encourage feedback loops that might stray over time from your intended outcomes? How will you monitor that? What benchmarks will you use to flag issues and make appropriate tweaks to the algorithm in response? What kind of feedback can you seek from users on how they *feel* about your product?

**Designing for feelings**

The more algorithmic systems proliferate in our social world, the more social power they will exert on relationships, identity, the self, and yes, our feelings. Designing for things that are not easily measured is challenging because it's hard to tell when you're successful. But not designing for affect causes real human harm, not to mention negative front-page news stories. A/B testing (responsibly!), focus groups, and interview-based probing during design phases, are all good methods of discovering potential emotional impacts of your product *before* release. Likewise, designing feedback channels for customers as they engage with your product is a good idea.

Human-centred algorithmic design must be guided by a measure of the user's holistic well-being. This must include psychological, emotional, and social health. With algorithmic systems proliferating deep into our social lives, Bucher encourages us to pay attention to the affective power of these systems. From there, it's up to all of us to decide how we want algorithms to make us feel.

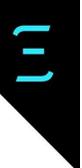

# Go Wide: Article Summaries (summarized by Abhishek Gupta)

## You Can Buy a Robot to Keep Your Lonely Grandparents Company. Should You?
([Original *Vox* article](#) by Sigal Samuel)

Social robots have been around for many years, albeit limited in their role and capabilities in the past. With the advent of more powerful AI-enabled robots and the pandemic related restrictions, we are seeing an increase in the deployment of robots in social contexts where human contact would have been the norm. For example, there is the case of robots being used to take care of the elderly. The article dives deep into understanding the benefits and drawbacks of offloading care responsibilities for the elderly to robots.

The gamut of companion robots runs from amiable pets like cats and dogs to more humanoid forms, some more functional than others, for example, ones that are designed to do heavy lifting or aid in other physical tasks. A particularly popular robot called Paro (that has the form of a baby seal) has been shown to reduce loneliness and depression among other factors in people suffering from dementia. An obvious benefit of these robots is that they are tireless and don't get flustered when dealing with human patients. They also won't defraud or abuse the people that they are taking care of, something that unfortunately happens with human caretakers for the elderly.

An argument against the use of such robots is that they can't replace genuine human contact and thus, might be a poor substitute for actual human contact. They also create a false sense of comfort in those who can now step away from having to care for the elderly. On the flip side, for some young people, the robots can help them improve their interactions with the elderly by providing a fun and common point of conversation.

Some elders might feel that they are able to preserve a higher sense of dignity in interacting with robots who don't understand that they are at their most vulnerable. Something that the elders might feel uncomfortable in sharing with family members and their caretakers.

The use of such robots can allow us the freedom to care for people by taking away some of the tedium, allowing us to engage in more meaningful interactions and help to avoid burnout. But, some of the researchers interviewed in the article caveated that by saying that there is the potential for burnout of a different form that now requires the caretakers to be "dialled to an 11" all the time because there is no tedium to punctuate their interactions.

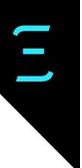

On the front of whether the elderly actually prefer the robots or not, they must have as unencumbered a choice as possible, mostly free from the influence that tech firms have in pushing people to purchase their products. On achieving this, if they prefer the companionship of robots over that of humans, there is still a way to respect their choices. This must be supplemented by routine checks and evaluations by healthcare professionals to ensure that they still recieve an adequate amount of human connection that is essential to maintain emotional well-being.

## UX for AI: Trust as a Design Challenge
([Original *SAP Design* article](#) by Vladimir Shapiro)

What will it take to get ubiquitous adoption of intelligent assistants? Trust is a key design element without which we risk relegating the Alexas of the world to being nothing more than kitchen timers and occasional interfaces to pull up search results using voice.

The current failure in this process is that we perhaps have too high expectations of the system because of the way it's marketed to us. When failures occur, there aren't any fallbacks that can help us get to a solution. Merely defaulting to a web search result betrays our confidence in the system, especially if the system isn't able to take into account prior failures and how we would have preferred for it to respond. So a failure in both the "intelligence" of the system and the lack of a feedback mechanism that can make it more aligned to the needs of the user.

As an example, a human assistant wouldn't just come up to you with a printed page of web results, but will mention some of the things that they tried and what worked and what didn't—allowing us to give them feedback on what is useful. Perhaps, given our high expectations, the current crop of systems fail to meet that standard and dissolve some of the trust that we might have otherwise placed in these systems in actually assisting us rather than just being a new interface to perfectly time a boiled egg. From a design perspective, being fully transparent in its limitations is the way to go until we get to a place where the systems have much stronger capabilities that match our expectations.

## Why People May Not Trust Your AI, and How to Fix It
(Original *Microsoft* article by Penny Collisson, Gwen Hardiman, Trish Miner)

Humans have a tendency to choose another human even when they are shown to be wrong repeatedly, something that might come as a surprise to those designing interfaces for AI systems in the hopes of facilitating a transition away from humans. Borrowing from design principles, the article talks about empathy as a core facet of getting the design of the system right—essentially, making sure that you walk a mile in the shoes of the user to understand how they might experience the service rather than rely solely on the engineers' point of view.

Working with real-world data will help bring to the forefront cases that you can't always simulate with limited training data. Keeping the user central to the experience in a way that gives them agency and control is essential in the success of the system's adoption vs. the approach of limiting the perception of control that the user has.

Building on established design methodologies, capturing data about what is working and what isn't is also deemed to be important. In addition, keeping in mind the feelings that are evoked in the user during their experience and making sure that they align with the purpose of the application is key. For example, building a ladder of trust in the case we want users to feel comfortable is another essential idea in successful AI systems that interact with humans. Finally, when collecting data about how the user is experiencing the system, make sure to account for "researcher pleasing," the notion that users might not truly convey their feelings and attempt to align with what they think is expected of them.

## What Buddhism Can Do for AI Ethics
(Original *MIT Tech Review* article by Soraj Hongladarom)

In this article, the author highlights some of the values from Buddhism that can help us shed new light on how to approach the building of responsible AI systems. In particular, the values of self-cultivation as an underpinning of compassion, accountability, and no-harm is a key contribution to this thought process.

While the overarching values are the same because we are all human and share a lot of ideals, the value of self-cultivation as fuel for pushing forward the efficacy of the other values is something that everyone can learn from. As we've seen with a lot of principles, it can be hard to translate them into practice and this is where I think such an approach can actually make them more actionable. One might ask if there are simple rules that we can utilize to start this journey, and taking the example of facial recognition technology and applying the principle of no-harm, the article argues that it should only be used in case we are able to show that it does actually reduce suffering rather than being used to surveil and oppress people.

Mostly, my takeaway from the article is that there is a lot that can be learned from different schools of thought, even (or perhaps especially) if they aren't directly related to AI.

### Forget User Experience. AI Must Focus on 'Citizen Experience'
(Original *VentureBeat* article by Jarno M. Koponen)

The core argument raised in the article is that there isn't enough attention paid to the needs and rights of citizens when designing and developing AI systems, which leads to misalignment in the expectations and harm inflicted on them by way of violating their rights. The article provides some basic ideas, such as utilizing a multidisciplinary team; integrating these considerations throughout the lifecycle; co-developing solutions with those who are going to be impacted by the systems; and placing citizens at the center of the design process rather than thinking of them as yet another data point that feeds your machine.

We need more integration of concerns around the legal validity and compliance with best practices that limit the negative impacts of a technological application. We also need to educate people on the capabilities and limitations of the systems they're using so they're better equipped to protect themselves.

### Snow and Ice Pose a Vexing Obstacle for Self-Driving Cars
(Original *Wired* article by Will Knight)

Self-driving cars struggle in places where there is inclement weather, upping the risks for misidentifying vehicles that are covered in snow on sidewalks or pedestrians who look like little balls with all their winter clothing on. Such disparities in the ability of self-driving cars pose challenges in terms of where they might be deployable and also point to the tremendous amount of work that remains before they can be widely used.

Publicly available datasets are going to be one of the key instruments in improving the state of the art in these systems. We need various incentives including funding to encourage researchers to collect and annotate such data and to also encourage industry players to make their data more widely available. That way, all the players in the ecosystem can work together in putting forth vehicles that are safer and more robust in challenging weather conditions.

### Artificial Intelligence Has Yet To Break The Trust Barrier
(Original *Forbes* article by Joe McKendrick)

The study referenced in this article examines situations where trust in machines is higher and subsequently what might be the best place to deploy it. For example, based on the study conducted, the authors found that people trusted machine recommendations more when those recommendations were considered utilitarian, like picking a computer (how meta!). But, for experiential things like flower arrangements, they tended to side with human recommendations.

Coining a new phrase, "word-of-machine" the authors seek to explain why machine recommendations are preferred when it comes to utilitarian items vs. those that are experiential. But, they say, all hope is not lost for those seeking to make recommendations using machines for experiences. Having hybrid teams where there is a human-in-the-loop to further handpick recommendations from those that have been shortlisted by a machine might be the way to go.

This is reminiscent of the Guardian article that was written by the Generative Pre-trained Transformer 3 (GPT-3) language model, only after human editors picked from five to six outputs generated by the system—perhaps not all that different from the process that human writers undergo with their human editors at the moment.

### China's Big Tech goes for the Big Grey
(Original *Protocol* article by Shen Lu)

Technology is quite ageist at the moment, meaning that current design practices do not always take the needs of the elderly into account. A few startling examples quoted in the article talk about the spending power of this demographic, making a strong business case for including design elements that address the needs of the elderly. This demographic's rate of saving is one of the highest, as is their spending power—but companies are only just now realizing that, at least in China. Indeed, the rest of the world has some catching up to do. Reports by Alibaba and other companies all attest to this trend.

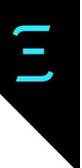

Dubbed the "silver hair economy" in China, the article makes a valid point that for local companies this demographic is perhaps the last bastion of new users as they have saturated all other user bases within the Chinese market. Apps like Meipian and Tangdou that offer easy ways to collage photos and teach elderly women to dance raised serious funding showing how the market is ripe for these opportunities.

The article also bears reading for those in markets outside of China as we are seeing a gradual shift in the demographic composition of the rest of the world. Instead of perceiving the elderly as technologically less competent, re-envisioning design practices and meeting head-on their needs will boost the ability of technology to uplift the lives of everyone. Sociologists, however, warn that profit motives and opportunism in this space can have negative effects, like privacy invasion. This is especially true for a demographic that might not be fully aware of how their data is being used or might not have the ability to provide adequate consent in the face of declining cognitive ability. These are issues to keep in mind as we try to make technology accessible to all.

**How AI Could Spot Your Weaknesses and Influence Your Choices**
([Original *The Next Web* article](#) by Jon Whittle)

Leveraging experiments conducted by Data61 from Australia, the article points to how AI systems can detect patterns in human behaviour and steer them in a direction that can be used to maximize the achievement of its own goals. Through simulated games where the AI acts as a trustee receiving money from a human participant and providing patterns of shapes where the human has to click when presented with something, the system was able to find patterns that caused the humans to make mistakes more frequently.

Dark design patterns do the same thing from a user interface and user experience standpoint. Examples are auto-scrolling and auto-play of videos that compel you to continue spending time on platforms beyond what might be in your own interest. The use of AI may accelerate the use of such approaches to achieve goals that are in the interest of those who are building these systems as opposed to the users. It strengthens the case for higher levels of transparency and accountability in the design, development, and deployment of these systems.

# 5. Privacy

**Opening Remarks** by Soraj Hongladarom (Professor of Philosophy and Director, Center for Science, Technology and Society at Chulalongkorn University in Bangkok)

I have just watched a recent YouTube video on individuals in China being monitored for their emotions. Yes, for their emotions. The idea is that if you appear moody or upset, chances are that the ubiquitous video cameras will catch your expression and the incident will be recorded on your individual profile that the Chinese authorities possess. They could use that information — the fact that you appear upset on a certain day at a certain location — in a variety of ways. So, if you are a Chinese citizen you better be smiling all the time, lest you incur the suspicion of the authorities.

This looks like a nightmare scenario, and the YouTube video might be a fake for all we know. But that is not the point. The point is that there is a belief for the current state of technology to record your facial expression, and to process that information in order to predict your next move. In other words, we are being turned into some kind of automaton where our actions and behaviour can be predicted by machines.

According to the Chinese authorities, it is also possible that all this can be done for beneficial purposes. For example, they might install a system where those who express negative emotion will be targeted for "help" soon thereafter. It might be possible that certain types of disorders can be predicted based on the expression of particular emotions. Moreover, the system might be in place in order to create a happier society where everyone smiles all the time. Those who are upset or grumpy will be identified and their problems solved. The same system could even detect the early onset of serious diseases, such as cancer, through an analysis of facial expressions.

Moving the technical considerations aside, the issue is why we should care for our privacy in the first place. Perhaps there are tangible benefits to a system that analyzes our faces in public. However, what seems to happen in China is that it is presumed that everyone consents to having their faces analyzed. And apparently, there is no way to opt out. The example of the system that analyzes our emotions may make us uncomfortable. Why should we trust our political authorities to let them handle such sensitive, personal information? Or, for that matter, why should we trust the giant global companies that run the websites that we visit every day?

This is why privacy is very important, and with the advent of this kind of technology, it has become even more crucial to understand the consequences. Although it is possible that certain benefits could occur as a result, we should be informed of what the authorities or the corporations are doing with our information.

There seems to be a widening gap between ourselves, the citizens, and either the political authorities or the multinational corporations. They possess and control our data. Even in the East, I can affirm that these values are also cherished. It is based on the universal conception of humanity as being equal to one another. Those who govern do so according to the consent of the governed. This is also acknowledged in China, where there is a saying that the people are like water, and the emperor is like a boat. Just as water can become turbulent and sink the boat, so the emperor—and by comparison, anybody with political power—can sink when the people do not consent to their rule. Without effective ethical guidelines or legal mechanisms, our trust in political authorities and corporations will remain on shaky ground.

Thus, I very much welcome the chapter on privacy by the Montreal AI Ethics Institute. The threat to our privacy rights is real, and we need effective, high-quality research and systematic thinking in order to understand all the ramifications. Guidelines can only be effective if they are based on such research. It is often said that data is the lifeblood of contemporary AI. I would like to add that trust is the component that makes the lifeblood viable, and trust is not possible without privacy.

---

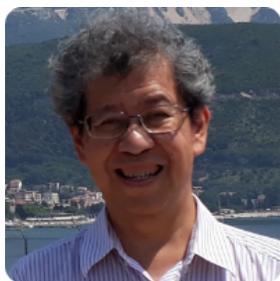

**Soraj Hongladarom**
Professor of Philosophy and Director
Center for STS at Chulalongkorn University in Bangkok

**Soraj Hongladarom** is a Professor of Philosophy and the Director of the Center for Science, Technology and Society at Chulalongkorn University in Bangkok, Thailand. His work focuses on bioethics, computer ethics, and the roles that science and technology play in the culture of developing countries. His concern is mainly on how science and technology can be integrated into the life-world of the people in the so-called Third World countries, and what kind of ethical considerations can be obtained from such relations. He is the author of *The Ethics of AI and Robotics: A Buddhist Viewpoint* (Rowman and Littlefield), *The Online Self* (Springer) and *A Buddhist Theory of Privacy* (Springer).

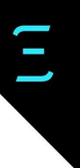

# Go Deep: Research Summaries

**Mapping the Ethicality of Algorithmic Pricing**
([Original paper](#) by Peter Seele, Claus Dierksmeier, Reto Hofstetter, Mario D. Schultz)
(Research summary by Shannon Egan)

**Overview:** Pricing algorithms can predict an individual's willingness to buy and adjust the price in real-time to maximize overall revenue. Both dynamic pricing (based on market factors like supply and demand), and personalized pricing (based on individual behaviour) pose significant ethical challenges, especially around consumer privacy.

Introduction

Is it moral for Uber's surge pricing algorithm to charge exorbitant prices during terror attacks? Using automated processes to decide prices in real-time (i.e. algorithmic pricing) is now commonplace, but we lack frameworks with which to assess the ethics of this practice. In this paper, Seele et al. seek to fill this gap.

The authors performed an open literature search of Social Science and Business Ethics literature on algorithmic pricing to identify key ethical issues. These ideas were filtered into an ethics assessment – categorizing the outcomes of algorithmic pricing practices by the level of society which they impact. "Micro" for the individuals, "meso" for intermediate entities like consumer groups, or industries, and "macro" for the aggregated population. The outcomes were further sorted as morally "Good", "Bad", or "Ambivalent" from an ethics standpoint, forming a 3×3 table that can be used to generalize the ethics assessment.

For all levels of the economy, the authors identify good, bad, and ambivalent outcomes that are likely common to most implementations of algorithmic pricing. Personalized pricing presents additional ethical challenges, however, as it requires more invasive data collection on consumer behaviour.

**What is algorithmic pricing?**

The idea of dynamic pricing has been popular since the 1980s, but the practice is increasingly powerful and profitable in the age of the internet economy. While it is easy to define algorithmic pricing in technical terms, the authors identified the need for a definition that is useful in a business ethics context. The result is as follows:

"Algorithmic pricing is a pricing mechanism, based on data analytics, which allows firms to automatically generate dynamic and customer-specific prices in real-time. Algorithmic pricing can go along with different forms of price discrimination (in both a technical and moral sense) between individuals and/or groups. As such, it may be perceived as unethical by consumers and the public, which in turn can adversely affect the firm."

While there are a variety of approaches, all pricing algorithms typically have the same mandate: predict the consumer's willingness to buy at a given price, either on aggregate (dynamic pricing) or at the individual level (personalized pricing) in order to maximize profit. The use of cutting-edge algorithms like neural networks and reinforcement learning, as well as increased tracking capabilities via browser cookies, enable companies to do this in an increasingly sophisticated way.

**Ethical concerns**

Although algorithmic pricing primarily alters the individual consumer's experience with a merchant (micro), the ripple effects of this practice extend upwards to the organization level (meso), and further to society and the entire economic system (macro).

Evidently, firms benefit from the increased sales that algorithmic pricing facilitates, but could this practice also contribute to the common good? The answer is yes, with some caveats. Algorithmic pricing doubles as a real-time inventory management mechanism. Good inventory management can lead to reduced waste in the product supply chain, thereby decreasing both costs to the firm and the carbon footprint of the production process. The firm will enjoy increased profits, which can also be considered a moral "good" if prosperity gains are shared with the rest of society; either through innovation, increased product quality, or wealth distribution.

The major ethical dilemma of algorithmic pricing comes from the collection of fine-grained consumer behaviour data, as well as the lack of transparency around that data collection. The driving force for algorithmic pricing models, especially personalized pricing, is tracking cookies. This data, which includes browsing activity such as clicks, and past page visits, as well as personal information entered on the site, can be used to finely segment consumers according to tastes, income, health etc. in order to display the most advantageous price for the merchant.

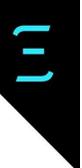

Many consumers are unaware that this information is even being collected, and merchants do not have to explicitly ask for consent to use tracking cookies for pricing purposes.  The onus is left to the consumer to protect themselves if they do not want to be targeted for personalized marketing.  This data collection creates a massive informational advantage for companies, which may offset the price advantage that a consumer can gain due to the ease of searching online.  It also limits a consumer's ability to plan future purchases, as prices constantly fluctuate.  These ethically "bad" features of algorithmic pricing may limit the economic freedom of the consumer, even if the moral "goods" tend to enable more choices.

Other outcomes of algorithmic pricing fall into a grey area.  Surge pricing is effective at managing demand to ensure that services/goods can be delivered to all who need them, but this method occasionally creates a trap of choosing between forgoing an essential service or paying an exorbitant price.

Ultimately, any application of algorithmic pricing requires a case-by-case treatment to ensure that the good outcomes are enhanced and the bad are appropriately mitigated.

**Between the lines**

Algorithmic pricing is already transforming our economy, and we need to adapt our understanding of economic systems to one where price reflects not only the value of the good but the consumer's perception of that value.

The problem of ensuring ethical algorithmic pricing has intersections with many existing domains: privacy, data sovereignty, competition law, micro and macroeconomics.  In some cases, existing laws and principles from these related fields can be extended to address algorithmic pricing.  However, brand new incentives and regulations specific to algorithmic pricing are also needed.  For example, policymakers should investigate limiting what time frame the information can be stored for, as well as granting consumers "the right to be forgotten" so that their ever more detailed consumer profile can occasionally be erased.

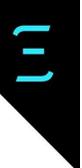

# The Epistemological View: Data Ethics, Privacy & Trust on Digital Platform

([Original paper](#) by Rajeshwari Harsh, Gaurav Acharya, Sunita Chaudhary)
(Research summary by Muriam Fancy)

**Overview:** Understanding the implications of employing data ethics in the design and practice of algorithms is a mechanism to tackle privacy issues. This paper addresses privacy or a lack thereof as a breach of trust for consumers. The authors draw on how data ethics can be applied and understood depending on who the application is used for and can build different variations of trust.

---

### Introduction

The role of data ethics is to value the concerns that humans have (privacy, trust, rights, and social norms) with how they can manifest in technology (ML algorithms, sensor data, and statistical analysis). Data ethics is meant to work in between and refine the approach of ethics towards the type of technology that it is being used for. What makes data ethics so important, especially for privacy concerns, is that it has been developed from macro ethics, so it can be tailored to focus on specific problems and issues, such as privacy and trust.

### Data ethics' two moral duties

The concern of data privacy is rooted in human psychology. Our concern for our data, such as name, address, community, and education, are essential features of the information that identify us as individuals. However, there is also a concern for group privacy. The article calls on data ethics to balance "two moral duties" such as human rights and improving human welfare. How we can do that is by weighing three variables regarding data protection: (1) individuals, (2) the society that the individual identifies/belongs to, (3) groups and group privacy.

To effectively address the moral duties presented above, it is necessary to understand the data ethics frameworks applied. There are three specific ethical challenges for which data ethics has a role in addressing. First is data ethics, which concerns research issues such as identification of a person or group, and de-identification of those people/groups through mechanisms such as data mining. As a result, the issue is group privacy, group discrimination, trust, transparency of data, and the lack of public awareness, which causes public concerns. The ethics of algorithms is the understanding of the complexity and autonomy of algorithms in machine learning applications. The ethical considerations are moral responsibility and accountability, the ethical design and auditing algorithms, and assessing for "undesirable outcomes." Individuals who could address these issues are data scientists and algorithm designers. And finally, there is ethics of practice which are the responsibilities of people and organizations responsible for leading data processes and policies. The concern areas for this problem are processional codes and protecting user privacy. Truly to address this issue, the data scientists and developers in these organizations need to be some of the first to bring up the concern.

**What we can do**

These ethical challenges are also present in artificial intelligence (AI). To effectively address the concerns brought up above, this paper proposes that AI needs to be developed and introduced by addressing trust, understanding ethics, and civil rights. To do so, AI needs to be designed using ethics, and there are three modules to do so proposed in this paper: ethics by design, ethics in design, and ethics for design. Ultimately, understanding how data ethics concerns privacy and, therefore, user/group trust, the opportunities to improve society are present. Technologies such as the internet of things, robotics, biometrics, facial recognition, and online platforms all require data ethics.

The paper concludes by addressing how trust is built-in technology, but more specifically in digital environments. The authors propose that ethics and trust work hand in hand; if one is not present, the other cannot have a meaningful effect. The two working together is how trust in digital environments can be present, which can occur through three situations:

1. The Frequency of Trust in Digital Environments: the quantification of communication of the individual in the environment is online trust. There are also two types of online trust: (1) general trust and (2) familiar trust.
2. The Nature of Trust in Tech: trust in technology must be differentiated from interpersonal trust.
3. Trust as 'Technology and Design': the notion of built-in trust technology is by humans; if the product/service fails to deliver an iteration of trust, that is a human fault.

The biggest challenge for data ethics to create trust is distributed morality, which questions the moral interactions between agents in a multi-agent system. Through distributed morality that "infraethics," the morally good action of an entire group of agents (privacy, freedom of expression, and openness).

In short, this article addresses the key challenges and normative ethical frameworks that data ethics harnesses to address trust and privacy. Understanding how trust and privacy are built-in data and data processes is one way to build ethical technology for individual and group use.

**Between the lines**

I believe that the perspective the authors take is important, and does to a degree, map out parts of the lifecycle of when data ethics should be considered. However, I would push the paper to discuss issues of how data is scrapped and thus that being an important privacy concern. The issue of consent, which may be a manifestation of moral action taken to build trust. Finally, I would push readers to consider the human element of data ethics, as to "who" is in the room choosing data sets, but even a step further, as to which groups are valued when considering data privacy.

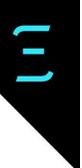

# Artificial Intelligence and the Privacy Paradox of Opportunity, Big Data and The Digital Universe

([Original paper](#) by Gary Smith)
(Research summary by Connor Wright)

**Overview:** Thanks to the pandemic, internet connectivity increasing, and companies more efficiently sharing our data, even our most private data…isn't. This paper explores data privacy in an AI-enabled world. Data awareness has increased since 2019, but the fear remains that Smith's findings will stay too relevant for too long.

---

With access to data fuelling the use of protestor tracking in Uttar Pradesh, India, as well as in the Myanmar demonstrations, privacy continues to hold a prominent position in the AI debate.

Where does data end up when we die? Will our data outlive us? What are the implications of that? With increased processing speeds and connectivity, companies are able to more accurately infer information from other websites about us. Increased data awareness has been able to slow down the speed at which this takes place somewhat, but I fear that Smith's findings may still continue to be relevant for a while.

What may even live past our own expiry date is our data itself, whereby I'll first touch upon Smith's questioning about where the data that is accumulated goes. In her op-ed about the lifecycle of AI systems, MAIEI staff member Alexandrine Royer refers to the 'AI junkyard', a place where now obsolete AI algorithms, software and hardware go to rest. As a result, given the huge amounts of data being collected due to the current pandemic, whether on coming into contact with someone who has tested positive, your medical history, or survey responses, Smith's question about where this all goes is still very much relevant today.

Such an observation then allowed me to think about whether when I eventually go to my own graveyard, will my data follow me? Smith's observation that our data has a very high chance of outliving us can be a chilling thought, especially given how we would have literally no control if a company refused to delete our data despite our passing. Even our most private data could remain in the hands of institutions, companies and other 3rd party actors who, without any legal fight being put up, are unlikely to go rooting for my data to then delete. Furthermore, even if they do take the time, my data has potentially already aided in influencing model behaviour meaning that its deletion is a mere formality. Well, at least my most private data dies with me, right?

According to Smith, this may not be the case. Encapsulated in his privacy paradox, even our most private data are not private. Given the interconnectivity of the internet, the possibility to infer certain qualities from our actions on different websites is increased through the sharing of data between different platforms (such as WhatsApp and Facebook). Having written in the midst of the Cambridge Analytica storm, Smith comments on the power of such companies to be able to extract data even without our knowledge (such as the 85 million Facebook users that formed Cambridge Analytica's database). Now with the slow entrance of 5G from companies such as Huawei, Smith notes how this interconnectivity is only going to increase, with slower processing times facilitating such data sharing actions.

Smith points out that mobile phones are "the modern-day version of the loyalty card without the perceived rewards" (p.g.151). We are gracing social media clients with our data custom and having our mobile phones 'stamped' each time we visit, but without an end in sight to reward us for our 'business'. This becomes even more interesting when Smith acknowledges how such social networking sites don't actually produce that much content on their platform. Rather, it's the users generating the content that keeps such sites engaging and vibrant, where such loyalty is still not 'rewarded' with a welcomed free coffee.

With faster processing speeds and even more data being required thanks to the pandemic, the never-ending expansion of the digital universe is very much still in full swing. There is now more awareness of privacy issues thanks to the Cambridge Analytica scandal in 2019, especially with businesses' legal obligation to now display their cookie policy and ask for consent to employ them as you enter their website. While such awareness goes a long way towards creating a space for privacy in our new digital universe, it's worrying how much of Smith's analysis still applies to our world in 2021.

**Between the lines**

One message that Smith had not made crystal clear in his paper and that is worth considering from my view is the need to own your data. It is widely touted that data has become the new oil, with AI basically being the mechanical equivalent of a human stranded in a barren desert without it. In this sense, true privacy comes through the medium of true control over your data in terms of who gets access to it and for what purpose. In this sense, privacy in 2021 would do well to focus on practical ways on how we can obtain such control, or even just begin to cultivate a business environment where this is encouraged. Without doing so, I fear Smith's privacy paradox will continue to hold steady for far too long into the future.

# Post-Mortem Privacy 2.0: Theory, Law and Technology

([Original paper](#) by Edina Harbinja)
(Research summary by Alexandrine Royer)

**Overview:** Debates surrounding internet privacy have focused mainly on the living, but what happens to our digital lives after we have passed? In this paper, Edina Harbinja offers a theoretical and doctrinal discussion of post-mortem privacy and makes a case for its legal recognition.

---

In January, The Independent revealed that Microsoft was granted a patent that would enable the company to develop a chatbot using deceased individuals' personal information. Algorithms would be trained to extract images, voice, data, social media posts, electronic messages and more personal information from deceased users' online profiles to create 2-D or 3-D representations; these digital ghosts would allow for continuous communication with living loved ones. Besides the gut reaction of creepiness raised by such chatbots, with Microsoft's Tim O'Brien admitting the bot to be disturbing, the company's patent pointed to the gaps in current legislation governing digital reincarnation.

Conversations on data privacy have tended to focus on the living, with fewer considerations of how to protect digital traces we leave behind after we have passed. Who has the right over our once private and personal information? Do the dead retain a right over their digital property, or can it be bestowed like physical possessions? Such legal ambiguities are at the center of Edina Harbinja's legal analysis of post-mortem privacy, defined as "the right of a person to preserve and control what becomes of his or her reputation, dignity, integrity, secrets or memory after death."

**Post-mortem privacy and autonomy**

The notion of post-mortem privacy is a relatively new concept, and legal scholarly attention has been slow to turn to the issue of digital assets and death in data protection. National laws are mainly inconsistent on the use of data after death, leaving companies like Google and Facebook to introduce their own policies for users to determine who can access accounts in the event of an untimely passing. The complexity of post-mortem privacy, and its consequences for user property, is compounded by the range of stakeholders involved, including interactions with other internet users, families, service providers, friends and family.

Through a brief theoretical discussion of the concept of autonomy in Western philosophy, Harbinja demonstrates how autonomy is deeply intertwined with notions of privacy, dignity and personhood. Citing legal scholar Bernal, who affirmed that 'privacy is a crucial protector of autonomy,' she aligns with his conception of internet privacy rights as entailing the concept of informational privacy. Seen through the legal and ethical rubric of autonomy, an individual's right to control their informational privacy should transcend their death.

To put it clearly, for Harbinja, "an individual should be able to exercise his autonomy online and decide what happens to their assets and privacy on death." The "real-world" assets and wealth accumulated throughout an individual's lifetime can be seen as analogous to their online assets. Freedom of testation- a person's right to decide how their estate should be distributed upon death- can be extended to the online environment. While freedom of testation appears as a straightforward legal solution, it may run counter to country-specific definitions of legal personality, with Harbinja noting there is "no clear-cut answer to when the legal personality dies". There are, however, legal examples of a person's moral rights extending beyond their death, such as in copyright law.

**Providing a legal and coded framework**

Despite post-mortem privacy being aligned with the rights of autonomy present in most North American and European judicial systems, there remains a few obstacles to its legal recognition. The principal argument against the legal recognition of post-mortem privacy is a lack of actual harm to the user, meaning "the deceased cannot be harmed or hurt." For Harbinja, such a line of reasoning is logically inconsistent with the legally enshrined principle of freedom of testation. Denying an individual control over their online data on the grounds of no harm caused would be akin to deny the rights of testament, as "the deceased should not be interested in deciding what happens to their property on death as they would not be present to be harmed by the allocation."

There are also conflicting levels of legislation that protect some aspects of post-mortem privacy, varying from laws of confidence, breach of confidence, and succession, but not the phenomenon as such. In the US, while federal law does not guarantee post-mortem privacy, certain states allow for the protection of certain 'publicity rights'- rights to name, image, likeness- up to seventy years after a person's passing. A similar situation is encountered in Europe. The EU's data protection measures, famously the GDPR, apply solely to living persons, but 12 members have introduced national legislation that protects the deceased's personal data. As highlighted by Harbinja, one notable advance in post-mortem privacy was the formation of the Committee on Fiduciary Access to Digital Assets by the Uniform Law

Commission in the United States, which proposed amendments to previous acts to allow fiduciaries to manage and access digital assets.

While certain North American and European lawmakers have yet to legislate the transmission of digital assets, many companies have begun to implement coded solutions to protect post-mortem privacy. Google launched the 'inactive account manager' back in 2013, which enables users to share "parts of their account data or to notify someone if they have been inactive for a certain period of time". For Harbinja, the main issue with IAM is the verification of trusted contacts, which happens through phone numbers, and the transfer of online content to beneficiaries- which might be individuals known solely through the digital community. Individuals would need to have their beneficiaries over digital assets be explicitly expressed in a digital or traditional will. Facebook implemented a similar measure with its option of a 'Legacy Contact', with the platforming allowing US users to designate a person to be their Facebook estate executor after their passing. Although, as Harbinja notes, this solution falls short when it comes to clarifying the rights of a designated legacy contact over the rights of heirs/kin.

**Law and life after technology**

The issue of post-mortem privacy allows us to reflect on how digital lives persist after death, and what aspects of our digital universe we wish to bestow to loved ones. With the deceased accounts still virtually present, who gains control and readership over the masses of digital traces we have left behind? As stated by Harbinja, "post-mortem privacy rights need to be balanced with other considerations, including the same privacy interests of others and the social and personal interests in free speech and security".

Beyond the legal considerations, post-mortem privacy merits a broader ethical conversation- one that is not entirely dictated by Euro-American norms and values. As our online personas become increasingly interconnected with individuals scattered across the globe, we must ensure that the legal recognition of post-mortem privacy rights does not come to the detriment of individuals residing in other societies. Indeed, how do we begin to determine ownership of digital assets, especially when it comes to the murky areas of shared messages, comments, retweets, and so forth? Further scholarly research is needed to allow for reflections on post-mortem privacy beyond the scope and principle of autonomy.

# Data Capitalism and the User: An Exploration of Privacy Cynicism in Germany

([Original paper](#) by Christoph Lutz, Christian Pieter Hoffmann, Giulia Ranzini)
(Research summary by Sarah P. Grant)

**Overview:** This paper dives deeply into the many dimensions of privacy cynicism in this study of Internet users in Germany. The researchers examine attitudes of uncertainty, powerlessness, and resignation towards data handling by Internet companies and find that people do not consider privacy protections to be entirely futile.

---

**Introduction**

For many people, social media is a critical avenue for social participation. However, users might feel powerless if they perceive that they have no choice but to give up their personal privacy in exchange for digital inclusion.

Lutz et Al examine these types of attitudes in this study, which analyzes data from a 2017 survey of 1,008 respondents in Germany. Their aim is to build on previous work related to the "privacy paradox"–the observation that privacy concerns do not always align with protection behaviours–and explore the phenomenon of privacy cynicism.

The authors define privacy cynicism as "an attitude of uncertainty, powerlessness, mistrust, and resignation toward data handling by online services that renders privacy protection subjectively futile." They assert that privacy cynicism is more than an attitude or belief–it's also a coping mechanism. As the authors maintain, this is the first study to contribute quantitative empirical evidence to online privacy research.

**Data capitalism**

The authors place their quantitative findings within the broader context of an interdisciplinary literature review. They reference the work of scholars who reflect on surveillance capitalism, data capitalism specifically, and emphasize that research in this area focuses on how data extraction is a central component of digital platforms' business models. A common thread in these critiques of digital platforms is that they "challenge user agency," in that users feel they have to choose between having meaningful social relationships or maintaining their privacy.

**The privacy paradox**

Their review of privacy paradox literature goes as far back as 1977, when the assertion was made that people acquire optimal privacy levels through their ability to control personal interactions. They reference several studies and note that, while it has been widely discussed, empirical evidence of the privacy paradox is actually weak.

**Surveillance realism, privacy apathy, and privacy fatigue**

The authors also review approaches to digital inclusion coping mechanisms other than privacy cynicism, which include privacy fatigue, surveillance realism, and privacy apathy.

Surveillance realism, for example, covers both an unease about data collection and "normalization" that distracts people from envisioning alternatives. Privacy apathy refers to a lack of privacy protection in the US, while privacy fatigue is a "negative coping mechanism, where individuals become disengaged and fail to protect themselves." Lutz et al. observe that cynicism is a core component of privacy fatigue.

**Privacy cynicism**

The authors of this study chose to focus on privacy cynicism because it has roots in social psychology and is tested by more generalizable data. They describe cynicism in great depth, noting that it is about assumptions of self-interest. Cynicism is also about powerlessness: when one of two participants in a relationship have little control over decision making, they grow cynical. Risks are therefore perceived as inevitable because they are out of the person's control.

They argue that the combination of data capitalist business models along with design goals of maximizing user engagement might make it too complicated for users to consider their desired level of disclosure for specific situations.

**Results**

The quantitative study tests several hypotheses and yields many key findings. In general, the research reveals that German users feel "quite powerless and distrustful" but do not harbour widespread resignation. Internet skills mitigate privacy cynicism, but do not eliminate feelings of mistrust. People tend to be more cynical after they have had a privacy threat experience, but mistrust does not appear to stem from experience.

Powerlessness is the most prevalent factor associated with privacy cynicism, while resignation that produces the perceived futility of privacy-protecting behaviours is the least prevalent factor.

One important finding is that privacy concerns have a positive effect on privacy protection behaviour. Therefore, the researchers find no evidence for the privacy paradox in this study.

**Implications for public policy and future research**

The authors state that "lacking control over the sharing of personal data online appears as the most salient dimension of privacy cynicism" and therefore policy and other interventions should focus on giving agency back to users. They also state that future research could look at powerlessness in relation to the kind of business models as described by researchers who focus on data capitalism and surveillance capitalism.

It is important to note that the work of Lutz et al. also provides the foundations for further investigations beyond privacy, setting the stage for future explorations into whether an awareness of social media's impacts on personal well-being and democracy contributes to user cynicism.

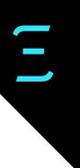

# Go Wide: Article Summaries (summarized by Abhishek Gupta)

**Why Getting Paid for Your Data Is a Bad Deal**
([Original *EFF* article](#) by Hayley Tsukayama)

Many arguments on privacy protections associate a remuneration to support the provision of data by people to organizations, but this article makes the well-reasoned claim that this leads to the commoditization of privacy. Additionally, there isn't even a fair way to arrive at a compensation structure for your data that wouldn't continue to provide huge advantages to the organizations to the detriment of individuals, especially those who are vulnerable. Providing an apt analogy with the freedom to speak, you would (at least rationally) not want a price tag on that.

One of the places where the argument for providing data dividends falls flat is in making the value determination of that data. Organizations using the data have the most insight into the life-time value (LTV) of a piece of data, and hence, they have an asymmetric advantage. Even if a third-party arbiter is to determine this, they still will not be able to really ascertain the value of this data. Seemingly innocuous pieces of data, which people might be willing to accept pennies for, when combined with troves of data that organizations have, can really augment its value by orders of magnitude.

Some organizations like AT&T make offers to knock off a few dollars of the monthly bills of their subscribers if they would be willing to watch more targeted ads. This exploits exactly the most vulnerable people since they could benefit from saving some money but in the process have to trade-in their privacy. Such an ecosystem that encourages data dividends necessitates the creation of a privacy class hierarchy, reinstituting larger existing social hierarchies. Thus, enacting privacy protections that are universal and don't make compromises are essential if we are to treat privacy as a right and not as a commodity. We need to move away from treating privacy as a luxury, affordable only to those who have the means to pay for it and leaving the rest behind as second-class citizens.

### Inside NSO, Israel's Billion-dollar Spyware Giant

(Original *MIT Tech Review* article by Patrick Howell O'Neill)

NSO is the company behind the Pegasus software that is utilized by law enforcement and other intelligence agencies around the world to target an individual by infecting their phone and gaining complete control over it. The article provides a fair amount of detail on the background of the company and what their software's impact has been on people who are misidentified as terrorists or persons of interest. It also details their responses to allegations and cases that large technology companies are bringing against it when they utilize their infrastructure to mount attacks.

Describing the horrific events that someone in Morocco had to endure because their phone was compromised by Pegasus, the article also mentions cases in Spain and Mexico where the software has been used for different purposes. The legal counsel for NSO claims that they are only manufacturers of this software and don't actually use it, so they should not be held liable for any harms that arise from it. This is essentially the same argument put forward by anyone that is building a tool that can be misused, of which there are many, many examples. It does not really allow them to obviate their moral responsibility. Especially when the controls that are currently present at NSO, both technical and organizational, seem to be insufficient to properly handle misuses. In particular, what is concerning is the reactive stance taken to any allegations brought against them rather than being proactive in addressing the concerns.

A lot of their work is shrouded in secrecy because of its association with the Israeli national government which further limits the amount of information available to the public in terms of how it operates. The Israeli regulators also don't hold themselves accountable to any misuses, which further diminishes the power of regulatory controls. They do have some technical guardrails in place, such as the prohibition around infecting US phones and self-destruction of the software on US soil, but those are few and far between. Calls from the global community for stronger regulation aren't radically new and more than the financial damages, NSO having to go through the discovery process in legal proceedings might actually pose the bigger threat to their modus operandi.

## Privacy Considerations in Large Language Models
(Original Google AI Blog article by Nicholas Carlini)

Large scale language models offer massive utility, but what are the tradeoffs of using such large-scale data and models? In this work from researchers, they point to privacy breaches that can emerge from such scenarios when we use very large corpora to train systems that might inadvertently sweep up personally identifiable information (PII).

Prompting the model with specific pieces of data, the researchers were able to extract data that constitute private information. Though they obtained requisite permissions from those who were affected by this, the possibility still looms large that anyone with access to the resources and know-how can do so as well and might not be as scrupulous in their approach. They essentially did this through a technique called 'membership inference attack', which through looking at the degree of confidence that the model has on its predictions when given certain inputs, can be used to identify in which cases there was memorization and hence potential sources for leaking private information.

The primary contribution of this work is that the researchers are able to sift through millions of possible inputs and outputs to select those that will prompt the model to output memorized and potentially private information from its training data. The researchers also found that the larger the trained models, the more the likelihood of memorization and hence the potential for it to be leaked when presented with well-crafted inputs. Techniques like differential privacy which can be utilized off-the-shelf from the frameworks in the case of TensorFlow, PyTorch, and Jax by replacing the optimizers with differentially private ones is a way to go about protecting the privacy of the data that models are trained on. Though, the researchers point out that if there are sufficient occurrences of that information, even differential privacy cannot protect us from the leakage of information.

## How Your Digital Trails Wind Up in the Police's Hands
(Original Wired article by Sidney Fussell)

There was a case in the US where smartphone data was used as the primary source of information to provide evidence about a crime that took place based on keyword search warrants. The warrants first yield lists of anonymized users, and then the police can ask for the results to be narrowed down so that they can hone in on the suspects. This is also supplemented by geofence warrants that provide a lot of spatially sensitive information about the goings-about of people.

Another consideration is the risk that such warrants pose to the privacy of those who are not ultimately deemed suspects, but whose information was shared nonetheless as a part of the investigation. Additionally, such approaches supercharge the ability of law enforcement to use existing legal instruments in novel ways to exercise more power than is perhaps mandated by law.

Data collected for innocuous purposes, like showing you the local weather, could be weaponized against individuals to figure out their visitation and travel patterns. This is only exacerbated by the fact that there is a lack of clear guidelines on how personal data should be handled and used. In speaking with developers, researchers and journalists found that most are quite unaware of the downstream uses of data by third parties that they integrate to provide services within their apps. A call for transparency on the part of the companies when they are queried to share information about their users is a starting point to balance the asymmetry of power in the information ecosystem.

### Here's a Way to Learn if Facial Recognition Systems Used Your Photos
([Original *NY Times* article](#) by Cade Metz, Kashmir Hill)

This article talks about a tool called Exposing AI that helps people determine if the pictures that they posted on Flickr were used to train facial recognition technology. Flickr is not as popular as it once was, but the permissive licenses for the images that were uploaded to the website continue to have large-scale implications for facial recognition technology today. Especially as they constitute training datasets in a lot of the modern-day systems, including those that have been used for surveillance in China.

One of the things that I appreciated in the creation of this tool was that they took security seriously to avoid people using this as a querying tool to eke out information about individuals and performance on those individuals in terms of the deployed facial recognition technology systems. It only allows you to query for images that are already publicly available and those that have a URL pointer further decreasing the possibility that someone can just upload an image and run a query against the tool that they have created.

The article points to how other large-scale datasets, some of which have been assembled illegally, like one from the University of Washington called MegaFace which was built in earnest to help researchers in the community and had an associated competition, might have been co-opted into something that now is being misused. This is characteristic of creating and making public datasets, especially in cases without consent, where we lose the ability to control who uses it in the future and how to control the distribution of it even if it is taken down, as was the case with MegaFace. Copies of that data still circulate online and are continuing to perhaps be used to fuel more surveillance applications.

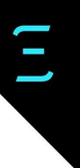

### This Is How We Lost Control of Our Faces
([Original *MIT Tech Review* article](#) by Karen Hao)

An article that talks about a comprehensive study recently published on the datasets that go into the making of facial recognition technology points out the devolution of the health of the ecosystem in terms of consent and respect for the privacy of individuals. The authors of the paper argue that in the early days of facial recognition technology, data was collected and annotated through highly manual and labor-intensive methods which made data collection quite limited and obtaining consent the norm. However, as more people realized the potential that this technology offered, they sought to overcome the limitations of the techniques at the time by utilizing larger and larger datasets.

At first, larger datasets were manually collected. To fully leverage the advantages that deep learning systems offered, the researchers needed to move beyond the limits imposed by human collection methods and relied on automated scraping of web data to find more faces. This inadvertently sucked up photos of minors, and of course consent wasn't a major concern as the researchers were focused on improving the performance of the system. Datasets like Labeled Faces in the Wild among others contain pictures of people who have not necessarily consented to all the uses of that data. Consumers are more sophisticated now; they want to know how their data is being used, and they unapologetically call out companies when they're collecting and using user data without proper consent.

Yet, as we know with data on the internet, once it is out there, we have very little control in terms of how it might be used. Therefore, prevention and proper consent mechanisms are going to be key if we're to achieve an environment where the rights of people are placed first and foremost over research interests and commercial applications.

## Clubhouse Is Suggesting Users Invite Their Drug Dealers and Therapists
(Original *OneZero* article by Will Oremus)

Clubhouse is the talk of the (virtual) town and even hardened privacy-minded folks have succumbed to the allure. I am guilty of the same when I saw that some other AI ethics folks were hosting conversations there (ironic!) and I wanted to listen in. Luckily for me, I don't have an iPhone so I was unable to join. But, after reading this article, I felt that it was a bullet dodged. The article highlights some of the insidious ways that information about your social graph is leaked once you join the app.

It employs dark patterns like strong nudges to get you to share your contact list from your phone, something that you can choose not to do, but then it strips away your ability to send out more invites. Additionally, it shows you a ranked list of your contacts in terms of how many other people on Clubhouse are also connections there. This can lead to exposing more data than you may like, like how many other people have the same lawyer, doctor, etc. as you.

Another interesting aspect pointed out by the author in the article is that a lot of folks who were going to join the app may have presumably done so by now (the more tech- and marketing-savvy folks) and those who are still listed as the top potential people that you should invite are those who are ardently refusing to do so and might give more information about them. Just as Facebook many years ago built shadow profiles on users who were not on the app but through the others on the app and their related contacts it was able to infer things about this person, Clubhouse now has the same ability. Until we get more information about their data storage practices and how they comply with lawful intercepts, perhaps it is best to stay away from such an app that has severe privacy issues.

## Facial Recognition Technology Isn't Good Just Because It's Used to Arrest Neo-Nazis
(Original *Slate* article by Joan Donovan, Chris Gilliard)

When we weigh the positives of facial recognition technology against the negatives, the scales are tipped unequivocally towards the negatives. With the tragic incidents at the US Capitol in January 2021, some people resurfaced the debate mentioning how the use of facial recognition technology helped us arrest some of the insurrectionists and bring them to justice. One of the points made in this article that hasn't been explicitly called out in many places is that facial recognition technology was only a piece of the puzzle, and a lot of the evidence also came from posts that were made on Instagram and Facebook, which helped to bring people to justice.

Time and again, researchers, activists, and others have demonstrated that facial recognition technology is particularly biased and has many flaws, yet the debate keeps popping up again. Even when it might be used in a positive way, facial recognition's downsides are too big to ignore. Some argue that the only way to combat the use of facial recognition technology is by also giving it into the hands of everyday people. Take, for example, the case in Portland where a person used it to identify erring police officers.

But even if we have bans on the use of this technology in a particular regime because of the amount of money that organizations stand to make from selling these technologies, we need to make sure that there is collective action at a global level that stops the development and deployment of this technology. As global readers, we see you as a core actor in the call for addressing the points of this debate in an informed manner.

## How One State Managed to Actually Write Rules on Facial Recognition
([Original *NY Times* article](#) by Kashmir Hill)

- What happened: A bill that will be enacted in July in Massachusetts will pave the way for having a meaningful ban on facial recognition technology to allow us to use it in a positive way while minimizing the harm that arises from its use, for example with racial biases. This balanced version came about due to significant efforts from Kate Crawford at the ACLU who was instrumental in helping the lawmakers understand the nuances behind the use of facial recognition technology, both from the positive and negative sides.

- Why it matters: While most existing efforts call for an outright ban (which is warranted in many situations), a balanced approach that puts in place mechanisms like the separation of obtention of a warrant and the execution of this technology to perform a search can help to build in accountability and mitigate harm from false matches and other problems that have plagued facial recognition technology.

- Between the lines: What was particularly great to see here was the degree of influence that non-governmental organizations can have in crafting bills and regulations that can bridge the gaps that lawmakers might have in their knowledge on the real capabilities and limitations of these systems and how to propose ways forward that are meaningful and those that leverage the best that this technology can offer.

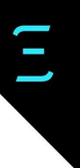

### China's Surveillance State Sucks Up Data. U.S. Tech Is Key to Sorting It.
(Original *NY Times* article by Paul Mozur, Don Clark)

While we might place a huge amount of scrutiny on the utilization of facial recognition technology in China, we often fail to recognize the underlying technology suppliers who are enabling that to take place. This is the classic example of dual-use technology like GPUs that can be used for examining protein folding using deep learning or serving the needs of an authoritarian state. However, in the case of both Nvidia and Intel supplying chips to Sugon, it seems that they were aware of what purposes the company would use the chips that they were selling to them.

In several pieces of marketing material from the firm, it was evident that they would use the technology for surveillance. In fact the suppliers also touted this particular case as a demonstration of the success of their chips. They have since retracted this, saying that they weren't aware that it might be used to violate the privacy and human rights of the Uighurs. Technology accountability ends up taking a backseat in the interest of profits and without firmer export controls and policies, we risk continuing to perpetuate harms.

A particularly chilling mention in the article that one might easily gloss over is the supposed development of ethics guidelines within Intel on the use of their technology. Interestingly, those were neither made public nor were the people associated with it willing to disclose their identities or discuss it in more detail. This exacerbates concerns around where else such suppliers might be enabling malicious actors to inflict harm on people, mostly outside the watchful eye of regulators, both internally and externally.

### China Is Home to a Growing Market for Dubious "Emotion Recognition" Technology
(Original *Rest of World* article by Meaghan Tobin, Louise Matsakis)

In this article, borrowing from work done recently by Article 19, a British organization, we get a peek into the pervasiveness of emotion recognition in classrooms in China. The limitations of systems attempting to recognize emotions by scanning physiological expressions are well-known. One reason for those limitations is the wide variation in how emotions are expressed based on culture and region — facial expressions often do not carry the same meanings universally.

Yet, in highly competitive environments like China and India where the school system offers a way to an improved standard of living, companies peddling these systems pander to the anxieties of parents to get schools to purchase and deploy these systems. With the pandemic in full swing, even in countries like the US, such systems have been used experimentally in school with the intention of boosting the productivity of students and teachers in the classroom.

Such systems suffer from similar biases and failures just like facial recognition technology systems. We need to be advising schools and other places to tread very carefully in the use of untested technology in the classroom on those who are unable to offer informed consent.

## How Regulators Can Get Facial Recognition Technology Right
([Original *Brookings* article](#) by Daniel Ho, Emily Black, Fei-Fei Li)

In this article, authors from the Stanford HAI Centre offer some concrete guidance for both technical and policy stakeholders to meaningfully regulate AI systems. Specifically, they look at the concepts of domain and institutional shift.

Domain shift is the notion that the performance of the system varies between what it was trained and tested on and how it behaves when the system is actually deployed and interacts with the real-world data. A lot of the training of these systems happens with well-sanitized inputs such as well-lit images, front-facing pictures but the real world has messy data where the lighting conditions are far from optimal and people might not be looking directly at the camera all the time. This has severe impacts on the performance of the system.

Institutional shift refers to how the system is actually used and how its outputs are interpreted by different organizations. An example of this is when you have police departments from different parts of the country that place different thresholds in terms of the confidence intervals required to flag someone as a match. This can have implications such as the high-profile case in the US earlier this year when someone was wrongly detained for 30 hours because of an incorrect match.

Combatting some of these challenges require: a higher degree of transparency from the manufacturers on what training data that they are using, allowing for 3rd party audits of their system and sharing those benchmarks publicly, and periodic recertification and assessments of the system. That way, we can verify that the system still meets the requirements and thresholds that we have mandated for it to operate in.

## Uganda Is Using Huawei's Facial Recognition Tech to Crack Down on Dissent After Anti-government Protests
([Original *QZ* article](#) by Stephen Kafeero)

An epitome of what we can imagine going wrong with the use of FRT is taking place in Uganda with the government actively using it to suppress anti-government sentiments amongst its people. It is also exploring the links of the extensive data gathering from this instrument with other government agencies, like the tax authorities and the immigration department, which will further impinge on the rights of the people in Uganda.

Many local organizations have called for a ban on the use of the technology but that has had limited to no effect. Whereas in places like the US and UK, where there is a backlash against the use of technology used for surveillance, Uganda has been welcoming of the support from these firms as they are helping them leapfrog older generations of technology.

This is a consequence of the uneven distribution of technology and the selective deployment of grants and resources to help nations which can push them into the arms of nation-state backed organizations that can subtly insert themselves into the affairs of a developing country with the aims of re-establishing colonialist patterns. Human rights violations are additional damage that these nations have to endure in this process.

## Facial Recognition Company Lied to School District About its Racist Tech
([Original *Vice* article](#) by Todd Feathers)

NIST, the US standards agency, has a benchmark that evaluates many of the facial recognition technology vendors for various factors, including their rates of accuracy on different demographic compositions. This article shines a light on the claims made by a particular facial recognition technology vendor that duped schools into buying its technology with the promise of better safety and minimal risks of bias.

One of the leading scientists who worked on the benchmark for NIST pointed to the discrepancy between the claimed performance of the system and what was found on the benchmark test in their published findings. What remains unclear are the rates of false positives and what the actual prevention of crime is. If parents are sold on the promise that their children will be safer through the use of such intrusive technology, then they are also well within their rights to demand that the efficacy of the system be shared with them.

As if this was not already a problem, some of the parents lamented that the use of the Smart School funds towards the procurement of this technology to the exclusion of other tools and upgrades has largely been rendered useless because of COVID-19 lockdowns. Specifically, some other schools chose to use the money for things like improved connectivity and new laptops, which would have been tremendously useful during the lockdown. This also raises the larger question around how we must be quite deliberate about fund allocation and not chase down shiny new pieces of technology without having ample evidence that they work as they claim to be, especially in a context where we have populations who are more vulnerable than the rest.

# 6. Outside the Boxes

**Opening Remarks** by MAIEI staff Shannon Egan, Muriam Fancy, and Victoria Heath

**Creating a Better Digital World Means Exposing the Physical One**

Humanity's knowledge and culture are increasingly collected, stored, and shared online. With every image posted to Instagram, article created on Wikipedia, or video uploaded to TikTok, the mundane moments to the extraordinary are recorded and shared across the globe, to billions. It's quite remarkable—and in many ways, wonderful. Every day, more people have access to our collective and individual histories, languages, and cultures than ever before. Despite the benefits of this shared digital archive, however, there are harms that we can't ignore.

This section of the State of AI Ethics Report explores, in various ways, how the illusion of "magic" leaves the digital world vulnerable to exploitation and misuse. Large swaths of the population are not aware of the physical infrastructure, from underwater cables to acres of server farms, that makes their ability to share photos, send emails, and access information possible. More concerning, the human labour behind our digital world remains hidden and ignored. This is especially true with artificial intelligence (AI). As Nazelie Doghramadjian explains in "Digital Archives Show Their Hand," "Companies design and implement AI systems in a way that purposely obscures the intimately linked human provenance and operation of those systems."

Further, when these systems are deployed, they interact with the physical world we inhabit. Researchers must anticipate how the world will impact these systems and vice versa. Before releasing a new machine learning algorithm, for example, developers should ask: Does the model work on even the messiest data it will encounter? Is it vulnerable to manipulation? The first question led a team at Microsoft Research to develop what they call "unadversarial examples," an approach to increasing robustness that modifies the training data, not the algorithm itself. The second question is at the forefront of conversations regarding machine learning security. Other technical highlights of this section include Green Algorithms, an open-source tool that estimates the carbon footprint of general computation, and a Stanford review on the current state of Natural Language Processing (NLP) technology and the impact it could have on society.

The State of AI Ethics Report, April 2021    150

The physical world also restricts access to the digital one. In "Where the Internet was Delivered by a Donkey," Zeyi Yang reminds us that the "magic" of the internet is grounded in a reality of physical infrastructure and natural boundaries. For example, the mountainous geography of Kyrgyzstan makes it extremely difficult to lay cables. Thus, most of the population doesn't have access to the internet. Luckily, humans are great at crafting makeshift solutions: the IlimBox, created by the Internet Society's Kyrgyzstan Chapter, stores 500 books, 250 videos, and 4 million Wikipedia articles and requires no internet access to use, making it a lifeline to students in the country.

Increasing global access to the internet requires more long-term solutions, however, especially as the digital world becomes increasingly concentrated. "Only seven companies control the vast majority of the internet's traffic and much of its infrastructure," writes Thomas Smith in his review of the Mozilla Foundation's 2021 Internet Health report. Worse, he writes, "four of the most-used platforms on the internet are controlled by a single company: Facebook." Thankfully, policymakers are waking up to this issue, but their efforts are not as effective as they need to be, particularly in regards to AI.

The ever-evolving capabilities of AI make it difficult to quantify its impact and touchpoints with both targeted and non-targeted users. Thus, policy needs to address conceivable consequences of AI, as is currently underway in many efforts to regulate (or outright ban) facial recognition technology. Several articles in this section of the report, however, outline how policymakers continue to fail in their foresight to account for future harms from AI. It's also becoming increasingly obvious that the degree to which national and international policies can be applied to AI is sector-dependent. Although there are legitimate reasons for why sectors need to be governed by specific policies, it's imperative that any policy considers power relations between developers and users and respects basic human rights.

To live in the 21st century is to be "connected"—reliant on a digital world that increasingly facilitates our relationships and houses our identities. Unfortunately, this world is frail and vulnerable to breaking, evident with the perpetuation of systemic harms. We have a chance to repair it, however. To begin, we must recognize the links between the physical world and the digital. We must erode the illusion of "magic" by peeking behind the curtain, exposing the complex realities underneath.

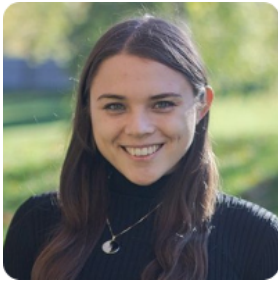

**Shannon Egan**
QM Research Intern
Montreal AI Ethics Institute

Shannon is a theoretical physicist-in-training whose research explores the properties of quantum materials. However, her interest in constructing frameworks to understand complex problems extends well beyond the microscopic realm. As a Quantitative Research Methods Intern at MAIEI, her role involves finding ways to quantify the impacts of AI on academia, industry, and humanity.

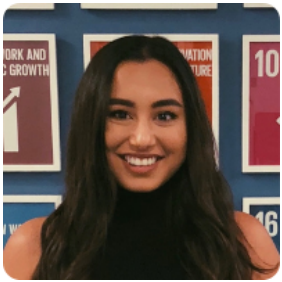

**Muriam Fancy (@muriamfancy)**
Network Engagement Manager
Montreal AI Ethics Institute

Muriam Fancy is the network engagement manager at the Montreal AI Ethics Institute. Muriam's research exists at the intersection of technology policy and design to examine the social implications of technology and the harms it can produce on racialized communities. Muriam's academic background is in global human rights policy and social innovation.

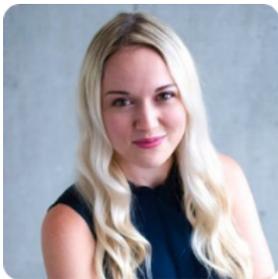

**Victoria Heath (@victoria_heath7)**
Associate Director of Governance & Strategy
Montreal AI Ethics Institute

Victoria is a researcher and storyteller working at the intersections of technology, science, human rights, and global policy. She has paired creativity and academic curiosity for over a dozen nonprofit organizations, academic institutions, and companies, such as the Institute for Gender and the Economy, MaRS Discovery District, the Campaign to Stop Killer Robots, Access Now, and Creative Commons. She is currently the Program Manager of Moon Dialogs for the Open Lunar Foundation and the Associate Director of Governance and Strategy at the Montreal AI Ethics Institute.

# Go Deep: Research Summaries

## Governance by Algorithms
([Original paper](...) by Francesca Musiani)
(Research summary by Connor Wright)

**Overview:** Through exploring the world of e-commerce and search engines, algorithms are no longer to be relegated to solely inputting and outputting data according to some specific calculations. Its intangibility, unquestionability, and influence over what we believe agency to be is explored in this paper, giving us the low-down on the influence such algorithms can and do, have in our lives.

---

As I'm sure is well-known by many, the influence of algorithms is very real. Algorithms are no longer portrayed as just being confined to turning input data into output data according to some specific calculations, with their effects expanding and influencing well beyond its bounds. To demonstrate this, I'll first draw on the paper's journey through e-commerce and the subsequent questions of responsibility that come up. I'll then look into how this too can then be seen in the topic of search engines. From there, it'll be worth noting the intangibility offered by algorithms, and the subsequent tabu around questioning its decisions despite such intangibility. Having considered the above, the question of algorithmic governance by who and over whom proves to be the next step in what is such an elusive topic.

**The realm of e-commerce**

The paper's venture into the role of algorithms in the field of e-commerce mainly centres on how algorithms are used in our everyday commercial lives. Here, Amazon's algorithmic uses are observed to have become a key player in "prescribing", with its algorithm having sorted through the endless amounts of data made available by our conduct on its website. For example, whenever we receive a 'other people were also interested in' upon buying an item, we have Amazon's algorithm to thank. Similarly, algorithms across the web space are being used to sift through the mountains of data available to it in order to track our engagements across different websites and thus 'personalise' our experience across each site (mainly done through cookies). In this sense, governance by algorithms in the e-commerce world is seen by the data separating being left up to the algorithm itself, proving to be a double automation as the resultant decision from said sorting also has to be made by the algorithm given how it sorted the data itself.

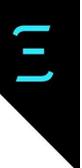

What then rises to the surface, are questions over responsibility and agency. Given how the algorithm has sorted through the data and made the decision on what to recommend you, what happens if you find the recommendation offensive? Can the human involved in providing the data to the algorithm or the designer of the algorithm be held accountable for what they weren't involved in? If not, this then leads us to the odd thought on whether we ought to grant the algorithm agency given the level of independence of its actions. Of course, since the algorithm doesn't actually have the capacity to realise what it's doing and is blissfully unaware of the consequences of its actions, this cannot be the case. Nevertheless, the potency of the algorithms is still demonstrated by the mere need to consider its agency, and especially if it cannot take the blame for its actions.

**The realm of search engines**

A similar situation is then found in exploring algorithmic involvement in search engines. Here, the order of results on our searches through search engines, whether on Google, Firefox or Yahoo, are determined through the sorting action of an algorithm. To give an example, Google's PageRank algorithm has been labelled by Masnick as a "benevolent dictator", benevolently sorting through the data and dictatorially prioritising what is being most engaged with on the internet itself. Such prioritisation then stems from the users bringing the content to the algorithm's attention through internet publications, thus making the space co-authored between the public and the algorithm. Hence, algorithms that form the basis of search engines are susceptible to being swayed by the public just as easily as humans. So, how can we be governed by algorithms in this space?

The main response to this question finds itself when talking about the public space. Here, how can the digital space actually be a public space if certain digital information is displayed more than others due to it being more prominent? There will be some conversations that never achieve such coverage, but are to be viewed as no less important. In this way, we can catch a glimpse of the invisible workings of algorithms and their ability to govern the digital space, captured eloquently in Masnick's quote.

**The intangibility of algorithms**

What then interests me most, as a result, is how despite only catching a glimpse of the workings of an algorithm, it is almost taboo to question its outcomes or decision-making process. To flesh this out, the paper mentions Gillespie's six dimensions of political valence which an algorithm influences: the "promise of objectivity" and the "entanglement with practice". Here, the assumption that algorithms guarantee objectivity due to their non-human touch and the subsequent adjustment of human processes to incorporate that, cedes even more control to the algorithmic governance process. Due to its perceived objectivity, any allegation against the algorithm gets quickly dismissed as owing to our own personal biases and distorting the truth presented by the algorithm. As a result, such questioning subsides and processes are altered in order to centralise the algorithm and its truth displaying ability. Step by step, such tabu around questioning the algorithm slowly slips into the increased governance of algorithms over human practice; the governance over the material by the immaterial.

This intangibility of the algorithmic process that gains such high repute starts to make it harder and harder to see its influence. The governance of algorithms takes on this 'cloak of invisibility', where its inner workings are hidden by an iron curtain thanks to the automation of its process (data sifting and subsequent decisions made). As seen in the e-commerce and search engine explorations, what makes these areas tick increasingly becomes more and more focused on the immaterial, rather than owing to the physical (such as a human agent). In this way, the influence of algorithms is there for us to observe, but we aren't even best suited to look for it.

As clearly shown in the arenas of e-commerce and search engines, governance by algorithms is widespread. Its elusiveness both in terms of appearance and in terms of its unquestionability means that algorithms are slowly being adhered to without a second thought. For me, this is where the true power of governance by algorithms lies. Its perceived objectivity in all cases and subsequent changing of the processes surrounding any algorithmic interaction is where governance by algorithms really takes form. Governance by whom and over whom is then a whole other story.

# Green Algorithms: Quantifying the Carbon Footprint of Computation

(Original paper by Loice Lannelongue, Jason Grealey, Michael Inouye)
(Research summary by Alexandrine Royer)

**Overview:** This paper introduces the methodological framework behind Green Algorithms, a free online tool that can provide a standard and reliable estimate of any computational task's carbon emissions, providing researchers and industries with a sense of their environmental impact.

---

In the past year, we have witnessed costly wildfires in Australia, increasing droughts in the continental US, heavy rainfall in India, and more natural disasters. No corner of the globe is left untouched by the direct consequences of our rapidly warming world, and there is mounting public pressure for industries to make their activities ecologically viable. In the tech realm, privacy, security, discrimination, bias, and fairness are the common buzzwords that surround AI, yet the word "green" is rarely present.

With the world facing an acute climate crisis, high-performance computing's carbon emissions continue to be overlooked and underappreciated. Lannelongue, Grealey, Inouye (the authors of the original paper) are hoping to change the conversation by introducing the free online tool Green Algorithms, which can estimate the carbon impact – or CO2 equivalent- of any computational task. By integrating metrics such as running time, type of computing core, memory used, and the computing facility's efficiency and location, the Green Algorithm tool will allow researchers and corporations alike to get a reliable estimate of their computation's environmental impact.

Climate policy is becoming a pressing issue for governments who are making new climate commitments. The EU has committed to reducing all its greenhouse-gas emissions to zero by 2050. Chinese president Xi Jinping has jumped on the green bandwagon and promised to reduce the country's carbon emissions to zero by 2060. While it is tempting to celebrate these governmental actions, the EU and China are significant investors in data centers and high-performance computing facilities, which release around 100 megatonnes of $CO_2$ emissions annually. When it comes to assessing an algorithm's environmental impact, we must consider both the energy required to run the system (i.e. the number of cores, running time, data centre efficiency) and the carbon impact of producing such energy (i.e. location and type of energy fuel).

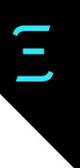

The authors note that while there have been advances in green computing, these are concentrated on energy-efficient hardware and cloud-related technologies. Power-hungry machine learning models have grown exponentially in the past few years. Although some studies have attempted to calculate such systems' carbon impacts, they rely on users' self-monitoring and apply to only particular hardware or software. As stated by the authors, "to facilitate green computing and widespread user uptake, there is a clear and arguably urgent, need for both a general and easy-to-use methodology for estimating carbon impact that can be applied to any computational task."

The Green Algorithm can calculate the energy needs of any algorithm by considering its "running time, the number, type and process time of computing cores, the amount of memory mobilized and the power draw of these resources." The model also accounts for the data centre's energy efficiency, such as lighting, heating or AC. To estimate the environmental impact behind the energy produced to run these systems, the authors employ a carbon dioxide equivalent to stand in for greenhouse gases' global warming effects. The environmental impact is assessed by calculating the carbon intensity, being the carbon footprint of producing 1 kWh of energy. The data centre location is also an essential factor, as the source of energy, whether it is hydro or coal or gas, will affect the calculation. The model further provides a pragmatic scaling factor, which multiplies the carbon impact by the "number of times a computational is performed in practice."

After entering the energy-related details of their algorithms, users will be provided with comparative figures such as the equivalent percentage of carbon emissions for international flights and the number of trees required to sequester the emissions of running their system. The Green Algorithm was tested by the authors for algorithms used in particle simulations and DNA irradiation, weather forecasting, and natural language processing, making it applicable to a wide variety of computational tasks. As summarized by the authors, "besides drawing attention to the growing issues of carbon emissions of data centres, one of the benefits of presenting a detailed open methodology and tool is to provide users with the information they need to reduce their environmental footprint." The results produced by the Green Algorithm should be taken as a generalizable and relative figure, as the authors note there are limitations to the tool, such as an omission of hyperthreading, the exact breakdown of energy mixing in a given country, and the lack of a standard approach to calculating power usage effectiveness.

The Green Algorithm is a welcome development in monitoring the environmental impact of AI-related advances. The use of such tools can help stir proactive solutions to mitigating the environmental consequences of modern computation. While the field of machine learning appears to be fueled by relentless growth, policymakers, industry leaders, and the public will need to consider whether the environmental costs of introducing these new systems are far-outweighed by the potential societal benefits.

# The Secret Revealer: Generative Model-Inversion Attacks Against Deep Neural Networks

([Original paper](#) by Yuheng Zhang, Ruoxi Jia, Hengzhi Pei, Wenxiao Wang)
(Research summary by Erick Galinkin)

**Overview:** Neural networks have shown amazing ability to learn on a variety of tasks, and this sometimes leads to unintended memorization. This paper explores how generative adversarial networks may be used to recover some of these memorized examples.

---

Model inversion attacks are a type of attack which abuse access to a model by attempting to infer information about the training data set. Effective model inversion attacks have largely been on extremely simple models such as linear regression and logistic regression, showing little promise in deep neural networks. However, generative adversarial networks (GANs) provide the ability to approximate these data sets.

Using techniques similar to image inpainting for obscured or damaged images, the GAN creates semantically plausible pixels based on what has been inferred about the sensitive features in the training data. A Wasserstein-GAN is used to set up a min-max problem as the loss function, and some auxiliary knowledge about the private images are provided to the attacker. This serves as an additional input to the generator. The generator then passes the recovered images to both the target network and a discriminator. The loss from both of these inferences is combined to optimize the generator.

Using facial recognition classifiers as a model, Zhang et al. find that generative model inversion is significantly more effective than existing model inversion methods. Notably, more powerful models which have more layers and parameters are more susceptible to the attack.

Zhang et al. also find that pre-training the GAN on auxiliary data from the training distribution helps recovery of private data significantly. However, even training on similar data with a different distribution – such as pre-training on the PubFig83 dataset and attacking a model trained on the CelebA dataset still outperforms existing model inversion attacks by a large margin. Some image pre-processing can further improve the accuracy of the GAN in generating target data.

Finally, Zhang et al. investigated the implications of differential privacy in recovering images. They note that differentially private facial recognition models are very difficult to produce with acceptable accuracy in the first place, due to the complexity of the task. Thus, using MNIST as a reference dataset, they find that generative model inversion can expose private information from differentially private models even with strong privacy guarantees, and the strictness of the guarantee does not impact the ability to recover data. They suggest that this is likely because "DP, in its canonical form, only hides the presence of a single instance in the training set; it does not explicitly aim to protect attribute privacy."

## AI Safety, Security, and Stability Among Great Powers

([Original paper](#) by Andrew Imbrie, Elsa B. Kania)
(Research summary by Abhishek Gupta)

**Overview:** This paper takes a critical view of the international relationships between countries that have advanced AI capabilities and makes recommendations for grounding discussions on AI capabilities, limitations, and harms through piggybacking on traditional avenues of transnational negotiation and policy-making. Instead of perceiving AI development as an arms race, it advocates for the view of cooperation to ensure a more secure future as this technology becomes more widely deployed, especially in military applications.

---

**What are some of the key problems?**

- Unsurprisingly, given the popularity of AI, military leaders are often excited by the potential of deploying this technology without complete consideration for the risks that might arise from its use.
    - Unexpected failures and emergent behaviour in a highly volatile environment like war presents very real concerns.
    - AI systems are vulnerable to new vectors of attack and the novel domain of Machine Learning Security is highly important to include in these discussions.

- The ability to use AI systems in warfare lends advanced capabilities to non-state actors who might not adhere to items like Article 36 that checks whether new weapons are consistent with the Geneva Convention posing an additional risk. Typically, state actors do follow these laws.

- From a policy perspective, a tradeoff that comes up frequently is the benefit that such coordination efforts can have in terms of minimizing miscalculation of the other's capabilities and reducing inadvertent escalation. But, this might also mean that we create more robust AI systems that are quicker and more effective in their deployments.

**How do we create pragmatic engagement?**

- Developing a shared vernacular: I've pointed out in my previous work with a colleague published at the Oxford Internet Institute that there is a dire need to have consistency in how we discuss the risks from AI systems. Specifically, without consensus and shared understanding we risk talking across each other.
  - Notably, the Chinese approach here has included societal impacts in addition to the technical considerations in the use of AI in the military.

- Shared evaluation of each other's work: Even in trying times of geopolitical tension, one can embark on carefully selected initiatives to translate and interpret work being done by others in an attempt to develop a shared understanding. Taking the example of the USSR-US collaboration on the Apollo-Soyuz project during the Cold War stands as an example of how diplomacy can be advanced through scientific endeavors.
  - In particular, this has implications for the kind of collaboration that might take place between China and USA, the two major forces in the use of AI in a national security context. Translation of each other's work will help avoid misunderstanding.

- Utilizing Track 2 and Track 1.5 mechanisms in addition to primary channels to achieve diplomacy is an effective approach to diffusing tensions and discussing policies and security considerations that might be mired amongst other issues in Track 1 discussions.

**What is Track 1.5 and Track 2?**

At major policy negotiations and conferences, these tracks are venues where supplemental agendas are discussed, often with the presence of domain experts and those operating in assistance capacities to the official delegates. It is an avenue for advancing goals like the ones being discussed here that are sometimes nascent and not as immediately included in the primary agendas of the gathering.

**Actions that people can take**

- Creating shared standards for testing, evaluating, verifying, and validating (TEVV) of these systems to compare capabilities and limitations across deployments is essential.

An example benefit of this would be in judging whether the systems are adequately able to separate military and civilian targets. Also, the degree to which they are able to assure confidence in their results.

- While the inclusion of AI into nuclear security can have benefits in terms of higher precision in targeting, etc. we must also be conscious of destabilization because of inherent uncertainty in the use of these systems. This further strengthens the case for effective TEVV approaches to be used and adopted across countries.
- A shared understanding on the relative weighting of the false positives and false negatives by different regimes will also help to calibrate the abilities of the systems in their usage across different regions.

**Open sky revival for AI systems**

While the Open Sky treaty has faced considerable flak from the policy community and a recent announcement from the US represents an unfortunate development in the space. But, in terms of soft enforcement and monitoring, it is an essential mechanism for accountability. It also serves to reinforce a more representative understanding of the capabilities and limitations of AI from different countries.

**Better communication**

In a field like AI that prides itself on open-source and open-access policies in terms of research and development, we ought to extend this to the field of policy as well. There is some risk that such an initiative might be one-sided, but taking an iterative approach to building trust can assess the viability of such an approach.

**Lessons learned**

- Sometimes bilateral sessions have points of friction that are hard to overcome, utilizing multilateral fora can help ease those points of friction.
- Starting with small, concrete, tractable issues will help to incrementally build trust to tackle larger issues later on.
- Gathering a diverse set of stakeholders, appropriate for the stage of conversation is important rather than having a blanket set of people to approach and talk to.
- Having a high degree of transparency in the operation of these initiatives and their goals along with a firm expectation of reciprocity will also help in the success of these initiatives.
- Mitigating the risks of counter-intelligence are also important, especially for those who are invited to these fora.
- Track 2 conversations should become routine and tracking their efficacy through metrics and outcomes can help justify their existence rather than having them as one-off events.
- Related to the above point, having tight feedback loops between Track 1 and other tracks will help to keep each other abreast of the relevant issues and their severity.

- A discussion on the seriousness of issues, especially those that might not be raised at Track 1 in the service of achieving other goals shouldn't deter their discussion in other tracks. This will be essential for places where for example there might be human rights implications, say in the case of persecution of Uighurs in China aided by the use of facial recognition technology.

**Conclusion**

There are many shortcomings in the way AI safety is discussed at an international level at the moment and without more coordinated efforts that build on existing policy making and negotiation instruments, we risk creating a fragmented ecosystem that can lead to unintended consequences in terms of assessing each other's AI capabilities and mitigating the risks that arise from its use.

As a practitioner, this means that we have a responsibility to communicate the impacts of our work more clearly to those who might be involved in policy making, both at a domestic and international level. Specifically, I envision working with others to create a shared commons.

# Understanding the Capabilities, Limitations, and Societal Impacts of Large Language Models

([Original paper](#) by Alex Tamkin, Miles Brundage, Jack Clark, Deep Ganguli)
(Research summary by Abhishek Gupta)

**Overview:** This paper provides insights and different lines of inquiry on the capabilities, limitations and the societal impacts of large-scale language models, specifically in the context of the GPT-3 and other such models that might be released in the coming months and years. It also dives into issues of what constitutes intelligence and how such models can be better aligned with human needs and values. All of these are based on a workshop that was convened by the authors inviting participation from a wide variety of backgrounds.

---

**Technical capabilities and limitations**

There is no transparency around large-scale models (like the GPT-3) being used by corporations — we just don't know when they're the backbone of a given solution. And based on the challenges that have been encountered in terms of biases in the GPT-3 model, we need to demand that the datasets that these models are trained on be evaluated. We also need to demand that the companies take protective measures against some of the potential harms that they can cause.

**Scale leading to interesting results**

The paper points to how the simple idea of scaling the number of parameters and the amount of data ingested by the system led to a lot of interesting properties in the system which is functionally similar to the smaller GPT-2 model. The participants in the workshop pointed to similarities with the laws in physics and thermodynamics.

We talk a lot about interdisciplinarity between domains if we're to build responsible AI systems, but something that jumped from the discussion as presented in the paper was the requirement to have many different technical subdomains also working together so that the system can be fielded responsibly including those who manage the infrastructure that helps to run such large-scale models in the first place.

**What do we mean by intelligence and understanding?**

The current methodology for judging the performance of a system stems from various accuracy and performance metrics that seek to allocate a score based on how frequently the system is right. An important distinction that participants pointed out was when you have systems that are going to be used in important domains like healthcare, etc. we might need to reconsider if being "mostly right" is going to be adequate. Specifically, we would need to consider if there are important examples that the system is getting wrong and have higher penalties in those situations to align the system more with what matters to humans.

Another interesting point of discussion was whether this obsession with understanding is even important for exhibiting intelligence as they pointed to how a recent Scrabble champion in French knew very little French in the first place as trained himself solely for the competition.

Acquiring more from less data might require multimodal training datasets that could accelerate the learning acquired by the system.

**Alignment with human values**

While what constitutes human values itself varies across different cultures and regions, there was a strong emphasis on involving people from many different domains to gain a more holistic sense for this. In addition, working towards techniques that are explicitly tasked with ensuring the alignment in the first place was another requirement that participants emphasized as we venture into ever-larger models that have the potential to start exhibiting what some might call general intelligence.

**Societal impacts**

As GPT-3 has exhibited many different capabilities, taking basic task descriptions like the natural language description of a website and then generating the boilerplate code for that or in parsing legal documents and sharing summaries of them, the questions about the societal impacts of such models loom large.

**Controlled access**

A point made by the participants was that given that OpenAI has constrained access to the model behind an API, there is a higher degree of control in preventing misuses of the model. But, a related concern is that there is a lack of transparency when it comes to how such requests for access are processed in the first place and who is and who isn't denied access in the first place.

**Deploying such models**

Arguments for the societal benefits and harms can be made in many different fora, but if we are to adopt a more scientifically rigorous approach, we need to come up with more defined metrics against which we can benchmark such systems. While no organization can maintain a lasting advantage when it comes to such models, there is also a responsibility that such organizations should undertake in setting responsible norms that others can follow, especially when they are large organizations and they can dedicate efforts towards doing so.

**Potential for accelerating disinformation**

This has been one of the most cited harms when it comes to large language models that can produce coherent and persuasive text. To that end, investments in being able to discern between the two and helping people recognize them in the wild is also going to be an important skill that needs to be built up over time. Cryptographic methods that can ascertain the authenticity of content and verify its provenance might prove to be effective ways to ensure a healthy information ecosystem.

**Bias**

This has been demonstrated through many ad-hoc and rigorous studies that GPT-3 exhibits bias, some of which is a function of the text corpora that it is trained on.

A very important distinction that was made by the participants here was the emphasis that we must place in carefully defining what is biased and what isn't for large language models such as GPT-3 that might be used in wildly differing contexts and being inherently multi-use.

Red-teaming and combating bias through the use of expansive and publicly shared bias test suites were the two measures that caught my eye in the paper in addition to some of the other well-known approaches to detecting and mitigating biases.

**Economic impacts**

Expanding the access to the model through the API can become a way to gather signals in terms of how the use of the model is impacting people from an economic standpoint as it would help to surface new ways that people employ the model for automation.

**Future research directions**

For anyone that wants to keep an eye out for where the field might be headed, the research directions section of the paper provides some good jumping off points. The one that caught my eye here was creating a more thorough understanding of the threat landscape in a way that emphasizes more rigorous research in making these models more robust.

**Conclusion**

The paper provides a comprehensive but abbreviated overview of the current capabilities and limitations of large language models. More importantly, it provides potential areas of investigation to improve the stature of how we conduct research into the societal impacts of these models and how we might do better.

Staying abreast of how large-scale models are built, the challenges they face, and what we can do to perform better on that front is going to be an important consideration when thinking about which ethical measures to put in place.

In particular, being cognizant of the areas where such systems can fail will also be essential if you're to build more robust systems.

# Go Wide: Article Summaries (summarized by Abhishek Gupta)

### Three Mayors on Their (Very Real) Challenge to Silicon Valley's Dominance
(Original *OneZero* article by Alex Kantrowitz)

- What happened: With the pandemic making remote work the norm, many tech workers have moved to cities outside of San Francisco to benefit from lower costs of living and to perhaps be closer to family. Mayors from Austin, Madison, and Miami talk about what the challenges facing San Francisco were (not just astronomical rents and dominance of tech firms) and how their cities have offered respite from some of those challenges, including what they are doing to attract people to these cities.

- Why it matters: What was heartening to read here was the fact that there is so much diversity that is present in other cities as well and decentralizing technology development and innovation away from a few major hubs will not only help people get closer to problems that they are trying to solve, but also open their eyes up to different problems that tech can help solve rather than just doing so for the 1%.

- Between the lines: The rise of virtual spaces like Clubhouse (audio) and Twitter Spaces will perhaps help to bring back some of the serendipity that arises in the collisions that happen in Silicon Valley that make it an attractive place to work in. But beware the privacy implications.

### When AI Reads Medical Images: Regulating to Get It Right
(Original *Stanford HAI* article by Katharine Miller)

The role of the FDA will need to expand to address AI's role in healthcare since their current expertise revolves around drugs and hardware related to medical technology. In particular, akin to the idea of robustness in the field of machine learning security, the software needs to be able to reliably perform the task that it claims to do and signal to the medical practitioner in case it is about to fail or isn't sure of the best course of action.

In terms of the standards that should be followed, what is quite clear is that from a medical perspective, they should be developed by medical practitioners rather than being forced in by software manufacturers who may not be aware of all the best practices in the field and might be skewed towards advocating for things that benefit their products and services. Premature enshrining of differing standards and definitions for what is effective will lead to problems in benchmarking different systems and reinforcing the normative core of the system.

Medicine typically has a 4-stage approval process that looks at feasibility, capability, effectiveness, and durability. This stages the release and testing process to arrive at something that is generally going to work well and minimizes the case that there might be harm.

When it comes to the notion of reliability it is much easier to make a system that is usually right rather than very rarely wrong. That difference might be quite crucial when it comes to medicine because there is a direct impact on human lives. A final suggestion from the authors is that we need commons on which we can evaluate and benchmark these systems, having a third-party entity may be the best approach in this case.

## The Tech Industry Needs Regulation for Its Systemically Important Companies
([Original *Brookings* article](#) by Mark MacCarthy)

The 2008 financial crisis brought the term "too big to fail" in the minds of the public consciousness and rightly so when the maloperations in an industry decimated the global economy because of the interlinkages and systemic importance of the financial system. Some argue now that the same might be the case with our digital infrastructure as well which has become just as crucial as public utilities.

Precipitated by incidents in recent months, like the hacking of prominent Twitter accounts that led to a small pandemonium in the cybersecurity world, researchers and activists advocate adopting a systemically important designation for technology companies that should be held to higher standards when it comes to their cybersecurity and resilience to failure since they are enmeshed centrally within the rest of the operations of the global economy. We already have that with telecommunications as an example that form the underpinnings of successful operations for a lot of the other utilities. Bringing technology companies under a similar fold with the understanding that they are systemically important will offer protections that will bolster the resilience of other fields as well.

By doing so, we will develop the mindset that regulation is essential to these companies rather than it being an afterthought. Certainly, something to think about when the social media companies push for self-regulation.

## Mozilla Took the Internet's Vitals. And the Results Are Concerning.
([Original *OneZero* article](#) by Thomas Smith)

Always an intriguing read from the folks over at Mozilla, the Internet Health Report is a good pulse check on everything that is going right and wrong with the internet and how we interact with each other over it. With some unsurprising results, namely that more people have started using the internet than the previous year, aided in part by the pandemic. The concentration of technology development and use was also held by a small set of companies, mostly American and a couple of Chinese companies. In addition, their influence spreads beyond the direct assets they own since they provide the fundamental building blocks like cloud computing that is used by other services that are increasingly becoming an essential part of our lives.

The report also pointed out how the distribution of such access is highly uneven and subject to a lot of surveillance and control, especially noting how there was an Internet Shutdown on each day of the year, sometimes those that lasted weeks or months that allow human rights violations to occur unchecked and unexamined. The mass deployment of technologies like facial recognition further risks the rights of people and there are a lot of activists both inside and outside companies that are opposing such rollouts from taking place.

Ending the report on a more positive note, the Mozilla Foundation gives the internet a positive prognosis based on the fact that we have more people rising up to assert their rights and powers. For example, the workers doing gig work to demand more basic protections like sick days, and other workers' benefits that are offered to those who hold full-time employment. It comes down to collective action and perhaps the more aware we become of some of these issues, the more we will be able to articulate our demands so that we have a healthier ecosystem for all.

## Postmates Drivers Have Become Easy Prey for Scammers. And Drivers Say the Company's Not Helping
([Original *The Markup* article](#) by Dara Kerr)

- What happened: Postmates delivery agents are subject to precarious employment arrangements, as evidenced by the case highlighted in this article where an agent was scammed out of his weekly earnings by being tricked into revealing account information that allowed the scammer to drain the agent's earnings for that week. Other phishing incidents are also mentioned in the article which showcases how crooks have tried to take advantage of vulnerable workers during COVID with the spike in the number of delivery requests.

- Why it matters: Gig workers in such cases have highly volatile earnings and rely on that money to meet basic needs. When platform companies refuse to offer help and are opaque with their redressal process, it harms the trust that these workers have in the system, disempowering and discouraging them from participating and offering their services.

- Between the lines: The platforms need to offer better cybersecurity education to the workers so that they can protect themselves from phishing attacks. Alerts that share common ways such attacks are mounted can also help workers. Finally, having high levels of transparency will also aid in building trust in the platform, especially because workers don't have the means to pursue legal action in case they are defrauded.

### Where the Internet Was Delivered by a Donkey
([Original *Rest of World* article](#) by Zeyi Yang)

Talking about an innovation called the Ilimbox, this article dives into the details of how a sliver of the internet is downloaded into a small, tissue-box-sized device and lugged to some of the most remote parts of the world where internet access is a challenge. The device is able to store articles from Wikipedia and educational content from YouTube and then physically transported atop a donkey by an organization called the Internet Society to remote villages in Kyrgyzstan that don't have access to the internet or electricity. The country is particularly challenging from a geographic standpoint because of its mountainous terrain.

COVID-19 made this need to reach remote parts of the world even more acute as traditional education faced headwinds. What caught my attention here is that the volunteers running this effort chose to download content in the local language in addition to English and Russian to make it more accessible to the students in those places. But, it only constitutes a small amount of content because of the lack of translation of many articles on websites like Wikipedia. Hopefully, as NLP progresses even more, we'd be able to open up even more content in local languages to accelerate educational efforts, harnessing AI in a positive way and overcoming the dominance of a single-language internet.

## Digital Archives Show Their Hand
(Original *Data & Society* article by Nazelie Doghramadjian)

An article that really shines a bright light on how the digital world around us is constructed, especially as we need the digitization of physical objects to meet the voracious AI systems head-on. It is reminiscent of the work from Mary Gray's Ghost Work, on the invisible labor that goes into creating magical AI systems (as MC Elish calls them) whereby the companies work tremendously hard to erase human "prints" away from the digitized objects to give us a highly sanitized view of the artifacts that we become accustomed to seeing online.

The article centres on the discovery of thumbs and hands holding open pages as they are digitized and what that means to the reader when they encounter that in digital archives. Drawing on some nostalgia (at least for those of us who remember having physical library cards with records of who borrowed the books before us), today's readers are sometimes shocked to see such artifacts when in fact some of us actually used to enjoy knowing who had borrowed a particular book before us, if we knew them, and if that changed anything about how we might interact with it. It seems antiquated now of course, but there was a slight tinge of excitement in discovering marginalia (little notes in the margins), bookmarks, and other nick-nacks in a book that were left behind, perhaps accidentally, by previous readers of those books.

But, coming back to the impact that such digitization of archives has on our lives, it is important to account for and pay due consideration to those who put in the hard labor of making those accessible to us in the first place. Erasing labor through corrective digital mechanisms just so that we get a "clean" version is problematic because it obviates the very real humans who are behind all these wonders that we now get to enjoy and take for granted in our daily lives.

## Preparing for the Future of Work
(Original *Stanford HAI* article by Sachin Waikar)

Talking about how the digitization of the economy has enlarged the pie, the panelists at an event hosted by the Stanford HAI pointed out that the distribution of the benefits has rarely been equitable with most of the gains accruing to those who already hold the keys to capital and power.

One popular argument is that the transformation of labor has been slower than what sensationalist media would portray them to be. In addition, the role that the government will play in how AI gets deployed in the economy can't be underestimated: take the case of China where some uses of AI have led to surveillance use cases but in other places like India, some bureaucratic processes might become easier allowing more people to access government services.

An important point around the role of the media in portraying more realistic scenarios and also providing positive and negative examples to help people make informed decisions will be essential. In addition, having people with deep knowledge of AI integrated within different government functions will also help. When talking about regulation, the tech companies have been pushing to have "lighter" regulation but it is not yet clear what that entails. On a geopolitical level, there is a huge decoupling of China and the rest of the world which will make it much harder to arrive at shared standards and principles. This is another place where the future of work might be jeopardized if we don't take adequate action.

## Twitter Joins Facebook and Youtube in Banning Covid-19 Vaccine Misinformation

([Original *Vox* article](#) by Rebecca Heilweil)

2020 certainly was a year filled with tremendous opportunities for misinformation to thrive. With many global catastrophes taking place simultaneously, it might seem like a huge problem that has no clear solution. Yet, inaction isn't a strategy and Twitter finally joined the fray with other companies in their commitment to curb COVID-19 related misinformation on their platform. The hesitation on the part of the platform had been that they were still trying to figure out what the right strategy was in addressing the misinformation.

Their tactics include labeling disputed information and outright removal of egregious content. The application of this two-pronged strategy will be an experiment — we don't know yet how successful this is going to be.

Combating misinformation will be an important consideration when dealing with anti-vaxxers during a pandemic. The platforms have a responsibility to ensure that they don't become instruments delaying the vaccine deployment efforts worldwide. Adopting a strategy that isn't just based on content, and one that looks at topological qualities of the network might also be an approach that can help accelerate combat efforts by the platforms. In fact, there's [some research](#) being done by network scientists who are trying to find non-content based methods to combat misinformation on the platform.

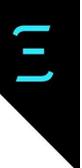

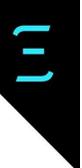

### How Social Media's Obsession with Scale Supercharged Disinformation
(**Original Harvard Business Review article** by Joan Donovan)

An unfortunate series of incidents finally precipitated action from the social media platforms that led to a ban on Donald Trump's account. It can be argued fairly well that this is a problem that has been building for years and this article makes a case for how the inherent structuring and incentives of social media platforms and the growth strategies that they adopted led to where we are today.

Achieving scale, a common Silicon Valley aspiration and venture capital requirement, meant that social media platforms optimized for anything that allowed them to bring on evermore users to the platform and keep them there through hacking their attention and evoking enough emotional responses that they wouldn't want to navigate away. Keeping the platform open and allowing anyone the ability to post content without much moderation in the service of meeting some of these "growth hacking" targets meant that user-generated content quickly grew to a scale where human moderation is no longer possible.

When advertisers realized the potential of these platforms, especially its political implications, not only did this bring in serious dollars but also boosted the prevalence of disinformation on the platform; to the point where it materially harmed the quality of the user experience. For years the platform companies evaded their moral responsibility to monitor the content on their platform. It all came to head in early January with the unfortunate and avoidable loss of lives as insurrectionists stormed the Capitol in the US. Hopefully, we have meaningful changes that get enacted to prevent such tragedies from occurring in the future.

### The High Price of Mistrust
(**Original *Farnam Street* article**)

While not specifically talking about the information ecosystem, I find that this article does a great job of laying down some of the fundamentals when it comes to addressing the problems that arise when we have a fragmented information ecosystem that is littered with problematic information and subsequently sows mistrust amongst people.

Drawing on economic theory, the article raises the point of the rise in transaction costs (and potential degradation of the user experience) as we have less and less to trust when people share information online. This comes in the form of a larger onus on the users to prove what they are saying is authentic and verifiable. The recent launch of BirdWatch from Twitter acts as a community policing mechanism, but it does add to that burden of interacting online. The article provides an example of how in the past you could rely on your neighbors for securing essential support and goods rather than having to purchase, say, your own tools since you could rely on generalized reciprocity, a concept like money that overcame some of the shortcomings of the barter system that required one-one immediate matching of needs for a transaction to take place.

In those smaller communities, there were also the mechanisms of reputation and repeated interactions that kept the levels of deviant behaviour to a minimum. In a world where we now have disposable identities online and the potential to engage in one-time interactions, it becomes much easier to flout those rules and social contracts in the favor of achieving perhaps narrow goals of causing harm in the near-term without the potential for any long-term consequences or accountability.

### Will Parler Prevail in Its Antitrust Case Against Amazon?
(Original *Knowledge @ Wharton* article)

One of the more definitive actions on behalf of large technology companies to curb the spread of hate speech online, the action by Amazon to de-platform Parler and revoke their access to use their cloud services put a halt to the pervasiveness of problematic information including hate speech that was spreading unchecked on Parler. It was swiftly followed by Parler trying to take legal action against Amazon alleging that they did so to favor Twitter over them and that they had a bias against conservatives.

Yet, as highlighted in the article here, the antitrust allegations are flimsy at best and require a strong backing by facts before the courts would consider the case as viable for adjudication. The defense that Amazon has mounted in the face of this case is that they had issued a sufficient number of warnings to monitor and remove speech that violated community guidelines.

Ultimately, this raises important questions on the power that large technology providers such as Amazon have and the responsibility that should go with that power (if there should be something to that effect) to help the technology ecosystem achieve a healthier posture. With lots of talk around Section 230 in the US and other pieces of legislation around the world, perhaps 2021 will be the year where there will be a clearer understanding of the roles and responsibilities that these companies have in helping us create an information ecosystem that supports our welfare and well-being rather than devolving into a toxic cesspool.

## Facebook Says "Technical Issues" Were the Cause of Broken Promise to Congress
(Original *The Markup* article by Alfred Ng, Leon Yin)

Through the Citizen Browser project, The Markup has been able to surface data on the level of recommendation activity from Facebook when it comes to political groups and their growing membership. This is something that was close to Election Day last year promised by Facebook as something that would be reduced given all the calls from Congress and the Senate in the US to ensure more fair elections, at least from an information diffusion standpoint.

But, as the results from this analysis show, Facebook has failed to deliver on that promise, pointing to technical issues as the primary culprit. The Citizen Browser is a unique initiative that gives insights into the kind of content that is recommended to people, something that is not possible to do otherwise because of the closed nature of the platform. While the results and the data from the study are linked in the article (and I encourage you to check them out), what is startling is that a small initiative like this from The Markup is able to unearth these problems and a well-funded team at Facebook is unable to spot and address them despite access to a large amount of data and resources.

Political polarization is a persistent problem that is harming the fundamental tenets of democracy and there isn't much that is being done by the platforms just yet in terms of weeding that out and creating a more healthy information ecosystem. There are justified concerns in terms of potentially snuffing out grassroots initiatives that use the platform to mobilize action - but at what cost? Perhaps, the inherent structure of the platforms and their associated incentives are the biggest culprits in the first place.

## How Covid-19 Overwhelmed a Leading Group Fighting Disinformation Online
(Original *Rest of World* article by Vittoria Elliott, Leo Schwartz)

- What happened: Debunk EU, a firm based out of Lithuania was heralded by a lot of people as the next stage in effectively fighting against disinformation using their hybrid AI-human approach that was supposed to scale to counter the threat of Russian troll and bot armies. But, with the pandemic in 2020 even they found themselves overwhelmed exposing how much more work is still required.
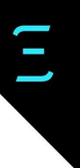

- Why it matters: What's noteworthy about this situation is how it serves as a good sandbox to see the effects of Russian-fueled disinformation given the relatively small size of the country and its history in dealing with such campaigns over many years. Yet, the citizens found themselves sharing that disinformation limiting the efficacy of the efforts of Debunk EU which focused more on foreign actors rather than domestic actors spreading problematic information.

- Between the lines: The disinformation problem is an inherently adversarial one and as I had highlighted here, there are many political and technical challenges to getting content moderation and platform governance right. AI-assisted efforts can help stem the flow of disinformation partially, but it is only the first line of defence in a long series of actions that need to be taken before we get to a healthier information ecosystem.

### Unadversarial Examples: Designing Objects for Robust Vision
(Original *Microsoft Research* article by Hadi Salman)

This research work flips the notion of adversarial examples on its head and talks about how objects might be designed in a manner that permits them to be robust to perturbation that can trigger misclassification.

Good design in general makes it easy for the intended audience to easily obtain the information that is required for them. Their approach allows one to make the objects more recognizable to the machines from a computer vision perspective. Think of how certain elements in nature are marked in a bright color to make them more recognizable. For example, brightly colored frogs indicate to predators that they are poisonous and should be approached with care. The motivation for making objects more recognizable is the increasing prevalence of computer vision systems in managing object detection and coordinating the activities of autonomous objects in our environment. Think about the case of a drone that is out of sight and needs to land on a pad in a place that has dusty or foggy conditions. It is not uncommon to lament that computer vision systems today aren't robust enough to operate independently in these conditions which partially restricts widespread use of these systems.

In experiments run by the researchers, they were able to apply "unadversarial" patches that made the landing pads much more obvious to the drone computer vision system making landings much more reliable. They also applied patterns and textures on cars and airplanes in a simulator that made them much more recognizable, reducing the rates of errors. Ultimately, this work has the potential for us to get much more reliable systems in practice. And reliability will be critical in the acceptance and trust from the user standpoint in the use of these systems where humans and machines co-exist.

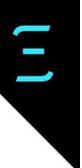

### Triggerless Backdoors: the Hidden Threat of Deep Learning
([Original *TechTalks* article](#) by Ben Dickson)

This article talks about a recent paper that looks at a new class of adversarial attacks that don't require explicit external triggers to get the model to behave in a way that the adversary wants.

Most of the current backdoor attacks rely on the fact that the adversary taints the training dataset in a way so that the model associates a certain type of example with particular target labels. The model behaves as normal for most cases, but when it comes across examples that were tainted, it gets triggered and behaves in a way that is deviant and serves the needs of the adversary.

While the classic backdoor attacks rely more so on "visible" interruptions in the data to trigger these sorts of behaviours, some argue that they might be more detectable by humans and are also more difficult to mount in practice in a physical context. This attack on the other hand relies on manipulating the dropout layers in the neural network and hence bakes in the deviant behaviour into the model architecture rather than relying solely on the data. But, this has some caveats like having even stronger adversary capabilities assumptions, probabilistic triggering of the deviant behaviour, and accidental triggering, though the paper does propose some guardrails against all of these.

While the article concludes that the attacks are a lot less feasible in practice, this approach definitely presents a new and interesting direction for research, which should lead to more robust AI systems in the long run.

### I Paid an AI to Get Me Instagram Followers. It Didn't Go Well
([Original Vice article](#) by Erik Galli)

The author of the article paid for a service to increase the number of Instagram followers for their account, which led to some pretty weird outcomes and created more hassles than not — antithetical to the role that automation is supposed to play in making our lives easier.

Because activities undertaken on your behalf by the automated agent operate largely outside of your purview, given the sophistication of the technique, there are potential negative consequences that can arise. In this case, the bot did some things that led to some nasty interactions with one of the author's exes on Instagram. This was despite the fact that there was a blacklist prohibiting certain actions. This is why effective guardrails are an important consideration in the fielding of AI systems.

This is also reminiscent of Google Duplex which was supposed to book hair salon appointments for us, sounding quite human in the process, to the extent of deceiving the other person into believing that they were interacting with another human. We are stuck in a place where such interfaces might be taking actions on our behalf without proper disclosure and violating well-established social norms and practices, potentially leading to more headaches and harm than good.

# 7. Community-Nominated Spotlights

[**Editor's Note:** The community-nominated highlights in this section are generous contributions, recommendations, and pointers from the global AI ethics community, with the goal of shedding light on work being done by people from around the world representing a diverse mix of backgrounds and research interests. We're always looking for more nominations, and you can nominate someone for our next report by emailing support@montrealethics.ai]

## Spotlight #1: Connor Leahy and Stella Biderman (Leaders of EleutherAI)

**The Hard Problem of Aligning AI to Human Values**

**A new paradigm in ML**

One of the most exciting recent developments in ML is the emergence of large, self-supervised pretrained language models (LMs), most notably OpenAI's GPT-3. These models are trained on large collections of text scraped from the internet, with the simple goal of predicting the next word from the previous text. This simple task may at first glance not seem likely to yield much beyond an academic curiosity, but in fact such models learn an astonishing variety of useful skills that are already being turned into useful applications. Models expanding this self-supervised pre-training regime to other modalities such as images are already emerging as well.

But with any new technology comes new negative side effects. The potential for misuse of large LMs, from the generation of hard-to-detect spam to reinforcing negative stereotypes, has been discussed in many outlets and academic papers, and solutions are clearly needed.

A common critique is that, since such models are trained on large, at best haphazardly curated, datasets, they learn to mindlessly parrot existing viewpoints and potentially negative or harmful content. Much of the content available on the free internet is not exactly the kind of content we would necessarily want to base society around. A common suggested mitigative strategy is to greatly increase the curation and filtering of training data used for such models. We believe this is not only not a solution, but implicitly covers over the actual hard questions that need addressing.

**Just the tip of the iceberg**

These issues are very real and very complex. We believe the complexity of these problems have, in fact, been severely underestimated by some practitioners in the field.

Our thesis is simple: If a model becomes dangerous by the mere exposure to unethical content, it is unacceptably dangerous and broken at its core. While gating such models (as OpenAI does with GPT3) behind an API with rudimentary automatic filters plus less rudimentary human moderation is a useful temporary patch, it does not address the underlying problem. These models are fundamentally not doing what we as humans want them to do, which is to act in useful, aligned ways, not just regurgitate an accurate distribution of the text they have been trained on. We need AI that is, like humans, capable of reading all kinds of content, understanding it, and then deciding to act in an ethical manner. Indeed, learning more about unethical ideologies should *enhance* one's ability to act ethically and fight such toxic beliefs.

In addition to the technical problem, there's also a sociological question that these conversations tend to omit: behind every deployed ML model is someone who said "this model is good enough to deploy." The determination that a model *is good enough to deploy* is the kind of act that can and should be judged on moral grounds. Ignoring this facet of the conversation moves criticism away from the people who actually have the power over what is and is not deployed. Technology companies want to have their cake and eat it too by painting the tools that they develop and sell for profit simultaneously as essential to modern life but also as grave dangers to society depending on which is more beneficial. For example, OpenAI wants you to believe that GPT-3 is too dangerous to let the public use, but not too dangerous to let Microsoft sell to companies for profit.

Jack Clark [makes a related point](#) in his newsletter ImportAI in the context of discussing of our ongoing project to produce GPT-3-type models, observing that Google hasn't publicly released models as large as the ones that we are building with the help of TFRC (a Google-sponsored initiative offering free TPU access to academics) but is supplying us with the compute to do so.

His succinct phrasing in his newsletter was:

> "Factories are opinions: Right now, it's as though Google has specific opinions about the products (software) it makes in its factories (datacenters), yet at the same time is providing unrestricted access to its factories (datacenters) to external organizations."

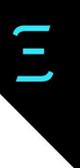

If we are to control the proliferation of dangerous technologies, we need to hold their creators accountable.

**An ambitious task ahead**

As laid out elegantly in [Alignment of Language Agents](#), a recent article on this topic from DeepMind, the challenges are immense. We do not currently have algorithms that are up to the incredibly difficult task of learning "good" behavior. But in our view, there is no acceptable alternative to this ambitious goal.

Of course questions of normative ethics have been a mainstay in human philosophical thought since time immemorial, and suggesting that clear, unambiguous answers could be achieved in this field seems extremely optimistic, to put it mildly. But as these technologies command an ever larger influence in our society, they must become more ethical, or the consequences will be disastrous. Finding robust ways to construct normative AI is, as Bostrom puts it, "philosophy on a deadline."

Proposals for how to conceptualize, nevermind formalize, such a problem are rare, and we do not claim to have solutions to this problem. But we believe that addressing the ambitious version of this problem, rather than various necessarily incomplete stopgap intermediate solutions, is the right path.

Some labs such as [Deep Mind](#) and [Open AI](#), and individual researchers such as [Abram Demski](#), [Vanessa Kosoy](#) and [Stuart Armstrong](#), have proposed concrete research agendas pointing towards the kinds of solutions necessary to align AIs to complex normative questions. More concrete techniques that might have promising qualities to make systems more interpretable and aligned that we find promising include [Imitative Generalization](#).

---

*Connor and Stella are leaders of [EleutherAI](#), an online distributed machine learning research collective. EleutherAI recently [published a 2.7B parameter GPT3-like model](#), the second largest open source autoregressive transformer, and significantly larger models are in training. EleutherAI is dedicated to not just creating, but more importantly studying such large self-supervised models and how to align them.*

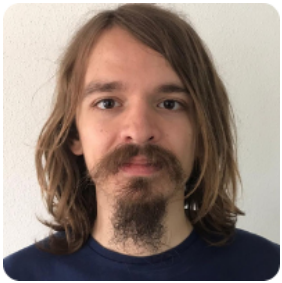

**Connor Leahy ([@NPCollapse](#))**

Connor works at [Aleph Alpha](#), where he researches empirical normative learning inspired by [Alignment by Default](#), [Steered Optimizers](#), [Model Splintering](#), and related work with a focus on near-future [model-based RL systems](#) and hopes to make progress both on practical techniques for understanding and [eliciting useful behavior from current models](#) and better conceptualizations of what the actual philosophical question is we're trying to solve, and how to approach it.

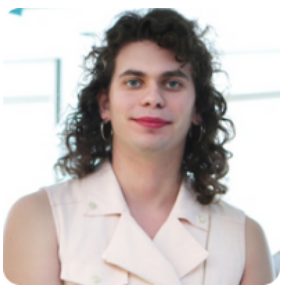

**Stella Biderman ([@BlancheMinerva](#))**

Stella's an industry researcher who devotes her free time to language modeling and alignment research. She leverages her background in logic and theoretical computer science to develop models of reasoning and decision-making, and methodologies for formal verification of AI technologies and research in strategic decision making. She is also a student in the College of Computing at the Georgia Institute of Technology where she is pursuing a Masters of Science in Computer Science and an organizer of the AI Village, a community of hackers and data scientists working to educate the world on the use and abuse of artificial intelligence in security and privacy.

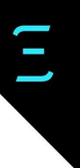

## Spotlight #2: Ruth Starkman (Lecturer, Stanford University)

**The gap my work fills in the AI Ethics discourse**

I'm an ethics and computer science educator with two equity projects.

First, I work with first-generation, low-income college students to ensure their success in computer science education and preparation for building better, more equitable algorithms. CS in higher education is becoming more inclusive, but there is still a need for curriculum and assessment revision, especially the high stakes testing that routinely benefits wealthier students, who've enjoyed more access to CS before college. Learning to code must become a more collaborative and inclusive experience for people from marginalized groups. From their arrival on campus to the tech interview in industry or grad school, first-gen students need to have better access to skill acquisition and more opportunities to ask questions which enable them to perform better on coding tests.

Second, I much admire my colleagues' efforts to teach coding and ethics together; not merely talking about ethical problems but engaging them in code.

For 30 years until 2015, I taught tech ethics by asking students to apply global frameworks to computing problems. These approaches produced great conversations about big issues, yet they rarely led to building better systems. Frameworks are important qualitative exercises, but students also need to investigate quantitative, actionable methods of determining ethical AI.

Now I ask students to develop new metrics for evaluation of their algorithms and build new projects that require them to rethink their code and write new datasheets, along the lines of Dr. Timnit Gebru's research before Google fired her. Getting engineers to debate and rebuild their code in specific ways that integrate the context of ethical discussions has been a much more productive mechanism. We can't just talk about the big issues in the abstract; we need to put them in relation to genuine coding projects. In addition, I encourage students to discuss ways for engineers to intervene and protect themselves and their peers who raise ethical questions, especially when they are working in industry but also in university contexts, which are hardly free from ethical challenges. Every tech company and every university should know that student and worker cooperation can be a formidable force of good.

**Why I was moved to do this work**

I hope to prepare all students, including engineers, for the ethical challenges they will encounter in their work lives. Teaching ethical frameworks and developing metrics help students tackle dilemmas, but the real obstacle remains the institution which employs them. I try to prepare students to help create a safer environment to question the algorithms they build. The fate of Google's AI Ethics team has proven that our best engineers, especially women of color, remain vulnerable when they speak up at their companies. Now is the time to create an atmosphere where hard ethical questions are embraced and those who ask them are supported and rewarded.

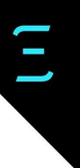

**Outlook for AI Ethics in 2021**

> We are a long way from safety for engineers who raise ethical questions. At the university and in the public sphere we need to make it clear to tech companies that if they silence their critics, they will encounter redoubled public pressure to address their ethical problems and protect those who speak up.

**Take action**

What needs to change:

- CS education is not yet inclusive for first-gens.

- The tech interview remains a hazing experience for people from marginalized groups.

- Most CS and ethics instructors are non-tenure line non-faculty at elite universities.

Fix these problems and more people from marginalized groups will become part of the AI workforce.

---

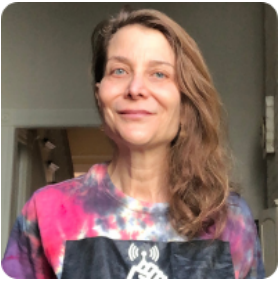

**Dr. Ruth Starkman (@ruthstarkman)**
Lecturer
Stanford University

Ruth Starkman is a nurse and engineer with PhDs in Comparative Literature and Philosophy. She teaches writing and ethics at Stanford University.

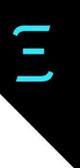

## Spotlight #3: Hubert Etienne (AI Ethics researcher, Facebook AI)

**The gap my work fills in the AI Ethics discourse**

My work intends less to fill a gap than to open a new path. The main questions discussed in AI ethics today are related to the potential threats and noxious applications of the technology; however, negative moral duties are only one part of ethics, and the fundamental question remains: "What is the right thing to do?" It is often difficult to identify the right decision, and reducing the scope of possibilities by using an apophatic definition can help avoid selecting the worst options. This, however, cannot alone satisfy our moral quest for doing the right thing(s).

I usually call 'defensive ethics' the moral reflection around the dangers of AI and 'offensive ethics' the investigation of how AI can be used in the right way to improve our world. My main research belongs to this second axis, which is not just about using AI 'for good' but building AI right for the right purpose.

I realized that there was a significant asymmetry between, on the one hand, the great complexity of machine learning models and the analytical methods used by computational social scientists and, on the other hand, the elementary reflection upon which data collection, labelling, and interpretation are based in the social sciences. This asymmetry is my workspace, as I am convinced that adding this other layer of complexity is necessary to develop more accurate and ethical models.

**Why I was moved to do this work**

I was never particularly interested in technology nor was I willing to make a significant change in the world before I started my Ph.D. I just wished to answer a few questions I had surrounding metaethics and action theories: Why do we do what we do? What prompts us to do it? Questioning our interactions with moral entities traditionally leads one from the investigation of moral agents to that of moral patients, and from animals to non-biological entities, especially robots.

By reading more transdisciplinary works, I also came to realize that philosophy alone could not provide me with satisfying answers but that once it was combined with relevant work in anthropology, literature, cybernetics, and neuropsychology, one may hope to build a more comprehensive paradigm to make sense of human interactions. After all, the best-performing machine learning models we have today are inspired by cognitive psychological findings!

I also have a profound respect for major thinkers and can easily take the needle when the thorny questions that they tried to address are reconsidered in a simplistic way by some computational scientists who assume that a large dataset, even when poorly built, is worth more than a brilliant philosophical argument. Raising awareness about the dangers of such reasoning was the primary goal of my research on the ethics of autonomous vehicles.

**Outlook for AI Ethics**

I believe at least three main factors underlie the development of AI ethics, and can serve as relevant forebearers of things to come.

The first is regulation. I believe that the eventual transcription of ethics principles into law should allow ethicists to pass the responsibility for dealing with issues such as privacy or transparency to legal scholars – as these concepts are primarily legal not philosophical – and thus focus on deeper questions.

The second is education. Tech employees constitute a key population and should be encouraged to distance themselves from their work, question their practices, and ensure we maintain an effective ethical pressure over the bigger structures. They have the information, the position, and from what I see also the genuine interest. It is now on ethicists to help them develop a sharper awareness to become engaged workers or whistle-blowers.

The third dimension is the market: we need to associate AI ethics with a business rationale so that ethical behaviour becomes, at the least, a strong competitive advantage, if not a condition *sine qua non* to survive on the market. If we fail here, AI ethics may just end up enriching the slogans and pictures that companies add on their prospectus to illustrate all the good they do during their yearly charity day.

**Take action**

I would like to request the community's support in spreading awareness around the dangers associated with the deployment of autonomous vehicles (AVs) and the clear lack of neutrality of many academics and industry experts concerning AVs and AV ethics. Such an instrumentalization of the ethical discourse for the purpose of avoiding regulation and distracting people from the real issues is unacceptable. The more I teach, the more I realize that research advances while people, even AI ethics experts, remain largely unaware of what is happening in this industry. Any help would be appreciated and relevant resources about this topic can be found on my research page ([www.hubert-etienne.com](www.hubert-etienne.com)).

---

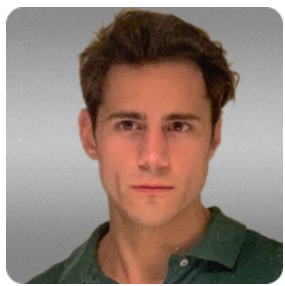

**Hubert Etienne**
AI Ethics Researcher
Facebook AI

Hubert Etienne is a French philosopher conducting research in AI ethics and computational social sciences at Facebook AI and Ecole Normale Supérieure. He is a lecturer in data economics at HEC Paris, a lecturer in AI ethics at Sciences Po, ESCP Europe and Ecole Polytechnique, as well as a research associate at Oxford University's Centre for Technology and Global Affairs.

# Closing Remarks

That was a whirlwind tour of what the staff at MAIEI found to be the most interesting highlights in research and reporting in AI ethics this past quarter!

I hope you had as much fun reading this report as we did in putting it together. It is truly a labor of love supported by our community who help us make such a resource open and accessible to all. The staff at MAIEI is persistently hard at work, finding topics, discussions, and resources that are underexplored in the domain today and those that demand higher scrutiny. We hold ourselves accountable to our community for all the work we do and strive to co-design and co-create accessible resources that lower the barrier to meaningful participation from all walks of life in this very important turning point in history as AI systems gain more widespread use.

In this edition, I was particularly touched by the community spotlights because they showcase how we are all working hard on varying aspects of these challenges and even though we might be miles away from each other, there is a sense of kinship that ties us all together. We invite you to be a part of this journey with us by signing up for our weekly newsletter The AI Ethics Brief.

If you have found a piece of literature, topic, discussion, or individual that you think should be included in the next report or have any other feedback for us, please don't hesitate in reaching out to us at support@montrealethics.ai.

Until next time, I wish you all the best in your journey towards making ethical, safe, and inclusive AI a reality. See you back here next quarter!

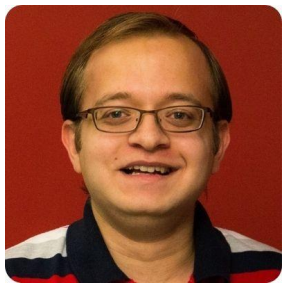

**Abhishek Gupta (@atg_abhishek)**
Founder, Director, & Principal Researcher,
Montreal AI Ethics Institute

Abhishek Gupta is the founder, director, and principal researcher at the Montreal AI Ethics Institute. He is also a machine learning engineer at Microsoft, where he serves on the CSE Responsible AI Board.

# Support Our Work

The Montreal AI Ethics Institute is committed to democratizing AI Ethics literacy. But we can't do it alone.

Every dollar you donate helps us pay for our staff and tech stack, which make everything we do possible.

With your support, we'll be able to:

- Run more events and create more content
- Use software that respects our readers' data privacy
- Build the most engaged AI Ethics community in the world

Please make a donation today at **montrealethics.ai/donate**.

We also encourage you to sign up for our weekly newsletter *The AI Ethics Brief* at **brief.montrealethics.ai** to keep up with our latest work, including summaries of the latest research & reporting, as well as our upcoming events.

If you want to revisit previous editions of the report to catch up, head over to **montrealethics.ai/state**.

Please also reach out to **Masa Sweidan** **masa@montrealethics.ai** for providing your organizational support for upcoming quarterly editions of the *State of AI Ethics Report.*

**Note:** All donations made to the Montreal AI Ethics Institute (MAIEI) are subject to our **Contributions Policy**.